\documentclass[12pt, letterpaper]{article}

\usepackage{amssymb}
\usepackage{amsmath}
\usepackage{verbatim}

\usepackage[nosort]{cite}
\usepackage[hyperref,bulletsep]{collect}

\mathversion{normal}

\makeatletter \@addtoreset{equation}{section} \makeatother

\newcommand{\gen}[1]{\mathfrak{#1}}
\newcommand{\alg}[1]{\mathfrak{#1}}
\newcommand{\superN}{\mathcal{N}}
\newcommand{\fldZ}{\mathcal{Z}}
\newcommand{\Real}{\mathbb{R}}
\newcommand{\sg}{\tilde{S}_{1,1}}

\ifx\genfrac\sdflkaj

\else

\fi
\newcommand{\sfrac}[2]{{\textstyle\frac{#1}{#2}}}
\ifx\half\sdflkaj
\newcommand{\half}{\sfrac{1}{2}}
\fi
\newcommand{\quarter}{\sfrac{1}{4}}

\newcommand{\comm}[2]{[#1,#2]}
\newcommand{\acomm}[2]{\{#1,#2\}}
\newcommand{\bigbrk}[1]{\bigl(#1\bigr)}

\newcommand{\state}[1]{\mathopen{\big|}#1\mathclose{\bigr \rangle}}
\newcommand{\bigstate}[1]{\mathopen{|}#1\mathclose{\rangle}}
\newcommand{\costate}[1]{\mathopen{\langle}#1\mathclose{|}}

\let\oldPhi=\Phi
\let\oldPsi=\Psi
\renewcommand{\Phi}{\mathnormal{\oldPhi}}
\renewcommand{\Psi}{\mathnormal{\oldPsi}}

\newcommand{\nln}{\nonumber\\}
\newcommand{\nl}{\nonumber\\&&\mathord{}}
\newcommand{\earel}[1]{\mathrel{}&#1&\mathrel{}}
\newcommand{\eq}{\earel{=}}
\newenvironment{myeqnarray}{\arraycolsep0pt\begin{eqnarray}}{\end{eqnarray}\ignorespacesafterend}
\newenvironment{myeqnarray*}{\arraycolsep0pt\begin{eqnarray*}}{\end{eqnarray*}\ignorespacesafterend}

\def\[{\begin{equation}}
\def\]{\end{equation}}
\def\<{\begin{myeqnarray}}
\def\>{\end{myeqnarray}}

\ifx\href\asklfhas\newcommand{\href}[2]{#2}\fi
\newcommand{\arxivno}[1]{\href{http://arxiv.org/abs/#1}{#1}}

\begin{document}
\thispagestyle{empty}
\begin{flushright}\footnotesize
\texttt{\arxivno{arXiv:0806.1786}}\\ 
\texttt{DAMTP-2008-52}
\end{flushright}
\vspace{.5cm}

\begin{center}%
{\Large\textbf{\mathversion{bold}%
Iterative Structure of the  $\superN = 4$ SYM Spin Chain}\par}\vspace{1.5cm}%

\textsc{Benjamin~I.~Zwiebel} \vspace{8mm}

\textit{ DAMTP, Centre for Mathematical Sciences \\ 
University of Cambridge\\%
 Wilberforce Road, Cambridge CB3 0WA, UK}%
\vspace{4mm}

\texttt{b.zwiebel@damtp.cam.ac.uk}\par\vspace{1.5cm}

\textbf{Abstract}\vspace{7mm}

\begin{minipage}{14.7cm}

We develop algebraic methods for finding loop corrections to the $\mathcal{N}=4$ SYM dilatation generator, within the noncompact $\alg{psu}(1,1|2)$ sector. This sector gives a 't Hooft coupling $\lambda$-dependent representation of $\alg{psu}(1,1|2) \times \alg{psu}(1|1)^2$. At first working independently of the representation, we present an all-order algebraic ansatz for the $\lambda$-dependence of this Lie algebra's generators. The ansatz solves the symmetry constraints if an auxiliary generator, $\gen{h}$, satisfies certain simple commutation relations with the Lie algebra generators. Applying this to the $\alg{psu}(1,1|2)$ sector leads to an iterative solution for the planar three-loop dilatation generator in terms of leading order symmetry generators and $\gen{h}$, which passes a thorough set of spectral tests. We argue also that this algebraic ansatz may be applicable to the nonplanar theory as well. 
\end{minipage}

\end{center}

\newpage
\setcounter{page}{1}
\renewcommand{\thefootnote}{\arabic{footnote}}
\setcounter{footnote}{0}



\section{Introduction}
While AdS/CFT provides a powerful weak-strong duality, finding the weak- to strong-coupling interpolation of unprotected physical quantities generically remains a very difficult problem. However, for planar $\mathcal{N}=4$ SYM and its string theory dual, integrability \cite{Minahan:2002ve,Bena:2003wd} leads to great simplifications. Here finding anomalous dimensions of single-trace local operators is equivalent to the spectral problem of an integrable spin chain \cite{Minahan:2002ve,Beisert:2003tq,Beisert:2003yb}. Due to integrability, the spectral problem at weak and strong coupling  can be reduced to solving a system of Bethe equations \cite{Beisert:2004hm, Arutyunov:2004vx, Staudacher:2004tk, Beisert:2005fw}. In fact, superconformal symmetry fixes the asymptotic Bethe equations up to an overall phase \cite{Beisert:2005tm, Arutyunov:2006ak}, which is constrained by crossing symmetry \cite{Janik:2006dc}. Following a proposal for the phase at large $\lambda$ \cite{Beisert:2006ib}, an all-order solution for the phase was found \cite{Beisert:2006ez} simultaneously and completely consistently with a four-loop gauge theory calculation of the cusp anomalous dimension \cite{Bern:2006ew}.  

Now through an integral equation \cite{Eden:2006rx,Beisert:2006ez}, the asymptotic Bethe equations apparently give the planar cusp anomalous dimension's  interpolation from weak to strong coupling  \cite{Benna:2006nd,Kotikov:2006ts,Cachazo:2006az, Alday:2007qf,Kostov:2007kx,Beccaria:2007tk, Roiban:2007jf,Casteill:2007ct, Basso:2007wd, Roiban:2007dq,Belitsky:2007kf,Kostov:2008ax}. The asymptotic Bethe equations also pass multiple tests in the near-flat-space limit \cite{Maldacena:2006rv,Klose:2007rz, Puletti:2007hq}. Furthermore, the asymptotic spectrum of BPS bound states \cite{Dorey:2006dq} is consistently reflected by the analytic structure of the phase \cite{Dorey:2007xn, Dorey:2007an}.  Finally, recent work has focused on the scaling function for the minimal anomalous dimensions of long operators with  Lorentz spin growing exponentially with twist \cite{Belitsky:2006en,Frolov:2006qe}. At strong coupling, the scaling function (in a more specialized limit) can be computed using a relation to the $O(6)$ sigma model \cite{Alday:2007mf}, as has been checked at two loops \cite{Roiban:2007ju}. From the asymptotic Bethe equations,\cite{Freyhult:2007pz} derived a generalized integral equation for the scaling function, which interpolates from weak to strong coupling \cite{Basso:2008tx} in perfect agreement with the previous results. For additional related work see \cite{Bombardelli:2008ah,Fioravanti:2008rv,Fioravanti:2008ak,Gromov:2008en}.

Despite these impressive results, there are questions that remain challenging even for the Bethe ansatz approach. Integrability is an assumption, and it seems that other methods will be required to verify that integrability is preserved by quantum corrections for all values of $\lambda$. Also, finite-length corrections are required both at strong \cite{Arutyunov:2003za,Park:2005ji,Schafer-Nameki:2005is,SchaferNameki:2006ey} and weak coupling \cite{Kotikov:2007cy}. These corrections are potentially addressable via thermodynamic Bethe ansatz methods \cite{Ambjorn:2005wa,Arutyunov:2007tc}. For recent studies related to the giant magnon \cite{Hofman:2006xt}, see \cite{Arutyunov:2006gs,Janik:2007wt,Hatsuda:2008gd,Minahan:2008re,Heller:2008at}. An alternative approach uses algebraic curve technology \cite{Gromov:2008ie}. However, even with recent progress the finite-length corrections still present a great challenge.

These considerations encourage additional approaches to the AdS/CFT spin chain. In this work, we will give evidence that the spin chain Hamiltonian\footnote{Gauge theory dilatation generator} and the other local spin-chain symmetry generators\footnote{Gauge theory superconformal symmetry generators} provide a promising direction for new insights.

Initially, a spin chain Hamiltonian-based approach appears daunting. Beyond the elegant complete one-loop gauge theory dilatation generator \cite{Beisert:2003jj}, the dilatation generator is known only to finitely many loops in subsectors. For compact sectors with finite-dimensional representations on each spin chain site, this includes the planar $\alg{su}(2|3)$ sector dilatation generator to three loops \cite{Beisert:2003ys} and the planar $\alg{su}(2)$ sector dilatation generator to four loops \cite{Beisert:2007hz}.  While these results for compact sectors have given essential input or verification for the Bethe ansatz, they reveal no simple structure for the Hamiltonian.  Furthermore, for noncompact sectors with infinite-dimensional modules on each spin-chain site it seems impossible to extend direct field theory calculations of the dilatation generator, such as \cite{Belitsky:2005bu}, beyond low-loop order. 

 Despite these apparent challenges,  the Hamiltonian-based approach is a viable path even for noncompact sectors because superconformal symmetry provides powerful constraints. Lie algebra (superconformal) constraints in $\mathcal{N}=4$ SYM have already been shown to be very strong in multiple cases. Above we referred to the most famous example of the AdS/CFT S-matrix, which is fixed up to an overall phase by two copies of extended $\alg{psu}(2|2)$ symmetry. Also, combined with some basic properties of Feynman diagrams, Lie algebra constraints completely fix the planar dilatation generator at three loops within the $\alg{su}(2|3)$ sector \cite{Beisert:2003ys}. In this work, we will ultimately conclude that the same statement likely applies at three loops even within the noncompact $\alg{psu}(1,1|2)$ sector.

Previous work \cite{Zwiebel:2005er} showed that the $\alg{psu}(1,1|2)$ sector Lie algebra representation has simple iterative structure at next-to-leading order. This has two main parts. First, the NLO corrections to $\alg{psu}(1,1|2)$ sector symmetry generators are  commutators of the leading order generators with an auxiliary generator\footnote{We have changed the normalization of some generators as well as notation, as explained in Section \ref{sec:alg} and Appendix \ref{sec:notation}.}:
\[
\frac{1}{\lambda} J_{\text{NLO}} = \pm \comm{J_{\text{LO}}}{\gen{X}} \label{eq:oldxcom},
\]
where $J_{\text{NLO}}$ includes a factor of $\lambda$ and the sign of the commutator is different for generators corresponding to positive or negative Lie algebra roots.  
Second, $\gen{X}$ is built iteratively from certain leading order supercharges, $\hat{\gen{Q}}^{\mathfrak{a}}$ and $\hat{\gen{S}}^{\mathfrak{a}}$, and an auxiliary generator $\gen{h}$, 
\[
\gen{X} = \half \varepsilon_{\mathfrak{ab}} \, \acomm{\hat{\gen{Q}}^\mathfrak{b}_{\text{LO}}}{\comm{\hat{\gen{S}}^{\mathfrak{a}}_{\text{LO}}}{\gen{h}}} + \text{ h.c.} \label{eq:olddefinex}
\]

We argue here that this is not a low order accident. The key result of this work is that \emph{algebraically}, this next-to-leading order solution naturally lifts to a consistent all-order solution. We simply replace (\ref{eq:oldxcom}) with an equation that is continuous in $\lambda$, 
\[
\frac{\partial}{\partial \lambda} J(\lambda) = \pm \comm{J(\lambda)}{\gen{X}(\lambda)}. \label{eq:xcom}
\]
In other words, $\gen{X}(\lambda)$ generates (plus/minus) translations in $\lambda$ for the local spin-chain symmetry generators\footnote{Actually only for raising or lowering generators. The action on the Cartan generator corresponding to the dilatation generator is given implicitly by commutators of the off-diagonal generators. The remaining Cartan generators are $\lambda$-independent.}. Correspondingly, the leading order result (\ref{eq:olddefinex}) lifts to the continuous version
\[
\gen{X}(\lambda) = \half \varepsilon_{\mathfrak{ab}} \, \acomm{\hat{\gen{Q}}^\mathfrak{b}(\lambda)}{\comm{\hat{\gen{S}}^{\mathfrak{a}}(\lambda)}{\gen{h}(\lambda)}} + \text{ h.c.} \label{eq:intodefinex}
\]
Note that $\gen{h}$ is also a function of $\lambda$, and the leading order expression (\ref{eq:olddefinex}) depends only on $\gen{h}(0)$. While we do not find an explicit algebraic solution for $\gen{h}$, we present two simple Serre-relation-like equations that are linear in $\gen{h}(\lambda)$. When satisfied, these equations ensure that all Lie algebra symmetry constraints remain satisfied after shifts in $\lambda$ generated by $\gen{X}$.  As anticipated in \cite{Beisert:2007sk}, the $\alg{su}(2)$ automorphism $\gen{B}$ of $\alg{psu}(1,1|2)$ plays a key role; the supercharges $\hat{\gen{Q}}^{\mathfrak{b}}$ and $\hat{\gen{S}}^{\mathfrak{a}}$ are doublets with respect to the $\gen{B}$ automorphism, and $\gen{X}$ consists of a $\gen{B}$-singlet combination.

Consequently, obtaining the perturbative corrections to the local symmetry generators reduces to a simple iterative procedure. At each step in the iteration (at each loop), we must solve the two equations for the next correction to $\gen{h}$. Substituting this solution into (\ref{eq:intodefinex}) gives the next contribution to $\gen{X}$. Then the next correction to the spin-chain symmetry generators follows from  integrating (\ref{eq:xcom}). This procedure gives the corrections to the dilatation generator completely in terms of leading order supercharges and $\gen{h}$.

The second main result of this work is a solution for $\gen{h}$ at next-to-leading order, which is sufficient to give a proposal for the three-loop planar $\alg{psu}(1,1|2)$ sector dilatation generator. This proposal passes thorough spectral tests, which is strong evidence that this is the correct field theory solution. Moreover, we find a simple homogeneous solution for $\gen{h}$ corresponding to the first nontrivial contribution from the Bethe ansatz phase, which appears at four loops in gauge theory. These results lead us to conjecture that the planar gauge theory spin chain realizes the \emph{all-order} algebraic proposal, at least asymptotically.

 Because the all-order solution is algebraic, the next two sections of this work do not assume a specific representation. In Section \ref{sec:alg} we review the extended algebra associated with this sector. In Section \ref{sec:solution} we present the all-order algebraic solution, and prove that it is in fact a solution. We apply this algebraic ansatz in Section \ref{sec:representation} to give the  representation for the $\alg{psu}(1,1|2)$ sector up to NNLO for the symmetry generators, including the three-loop planar dilatation generator. This section also includes the solution for $\gen{h}$ corresponding to the leading phase contribution.  The all-order proposal may be general enough even for the complete nonplanar symmetry generators, as described also in Section \ref{sec:representation}. Section \ref{sec:verification} is devoted to verifying our  proposal for the planar three-loop dilatation generator. As explained in Section \ref{sec:addalg}, it is possible to give two generalizations of the algebraic solution, still keeping the essential idea of some generator(s) of translations in $\lambda$, but not requiring the generator(s) to be built iteratively as in (\ref{eq:intodefinex}). These generalizations are not presented until then because they are not needed for  the planar $\alg{psu}(1,1|2)$ sector, at least not until four loops.  Finally, Section \ref{sec:conc} summarizes our results and discusses many questions that follow naturally. Appendix \ref{sec:notation} reviews the restriction of the full theory to the $\alg{psu}(1,1|2)$ sector and relates the notation of this work to that of the previous works  \cite{Zwiebel:2005er,Zwiebel:2007th,Beisert:2007sk}, and Appendix \ref{sec:Chevalley} gives the Chevalley-Serre basis for the Lie algebra. Some details of the proof of section \ref{sec:solution} are relegated to Appendix \ref{sec:extradetails}, the complete solution for $\gen{h}$ at NLO is given in Appendix \ref{sec:h1solution}, and Appendix \ref{sec:homsolution} presents a class of homogeneous solutions for $\gen{h}$ at NLO.

\section{The algebra \label{sec:alg}}
We begin with a review of the extended subalgebra of $\alg{psu}(2,2|4)$ that acts within the $\alg{psu}(1,1|2)$ sector, $\alg{u}(2) \ltimes \alg{psu}(1,1|2) \times \alg{psu}(1|1)^2 \ltimes \Real $. We use the notation of\cite{Beisert:2007sk}, with two key changes. First, this algebra admits a further extension by a triplet of central charges. Since these central charges vanish for the gauge theory, for simplicity we will rarely discuss them in this work\footnote{They are included in Appendix \ref{sec:Chevalley}, where they are required for the complete Chevalley-Serre basis of the algebra. Also, as discussed in Section \ref{sec:conc}, these extra central charges possibly would be useful for obtaining a natural embedding of the all-order algebraic solution.}.  Second, for the gauge theory representation of this algebra we rescale the $\alg{psu}(1|1)^2$ generators by $4 \pi g$ in order that all generators expand in even powers of $g$. Therefore, instead of $g$, we will mostly use the 't Hooft coupling, $\lambda = (4 \pi g)^2$.\footnote{In Section \ref{sec:verification} we switch back to $g$ to simplify checks of anomalous dimensions at three loops.} While the explicit gauge theory representation does not appear until Section \ref{sec:representation}, $\lambda$ enters the algebra relations through a relationship between the $\alg{psu}(1,1|2)$ and $\alg{psu}(1|1)^2$ subalgebras, as explained below.

\subsection{Lie algebra generators}
The extended algebra includes 
\begin{itemize}
\item A $\alg{u}(2)$ automorphism generated by $\alg{su}(2)$ generators $\gen{B}^{\mathfrak{ab}}=\gen{B}^{\mathfrak{ba}}$ and a $\alg{u}(1)$ generator $\gen{L}$
\item The nonextended $\alg{psu}(1,1|2)$ algebra generated by
\begin{itemize} 
\item $\gen{R}^{ab}= \gen{R}^{ba}$, which generate an $\alg{su}(2)$ subalgebra
\item $\gen{J}^{\alpha \beta} = \gen{J}^{\beta \alpha}$, which generate an $\alg{su}(1,1)$ subalgebra
\item 8 supercharges $\gen{Q}^{a \beta \mathfrak{c}}$, which transform  as doublets with respect to $\gen{R}^{ad}$, $\gen{J}^{\beta\epsilon}$ and $\gen{B}^{\mathfrak{cf}}$
\end{itemize}
\item Two $\alg{psu}(1|1)$ algebras, generated by $( \hat{\gen{Q}}^<, \, \hat{\gen{S}}^> )$  and $( \hat{\gen{Q}}^>, \, \hat{\gen{S}}^< )$. These supercharges transform as doublets under $\alg{su}(2)_\gen{B}$, $ \hat{\gen{Q}}^\mathfrak{a}$ and $ \hat{\gen{S}}^\mathfrak{b}$.
\item  $\mathcal{H}$, a shared central charge of $\alg{psu}(1|1)^2$ and $\alg{psu}(1,1|2)$
\end{itemize}
Note that we use Latin indices (or 1, 2) for $\alg{su}(2)_\gen{R}$, Greek indices (or +, -) for $\alg{su}(1,1)_\gen{J}$, and Gothic indices (or $<$, $>$) for $\alg{su}(2)_\gen{B}$. 

Throughout this work $\gen{L}$, $\gen{R}^{ab}$ and $\gen{B}^{\mathfrak{ab}}$ will be $\lambda$-independent. However, the following generators will be functions of $\lambda$:
\[
\gen{J}^{\alpha \beta}(\lambda), \, \gen{Q}^{a\beta\mathfrak{c}}(\lambda), \, \hat{\gen{Q}}^{\mathfrak{a}}(\lambda), \,   \hat{\gen{S}}^{\mathfrak{a}}(\lambda), \, \mathcal{H}(\lambda).
\]
Importantly, for the gauge theory representation  $\mathcal{H}$ is identified with the anomalous part of the dilatation generator, $\delta \gen{D}$. We therefore relate these two generators and the Cartan element of the $\alg{su}(1,1)$ subalgebra $\gen{J}^{+-}$,
\[ 
\lambda \, \mathcal{H}(\lambda) = \delta \gen{D}(\lambda), \quad \gen{J}^{+-}(\lambda)=\gen{J}^{+-}(0) +  \half \delta \gen{D}(\lambda). \label{eq:sharedcentralcharge}
\]
The factor of $\lambda$ appearing in the first equation is a consequence of the rescaling mentioned above. Anticipating perturbative expansions of later sections, we will write $\gen{J}^{+-}_0$ for $\gen{J}^{+-}(0)$, and use the subscript 0 similarly for other generators.
  
The non-Cartan generators can be split into raising and lowering generators. We use $J^+$  to represent
any of the following eight generators of the algebra, 
\[
\gen{R}^{11}, \, \gen{J}^{++}, \, \gen{Q}^{a+\mathfrak{b}}, \, \hat{\gen{Q}}^{\mathfrak{c}},
\]
and $J^-$  to represent any of the conjugate generators,
\[
 \gen{R}^{22}, \,\gen{J}^{--}, \, \gen{Q}^{a-\mathfrak{b}}, \, \hat{\gen{S}}^{\mathfrak{c}}. 
\]
In the gauge theory one encounters highest-weight representations. Then the highest-weight states are annihilated by all of the $J^-$, and descendants are generated using the $J^+$. According to this convention, the $J^+$ are the \emph{lowering} generators.

\subsection{Commutation relations \label{sec:comm}}
The $\alg{u}(1)$ automorphism generator $\gen{L}$ commutes with all but the $\alg{psu}(1|1)^2$ generators, which instead transform with charge $\pm 1$,
\[
\comm{\gen{L}}{\hat{\gen{Q}}^{\mathfrak{a}}}= \hat{\gen{Q}}^{\mathfrak{a}}, \quad \comm{\gen{L}}{\hat{\gen{S}}^{\mathfrak{a}}}= -\hat{\gen{S}}^{\mathfrak{a}}.
\]

Since all of the generators ($\gen{B}$, $\gen{R}$, $\gen{J}$) associated with the rank-one subalgebras have canonical transformation rules, we describe these commutators all at once. Let $J^{AB}$ be any $\alg{su}(2)$ or $\alg{su}(1,1)$ generator, and let $X^C$ be any generator carrying a single index with respect to the $\alg{su}(2)$ or $\alg{su}(1,1)$. Then 
\[
\comm{J^{AB}}{J^{CD}}=\varepsilon^{CB} J^{AD}-\varepsilon^{AD} J^{CB}, \quad \comm{J^{AB}}{X^C}=\half \varepsilon^{CA} X^{B} + \half \varepsilon^{CB} X^{A}.
\]

Since the extended algebra is a product, $\alg{psu}(1,1|2)$ and $\alg{psu}(1|1)^2$ generators commute. Also, due to the shared central charge (\ref{eq:sharedcentralcharge}), $\gen{J}^{+-}_0$ commutes with all $\alg{psu}(1|1)^2$ generators, including $\mathcal{H}(\lambda)$.  To complete the description of the nonvanishing commutators we need only to specify the anticommutators of the supercharges within each subalgebra. For $\alg{psu}(1,1|2)$ they are
\[
\acomm{\gen{Q}^{a\gamma\mathfrak{e}}(\lambda)}{\gen{Q}^{b\delta\mathfrak{f}}(\lambda)} =  \varepsilon^{\gamma\delta}\varepsilon^{\mathfrak{ef}} \gen{R}^{ab} - \varepsilon^{ab}\varepsilon^{\mathfrak{ef}} \gen{J}^{\gamma \delta}(\lambda),
\]
and the $\alg{psu}(1|1)^2$ relations are
\[
\acomm{\hat{\gen{Q}}^\mathfrak{a}(\lambda)}{\hat{\gen{S}}^\mathfrak{b}(\lambda)}=\half \varepsilon^{\mathfrak{ab}} \mathcal{H}(\lambda). \label{eq:psu11squaredalg}
\]
Note that anticommutators between the $\hat{\gen{Q}}$ vanish, as do those between the $\hat{\gen{S}}$.

Also, see Appendix \ref{sec:Chevalley} for the Chevalley-Serre basis of the $\alg{psu}(1,1|2)$ subalgebra.

\section{The all-order algebraic solution \label{sec:solution}}

Here we present the algebraic proposal and prove that it yields representations of the extended algebra described in the previous section.

\subsection{Details of the proposal}
Our ansatz for the solution involves three main steps.

\paragraph{I. Generator of $\lambda$-translations.} There is a $\gen{B}$ and $\gen{R}$ singlet $\gen{X}$ that generates positive (negative) translations in $\lambda$ for lowering (raising) generators: 
\[ \label{eq:step1}
\frac{\partial}{\partial \lambda}  J^{\pm}(\lambda) =   \pm \comm{J^{\pm}(\lambda)}{\gen{X} (\lambda)}.
\]

\paragraph{II. Iterative structure.}  $\gen{X}(\lambda)$ can be constructed simply from $\alg{psu}(1|1)^2$ generators and a single auxiliary generator, $\gen{h}(\lambda)$:  
\[ \label{eq:definex}
\gen{X}^{\mathfrak{ab}}(\lambda) =  \acomm{\hat{\gen{Q}}^{\mathfrak{b}}(\lambda)}{\comm{\hat{\gen{S}}^{\mathfrak{a}}(\lambda)}{\gen{h}(\lambda)}}, \quad  \gen{X}(\lambda)= \varepsilon_{\mathfrak{ab}} \gen{X}^{\mathfrak{ab}}(\lambda)  + \half \comm{\mathcal{H}(\lambda)  }{\gen{h}(\lambda)}.
\]
$\gen{h}$ also is  a $\gen{B}$ and $\gen{R}$ singlet, and  $\gen{h}$ commutes with $\gen{J}^{+-}_0$. This  ensures that $\gen{X}$ also has these properties. 

\paragraph{III. Equations for $\gen{h}$.}  $\gen{h}(\lambda)$ satisfies the following equations\footnote{Here and frequently throughout this work we suppress the argument $\lambda$.}
\begin{gather} 
\quad \acomm{\hat{\gen{Q}}^{\mathfrak{a}}}{\comm{\gen{Q}^{c-\mathfrak{b}}}{\gen{h}}} + \acomm{\hat{\gen{Q}}^{\mathfrak{b}}}{\comm{\gen{Q}^{c-\mathfrak{a}}}{\gen{h}}}=0, \quad \acomm{\hat{\gen{S}}^{\mathfrak{a}}}{\comm{\gen{Q}^{c+\mathfrak{b}}}{\gen{h}}} + \acomm{\hat{\gen{S}}^{\mathfrak{b}}}{\comm{\gen{Q}^{c+\mathfrak{a}}}{\gen{h}}}=0,  \notag \\
\quad \acomm{\gen{Q}^{a+\mathfrak{c}}}{\comm{\gen{Q}^{b-\mathfrak{d}}}{\gen{h}}}=\half \varepsilon^{\mathfrak{cd}} \gen{R}^{ab} -\half \varepsilon^{ab} \gen{B}^{\mathfrak{cd}}  -\quarter \varepsilon^{ab} \varepsilon^{\mathfrak{cd}} \gen{L} +  \lambda \, \varepsilon^{ab} \gen{X}^{\mathfrak{cd}}.  \label{eq:eqforh}
\end{gather}
We usually consider a Hermitian representation for the algebra with $\gen{h}=\gen{h}^\dagger$, or a representation related by a simple similarity transformation to a Hermitian one. For the Hermitian case, the second equation on the first line of (\ref{eq:eqforh}) is related by Hermitian conjugation to the first, and there are only two (multicomponent) independent equations for $\gen{h}$, as stated above. When we refer to the first equation for $\gen{h}$, therefore, this should be understood as either equation on the first line, and ``the second equation for $\gen{h}$'' will label the equation on the second line, with nonvanishing right side. 

Since $\pm \gen{X}$ is the generator of translations in $\lambda$ for the non-Cartan elements of the algebra, given a solution at $\lambda_0$ we can integrate to obtain the solution at different values of $\lambda$. In particular, setting $\lambda_0=0$ (for which the gauge theory representation of the algebra is known) leads to
\[ \label{eq:pathintegral}
J^{\pm}(\lambda)= U(\mp \gen{X}, \, \lambda)J^{\pm}_0 U^\dagger(\pm \gen{X}, \,  \lambda), \quad  U(\gen{X}, \, \lambda) = P \Big \{ \exp{\big[\int_0^\lambda \mathrm{d} \lambda' \,  \gen{X}{(\lambda ')}\big]} \Big\}.
\]
Path ordering is required since a priori $\gen{X}(\lambda)$ and $\gen{X}(\lambda')$ do not commute for $\lambda \neq \lambda'$. The element of the algebra not included in this definition that has $\lambda$-dependence is the dilatation generator (central charge), and it is defined implicitly since it is proportional to anticommutators of $\alg{psu}(1|1)^2$ supercharges.

\subsection{Proof that the algebra is satisfied \label{sec:proof}}
In this section, we simply prove that the structure of the algebraic proposal is sufficient to guarantee that $\gen{X}$ generates algebra-satisfying translations in $\lambda$ for the raising/lowering generators. Many of the steps used below were introduced in \cite{Zwiebel:2005er,Zwiebel:2007th}. The key difference here is that these steps are used now to prove the consistency of a proposal at all orders, rather than just at next-to-leading order.

It is useful to simplify $\gen{X}$ (\ref{eq:definex}) as
\<
\gen{X} \eq \acomm{\hat{\gen{Q}}^>}{\comm{\hat{\gen{S}}^<}{\gen{h}}} - \acomm{\hat{\gen{Q}}^<}{\comm{\hat{\gen{S}}^>}{\gen{h}}}  + \half \comm{\mathcal{H}  }{\gen{h}} \nln
\eq \acomm{\hat{\gen{Q}}^>}{\comm{\hat{\gen{S}}^<}{\gen{h}}} - \acomm{\hat{\gen{Q}}^<}{\comm{\hat{\gen{S}}^>}{\gen{h}}}  +\comm{\acomm{\hat{\gen{Q}}^<}{\hat{\gen{S}}^>}}{\gen{h}} \nln
\eq \acomm{\hat{\gen{Q}}^>}{\comm{\hat{\gen{S}}^<}{\gen{h}}} + \acomm{\hat{\gen{S}}^>}{\comm{\hat{\gen{Q}}^<}{\gen{h}}}. \label{eq:xsimp>}
\>
To reach the second line we used (\ref{eq:psu11squaredalg}), and applying the Jacobi identity and combining terms yields the last line.
Alternatively, $\gen{X}$ can be simplified to
\[
\gen{X}=-\acomm{\hat{\gen{Q}}^<}{\comm{\hat{\gen{S}}^>}{\gen{h}}} - \acomm{\hat{\gen{S}}^<}{\comm{\hat{\gen{Q}}^>}{\gen{h}}}. \label{eq:xsimp<}
\]

First, since  $\gen{X}$ is a $\gen{B}$ and $\gen{R}$ singlet, translations in $\lambda$ generated by $\gen{X}$ preserve these $\lambda$-independent $\alg{su}(2)$ symmetries. 
For the remaining commutators, we assume that the algebra relations are satisfied at $\lambda=\lambda_0$. We will then show that acting with $\partial/\partial \lambda$ on both sides of the algebra relations at $\lambda=\lambda_0$ yields identities. This guarantees that the integration described above (\ref{eq:pathintegral}) yields the $\lambda$-dependence of solutions. 

We will prove this by considering four types of commutators. The commutators not included in these groups are guaranteed to be satisfied if all of these groups are satisfied, as explained in Appendix \ref{sec:Chevalley}. 
\begin{itemize}
\item Commutators between two lowering (raising) generators
\item Commutators between a lowering $\alg{psu}(1|1)^2$ supercharge and a raising $\alg{psu}(1|1)^2$ supercharge
\item Commutators between a lowering (raising) $\alg{psu}(1,1|2)$ supercharge and a \\
raising (lowering) $\alg{psu}(1|1|)^2$ supercharge
\item Commutators between a lowering $\alg{psu}(1,1|2)$ supercharge and a raising $\alg{psu}(1,1|2)$ supercharge
\end{itemize}

\paragraph{Commutators between two lowering (raising) generators.} Consider the commutator\footnote{We use commutator to refer to a commutator or anticommutator depending on the statistics of the generators. The argument proceeds similarly in the fermionic case.} between  $J_i^+$ and $J_j^+$ at an initial value of $\lambda=\lambda_0$ such that the commutation relations are satisfied,
\[
\comm{J^+_i(\lambda_0)}{J^+_j(\lambda_0)} =f_{ij}^kJ^+_k(\lambda_0).
\]
 Substituting commutators with $\gen{X}$ for derivatives as specified by (\ref{eq:step1}) and using the Jacobi identity results in
\<
\frac{\partial}{\partial \lambda} \comm{J^+_i(\lambda)}{J^+_j(\lambda)}_{|\lambda=\lambda_0} \eq \comm{ \comm{J^+_i(\lambda_0)}{J^+_j(\lambda_0)}}{\gen{X}(\lambda_0)}
\nln
\eq f_{ij}^k\comm{J^+_k(\lambda_0)}{\gen{X}(\lambda_0)} 
\nln
\eq \frac{\partial}{\partial \lambda} f_{ij}^k J^+_k(\lambda)_{|\lambda=\lambda_0},
\>
as required. The proof for the commutator involving two raising generators proceeds similarly.

\paragraph{Anticommutators between $\hat{\gen{Q}}$ and $\hat{\gen{S}}$.} The $\gen{B}$-singlet component of these commutators defines $\mathcal{H}$, and the only algebraic requirement here will be that $\mathcal{H}$ is simply related to $\gen{J}^{+-}$ as in (\ref{eq:sharedcentralcharge}). This will be verified below. The other three independent commutation relations of this type (\ref{eq:psu11squaredalg}) form a $\gen{B}$-triplet :
\[
\acomm{\hat{\gen{Q}}^\mathfrak{\{a}}{\hat{\gen{S}}^\mathfrak{b\}}} = 0. \label{eq:btriplet}
\]
So, using $\gen{B}$ symmetry it is sufficient to consider the $>>$ component. Taking the derivative, evaluating at $\lambda_0$. and again substituting (\ref{eq:step1}) yields
\< \label{eq:psu11squaredproof}
\frac{\partial}{\partial \lambda} \acomm{\hat{\gen{Q}}^>(\lambda)}{\hat{\gen{S}}^>(\lambda)}_{|\lambda=\lambda_0}\eq 
\acomm{\comm{\hat{\gen{Q}}^>(\lambda_0)}{\gen{X}(\lambda_0)}}{\hat{\gen{S}}^>(\lambda_0)} - \acomm{\hat{\gen{Q}}^>(\lambda_0)}{\comm{\hat{\gen{S}}^>(\lambda_0)}{\gen{X}(\lambda_0)}}  \nln
\eq 2 \acomm{\comm{\hat{\gen{Q}}^>(\lambda_0)}{\gen{X}(\lambda_0)}}{\hat{\gen{S}}^>(\lambda_0)} - \comm{\acomm{\hat{\gen{Q}}^>(\lambda_0)}{\hat{\gen{S}}^>(\lambda_0)}}{\gen{X}(\lambda_0)} \nln
\eq  2 \acomm{\comm{\hat{\gen{Q}}^>(\lambda_0)}{\gen{X}(\lambda_0)}}{\hat{\gen{S}}^>(\lambda_0)}.
\>
We used the Jacobi identity to reach the second line, and the vanishing anticommutator of $\hat{\gen{Q}}^>$ and $\hat{\gen{S}}^>$ at $\lambda_0$ to reach the last line.
Next, substitute (\ref{eq:xsimp>}) in the commutator that appears on the right side of the last line, 
\< \label{eq:qhatcommx}
\comm{\hat{\gen{Q}}^>(\lambda_0)}{\gen{X}(\lambda_0)}\eq \comm{\hat{\gen{Q}}^>(\lambda_0)}{\acomm{\hat{\gen{Q}}^>(\lambda_0)}{\comm{\hat{\gen{S}}^<(\lambda_0)}{\gen{h}(\lambda_0)}}} 
\nl
+ \comm{\hat{\gen{Q}}^>(\lambda_0)}{\acomm{\hat{\gen{S}}^>(\lambda_0)}{\comm{\hat{\gen{Q}}^<(\lambda_0)}{\gen{h}(\lambda_0)}}} \nln
\eq  \comm{\hat{\gen{Q}}^>(\lambda_0)}{\acomm{\hat{\gen{S}}^>(\lambda_0)}{\comm{\hat{\gen{Q}}^<(\lambda_0)}{\gen{h}(\lambda_0)}}}.
\>
The last equality follows from the nilpotency of $\hat{\gen{Q}}^>(\lambda_0)$.\footnote{Any expression of the form $\comm{\hat{\gen{Q}}^>(\lambda_0)}{\acomm{\hat{\gen{Q}}^>(\lambda_0)}{S}}$ vanishes since upon expanding we obtain
\[
\comm{\hat{\gen{Q}}^>(\lambda_0)}{\acomm{\hat{\gen{Q}}^>(\lambda_0)}{S}} = (\hat{\gen{Q}}^>(\lambda_0))^2 S -\hat{\gen{Q}}^>(\lambda_0)S \hat{\gen{Q}}^>(\lambda_0) + \hat{\gen{Q}}^>(\lambda_0) S \hat{\gen{Q}}^>(\lambda_0) - S (\hat{\gen{Q}}^>(\lambda_0))^2 =0. \notag
\]
Similarly, the opposite statistics version $\acomm{\hat{\gen{Q}}^>(\lambda_0)}{\comm{\hat{\gen{Q}}^>(\lambda_0)}{J}}$ also vanishes  The vanishing of these expressions only requires the nilpotency of $\hat{\gen{Q}}^>(\lambda_0)$, so they generalize for any nilpotent supercharge.} Then, since $\hat{\gen{Q}}^>(\lambda_0)$ and $\hat{\gen{S}}^>(\lambda_0)$ anticommute,  (\ref{eq:psu11squaredproof}) vanishes by the nilpotency of $\hat{\gen{S}}^>(\lambda_0)$,
\<
2 \acomm{\comm{\hat{\gen{Q}}^>(\lambda_0)}{\gen{X}(\lambda_0)}}{\hat{\gen{S}}^>(\lambda_0)} \eq 2 \acomm{\comm{\hat{\gen{Q}}^>(\lambda_0)}{\acomm{\hat{\gen{S}}^>(\lambda_0)}{\comm{\hat{\gen{Q}}^<(\lambda_0)}{\gen{h}(\lambda_0)}}}}{\hat{\gen{S}}^>(\lambda_0)} \nln
\eq -2 \acomm{\comm{\hat{\gen{S}}^>(\lambda_0)}{\acomm{\hat{\gen{S}}^>(\lambda_0)}{\comm{\hat{\gen{Q}}^<(\lambda_0)}{\gen{h}(\lambda_0)}}}}{\hat{\gen{Q}}^>(\lambda_0)} \nln
\eq -2 \acomm{0}{\hat{\gen{Q}}^>(\lambda_0)} \nln
\eq 0.
\>
Therefore, by $\gen{B}$ symmetry all three equations of (\ref{eq:btriplet}) are invariant under shifts in $\lambda$ generated by $\gen{X}$. 

For later convenience, we note the three analogous equations to (\ref{eq:qhatcommx}) 
\begin{align} \label{eq:moreqhatcommx}
\comm{\hat{\gen{Q}}^<}{\gen{X}} & = -\comm{\hat{\gen{Q}}^<}{\acomm{\hat{\gen{S}}^<}{\comm{\hat{\gen{Q}}^>}{\gen{h}}}},  & \comm{\hat{\gen{S}}^>}{\gen{X}} & = \comm{\hat{\gen{S}}^>}{\acomm{\hat{\gen{Q}}^>}{\comm{\hat{\gen{S}}^<}{\gen{h}}}}, \notag \\
\comm{\hat{\gen{S}}^<}{\gen{X}} & = -\comm{\hat{\gen{S}}^<}{\acomm{\hat{\gen{Q}}^<}{\comm{\hat{\gen{S}}^>}{\gen{h}}}}. &.
 \end{align}
 These equations are satisfied for any value of the argument of the generators (which is suppressed), as they follow from the definition of $\gen{X}$ and the $\alg{psu}(1|1)^2$ algebra.

\paragraph{Anticommutators between $\gen{Q}^{-}$ and $\hat{\gen{Q}}$ (or $\gen{Q}^{+}$ and $\hat{\gen{S}}$).} There are 16 commutators of this form that must be satisfied,
\[ \label{eq:proofpsu112psu11}
\acomm{\gen{Q}^{a-\mathfrak{b}}}{\hat{\gen{Q}}^\mathfrak{c}}=0, \quad \acomm{\gen{Q}^{a+\mathfrak{b}}}{\hat{\gen{S}}^\mathfrak{c}}=0.
\]
We will only consider the first set of equations, since the proof for the second set is similar. Furthermore, for the first set of equations and by $\gen{R}$ and $\gen{B}$ symmetry it is sufficient to consider the requirement
\[
\acomm{\gen{Q}^{a-<}}{\hat{\gen{Q}}^>}=0.
\]
Following the now standard steps gives
\<
\frac{\partial}{\partial \lambda} \acomm{\gen{Q}^{a-<}(\lambda)}{\hat{\gen{Q}}^>(\lambda)}_{|\lambda=\lambda_0} \eq -\acomm{\comm{\gen{Q}^{a-<}(\lambda_0)}{\gen{X}(\lambda_0)}}{\hat{\gen{Q}}^>(\lambda_0)} 
\nl
+ \acomm{\gen{Q}^{a-<}(\lambda_0)}{\comm{\hat{\gen{Q}}^>(\lambda_0)}{\gen{X}(\lambda_0)}} 
 \nln
\eq 2 \acomm{\gen{Q}^{a-<}(\lambda_0)}{\comm{\hat{\gen{Q}}^>(\lambda_0)}{\gen{X}(\lambda_0)}}.
\>
We used the Jacobi identity and the vanishing of the anticommutators between $\gen{Q}^{a-<}$ and $\alg{psu}(1|1)^2$ generators at $\lambda_0$. Next substituting (\ref{eq:qhatcommx}) and again using the vanishing of the anticommutators between $\gen{Q}^{a-<}$ and $\alg{psu}(1|1)^2$ generators at $\lambda_0$ yields,
\<
2 \acomm{\gen{Q}^{a-<}(\lambda_0)}{\comm{\hat{\gen{Q}}^>(\lambda_0)}{\gen{X}(\lambda_0)}} \eq 2 \acomm{\gen{Q}^{a-<}(\lambda_0)}{\comm{\hat{\gen{Q}}^>(\lambda_0)}{\acomm{\hat{\gen{S}}^>(\lambda_0)}{\comm{\hat{\gen{Q}}^<(\lambda_0)}{\gen{h}(\lambda_0) }}}} \nln
\eq -2 \acomm{\hat{\gen{Q}}^>(\lambda_0)}{\comm{\hat{\gen{S}}^>(\lambda_0)}{\acomm{\hat{\gen{Q}}^<(\lambda_0)}{\comm{\gen{Q}^{a-<}(\lambda_0)}{\gen{h}(\lambda_0) }}}} \nln
\eq 0.
\>
To reach the last line we used the first equation for $\gen{h}$ of (\ref{eq:eqforh}), completing this part of the proof. 

\paragraph{Anticommutators between $\gen{Q}^{+}$ and $\gen{Q}^{-}$.} There are 16 equations of this form. However, under $\alg{su}(2)_{\gen{R}} \otimes \alg{su}(2)_{\gen{B}}$ they transform as 
\[
(\mathbf{2},\mathbf{2}) \otimes (\mathbf{2},\mathbf{2})= (\mathbf{3},\mathbf{3}) \oplus (\mathbf{3},\mathbf{1}) \oplus (\mathbf{1},\mathbf{3})  + \oplus (\mathbf{1},\mathbf{1}).
\]
Furthermore, a single commutator such as
\[
\acomm{\gen{Q}^{1+>}}{\gen{Q}^{2-<}}=\gen{R}^{12} + \gen{J}^{+-} \label{eq:q+q-}
\]
has components within all four of these irreducible representations. Therefore, it is sufficient just to consider this equation. Taking the derivative of the left side and evaluating at $\lambda_0$ yields an identity, as shown in detail in Appendix \ref{sec:extradetails}. This proof is mildly more complicated due to the nonvanishing right side of (\ref{eq:q+q-}), and it requires the second equation of the equations for $\gen{h}$ (\ref{eq:eqforh}). This completes the proof at the level of algebra that the all-order proposal is consistent. We now move on to consider its perturbative realization in $\mathcal{N}=4$ SYM.

\section{The spin chain realization to three loops \label{sec:representation}}
While we have given a consistent all-order algebraic proposal, we have not shown that the  representation for the $\alg{psu}(1,1|2)$ sector must take this iterative form. So, we will now use this proposal as an ansatz for the perturbative expansion of the planar gauge theory. This ansatz enables us to construct the next-to-leading order representation. We present this new result including the three-loop dilatation generator, after first reviewing general aspects of the gauge theory representation, and the leading order representation. The last parts of this section discuss homogeneous solutions to the equations for $\gen{h}$ including a solution for the leading Bethe ansatz phase contribution, and wrapping interactions and the possible application to the nonplanar theory.

\subsection{General considerations}
Here we briefly discuss key properties of the $\alg{psu}(1,1|2)$ sector spin chain. For more detailed discussion see \cite{Beisert:2004ry} for the entire $\mathcal{N}=4$ SYM spin chain, and \cite{Zwiebel:2005er, Zwiebel:2007th} for specifically this sector. We also point out requirements for $\gen{h}$ that follow from properties of the spin chain. 

The model includes cyclic spin-chain states with individual sites of the chain inhabited by a $\alg{psu}(1,1|2)$ module. This infinite-dimensional module is spanned by
\[
\phi_a^{(n)} \text{ or } \psi_{\mathfrak{b}}^{(n)}, \quad a=1,2, \quad \mathfrak{b}=<, >, \quad n = 0, 1, 2, \ldots
\]
The $\phi$ are bosonic, and the $\psi$ are fermionic. Equivalently, this module is generated by (leading order) lowering $\alg{psu}(1,1|2)$ generators acting repeatedly on the one-site vacuum 
$\state{\phi_1^{(0)}}$. We will call the superscript index the ``number of derivatives'' on a site, due to the gauge theory origin of these states as multiple covariant derivatives acting on a scalar or fermionic field. 

Next, we specify the action of the $\lambda$-independent generators. $\gen{L}$ simply counts the number of sites of the spin chain. The uncorrected $\alg{su}(2)$ generators act homogeneously on the spin chain; they act as if the spin chain is simply a tensor product of the modules on each sites. Their action on individual modules is \cite{Beisert:2007sk}
\[\label{eq:TransRB}
\gen{R}^{ab}\state{\phi_c^{(n)}}
=
\delta^{\{a}_c\varepsilon^{b\}d}\state{\phi_{d}^{(n)}}, \quad \gen{B}^{\mathfrak{ab}}\state{\psi_{\mathfrak{c}}^{(n)}}
=\delta^{\{\mathfrak{a}}_{\mathfrak{c}}\varepsilon^{\mathfrak{b}\}\mathfrak{d}}\state{\psi_{\mathfrak{d}}^{(n)}}.
\]

Let us consider several properties of the spin chain in relation to the all-order proposal. Charge conjugation symmetry of the gauge theory requires that Lie algebra symmetry generators have spin-chain parity-even interactions only\footnote{Parity acts on spin chain states by reversing the order of the sites of the chain, with a minus sign for each crossing of fermions and an overall factor of $(-1)^L$ for a length $L$ spin chain.}. This implies that $\gen{h}$ should be parity even. Also, only connected Feynman diagrams contribute to gauge theory anomalous dimensions. This translates to the requirement that Lie algebra generators act locally on the spin chain. In turn, this means that $\gen{h}$ should act locally, on adjacent sites in the planar limit\footnote{Like all of the Lie algebra generators, $\gen{h}$ acts homogeneously on the spin chain.}. Because commutators of local generators are still local, this is sufficient to be consistent with the gauge theory. Finally, to match the powers of $\lambda$ that arise from Feynman diagram perturbative calculations of anomalous dimensions, the $\mathcal{O}(\lambda^n)$ interactions of  $\gen{h}$ must act nontrivially on at most $n+1$ adjacent sites of the spin chain.

Also, there are two ways to generalize the all order proposal: similarity transformations and gauge transformations. However, they are unphysical since they do not affect anomalous dimensions. As a result, after giving a brief explanation here, we will rarely consider these transformations\footnote{A notable exception is in discussed in Section \ref{sec:Dilatation} and is very useful for the anomalous dimension calculations of  Section \ref{sec:verification}.}. A similarity transformation can be written as  
\[
J(\lambda) \mapsto e^{V(\lambda)}J(\lambda) e^{-V(\lambda)}.
\]
$V$ should be spin-chain parity even and have an expansion about $\lambda=0$ in local interactions with the same $\lambda$ dependence of interaction range as given above for $\gen{h}$, but is otherwise arbitrary. 

Gauge transformations, as appearing in the planar spin-chain description, are local interactions that cancel when summed over the length of cyclic spin-chain states\cite{Beisert:2004ry}. For instance, the commutators between $\alg{psu}(1,1|2)$ and $\alg{psu}(1|1)^2$ generators only vanish up to such gauge transformations. One could add infinitely many gauge transformation interactions to any solution for the Lie algebra spin-chain representation. However, this would have no effect on anomalous dimensions or on the closure of the algebra, since the gauge theory spin-chain model includes only cyclic states.  

Due to the reality of anomalous dimensions for $\mathcal{N}=4$ SYM, it is always possible to apply a similarity transformation to obtain a Hermitian representation for the Lie algebra generators. According to our convention, Hermitian conjugation simply exchanges initial and final spin-chain states, maintaining the same ordering of sites\footnote{For this convention to result in a Hermitian matrix when the dilatation generator is applied to cyclic states, certain $1/\sqrt{n}$ normalization factors for the states are needed.  Here $n$ is the largest integer such that a given state is invariant under cyclic rotations by $L/n$ sites, with $L$ equal to the length of the state \cite{Beisert:2004ry}.}. In the basis used in (most of) this work,  Hermitian conjugation relates generators as
\begin{align} \label{eq:hermiticity}
(\gen{J}^{+-})^\dagger &= \gen{J}^{+-}, & (\gen{B}^{\mathfrak{ab}})^\dagger &= -\varepsilon_{\mathfrak{ac}} \varepsilon_{\mathfrak{bd}} \gen{B}^{\mathfrak{cd}}, & (\gen{R}^{ab})^\dagger &= -\varepsilon_{ac} \varepsilon_{bd} \gen{R}^{cd}, \notag \\
 (\gen{J}^{++})^\dagger &= \gen{J}^{--}, & (\gen{Q}^{a+\mathfrak{c}})^\dagger &= -\varepsilon_{ad} \varepsilon_{\mathfrak{ce}} \gen{Q}^{d-\mathfrak{e}}, & (\hat{\gen{Q}}^{\mathfrak{a}})^\dagger & = \varepsilon_{\mathfrak{ab}} \hat{\gen{S}}^{\mathfrak{b}} .
\end{align}
If $\gen{h}= \gen{h}^\dagger$, it follows from the Hermiticity of $\alg{psu}(1|1)^2$ generators that $\gen{X} = \gen{X}^\dagger$. This self-consistently guarantees that the Hermitian structure of the generators is preserved by the all-order proposal.  Below we present a solution for the Lie algebra generators to NNLO and for $\gen{h}$ and $\gen{X}$ to NLO. This solution satisfies $\gen{h}= \gen{h}^\dagger$ and $\gen{X} = \gen{X}^\dagger$, which implies that the Hermiticity of the Lie Algebra generators (\ref{eq:hermiticity}) is preserved as well\footnote{However, the similarity transformation (mentioned in a previous footnote) that will be used in Section \ref{sec:verification} does not preserve the Hermitian structure.}.

Finally, consider the coupling constant transformation
\[
\lambda \mapsto \tilde{\lambda} (\lambda).
\]
This is clearly a symmetry of the commutation relations. However, to maintain consistency with the $\lambda$ expansion at weak coupling, the most general allowed transformation is
\[
\lambda \mapsto \lambda + c_2 \lambda^2 + c_3 \lambda^3 + \ldots
\]
This can easily be applied to the perturbative expansion of our ansatz. However, to three loops there will be no need to use such a transformation.

\subsection{The leading order representation}
The leading order representation of the $\lambda$-dependent $\alg{psu}(1,1|2)$ generators is a tensor-product representation. For the reader's convenience, we repeat the one-site representation given in \cite{Beisert:2007sk} for the $\gen{J}^{\alpha \beta}$ 
\[\label{eq:TransJ}
\begin{array}[b]{rclcrcl}
\gen{J}^{++}_0\state{\phi_a^{(n)}}
\eq(n+1) \state{\phi_{a}^{(n+1)}},
&\quad&
\gen{J}^{++}_0\state{\psi_{\mathfrak{a}}^{(n)}}
\eq\sqrt{(n+1)(n+2)} \state{\psi_{\mathfrak{a}}^{(n+1)}},
\\[3pt]
\gen{J}^{+-}_0\state{\phi_a^{(n)}}
\eq (n + \frac{1}{2}) \state{\phi_{a}^{(n)}},
&\quad&
\gen{J}^{+-}_0\state{\psi_{\mathfrak{a}}^{(n)}}
\eq (n+1)\state{\psi_{\mathfrak{a}}^{(n)}},
\\[3pt]
\gen{J}^{--}_0\state{\phi_a^{(n)}}
\eq n \state{\phi_{a}^{(n-1)}},
&\quad&
\gen{J}^{--}_0\state{\psi_{\mathfrak{a}}^{(n)}}
\eq \sqrt{n(n+1)} \state{\psi_{\mathfrak{a}}^{(n-1)}}, 
\end{array}
\]
and for the $\gen{Q}^{a\beta\mathfrak{c}}$
\[\label{eq:TransQ}
\begin{array}[b]{rclcrcl}
\gen{Q}^{a+\mathfrak{b}}_0\state{\phi_c^{(n)}}
\eq \sqrt{n+1}\, \delta^a_c\varepsilon^{\mathfrak{bd}}\state{\psi_{\mathfrak{d}}^{(n)}},
&\quad&
\gen{Q}^{a+\mathfrak{b}}_0\state{\psi_{\mathfrak{c}}^{(n)}}
\eq\sqrt{n+1}\,\delta^{\mathfrak{b}}_{\mathfrak{c}}\varepsilon^{ad}\state{\phi_{d}^{(n+1)}},
\\[3pt]
\gen{Q}^{a-\mathfrak{b}}_0\state{\phi_c^{(n)}}
\eq \sqrt{n}\, \delta^a_c\varepsilon^{\mathfrak{bd}}\state{\psi_{\mathfrak{d}}^{(n-1)}},
&\quad&
\gen{Q}^{a-\mathfrak{b}}_0\state{\psi_{\mathfrak{c}}^{(n)}}
\eq\sqrt{n+1}\,\delta^{\mathfrak{b}}_{\mathfrak{c}}\varepsilon^{ad}\state{\phi_d^{(n)}}.
\end{array}
\]
Note our convention that $J_n$ represents the $\mathcal{O}(\lambda^n)$ term in the expansion of the generators,
\[
J = \sum_{n=0}^\infty \lambda^n \, J_n.
\]

The leading order $\alg{psu}(1|1)^2$ generators in contrast do not act on the spin chain simply as if it were a tensor product. However, they still act homogeneously on the spin chain. The $\gen{Q}^{\mathfrak{a}}$ act by replacing each individual site with 2 sites as \cite{Zwiebel:2005er,Beisert:2007sk}
\<\label{TransQhat}
4 \pi \, \hat{\gen{Q}}^{\mathfrak{a}}_0\state{\phi_{b}^{(n)}}
\eq
\sum_{k=0}^{n-1}
\frac{1}{\sqrt{k+1}}\,
\varepsilon^{\mathfrak{ac}}
\bigstate{\psi_{\mathfrak{c}}^{(k)}\phi_{b}^{(n-1-k)}}
-
\sum_{k=0}^{n-1}
\frac{1}{\sqrt{n-k}}\,
\varepsilon^{\mathfrak{ac}} 
\bigstate{\phi_{b}^{(k)}\psi_{\mathfrak{c}}^{(n-1-k)}},
\nln
4 \pi \, \hat{\gen{Q}}^{\mathfrak{a}}_0\state{\psi_{\mathfrak{b}}^{(n)}}
\eq
\sum_{k=0}^{n-1}
\frac{\sqrt{n-k}}{\sqrt{(k+1)(n+1)}} 
 \, \varepsilon^{\mathfrak{ac}}
\bigstate{\psi_{\mathfrak{c}}^{(k)}\psi_{\mathfrak{b}}^{(n-1-k)}} 
\nl
+
\sum_{k=0}^{n-1}
\frac{\sqrt{k+1}}{\sqrt{(n-k)(n+1)}}  \, \varepsilon^{\mathfrak{ac}}
\bigstate{\psi_{\mathfrak{b}}^{(k)}\psi_{\mathfrak{c}}^{(n-1-k)}} 
\nl
-
\sum_{k=0}^n
\frac{1}{\sqrt{n+1}}\,
\delta^\mathfrak{a}_\mathfrak{b}  \varepsilon^{cd}
\bigstate{\phi_{c}^{(k)}\phi_{d}^{(n-k)}}.
\>
The factors of $4 \pi$ are due to the rescaling by $\sqrt{\lambda}$ mentioned in Section \ref{sec:alg}. 

The $\gen{S}^{\mathfrak{a}}$ act in a conjugate fashion, replacing two adjacent sites with one new site, as 
\<\label{TransShat}
4 \pi \, \hat{\gen{S}}^{\mathfrak{a}}_0\state{\phi_{b}^{(m)}\psi_{\mathfrak{c}}^{(n)}}
\eq
-\frac{1}{\sqrt{n+1}}\,
\delta^\mathfrak{a}_\mathfrak{c}\bigstate{\phi_b^{(n+m+1)}},
\nln
4 \pi \,  \hat{\gen{S}}^{\mathfrak{a}}_0\state{\psi_{\mathfrak{b}}^{(m)}\phi_{c}^{(n)}}
\eq
\frac{1}{\sqrt{m+1}}\,
\delta^\mathfrak{a}_\mathfrak{b}\bigstate{\phi_c^{(n+m+1)}},
\nln
4 \pi \, \hat{\gen{S}}^{\mathfrak{a}}_0\state{\psi_{\mathfrak{b}}^{(m)}\psi_{\mathfrak{c}}^{(n)}}
\eq
\frac{\sqrt{n+1}}{\sqrt{(m+1)(m+n+2)}}\,
\delta^\mathfrak{a}_\mathfrak{b}\bigstate{\psi_{\mathfrak{c}}^{(n+m+1)}}  
\nl+
\frac{\sqrt{m+1}}{\sqrt{(n+1)(m+n+2)}}\,
\delta^\mathfrak{a}_\mathfrak{c}\bigstate{\psi_{\mathfrak{b}}^{(n+m+1)}}, 
\nln
4 \pi \, \hat{\gen{S}}^{\mathfrak{a}}_0\state{\phi_{b}^{(m)}\phi_{c}^{(n)}}
\eq
\frac{1}{\sqrt{n+m+1}}\,
\varepsilon_{bc} \varepsilon^{\mathfrak{ad}}  \bigstate{\psi_{\mathfrak{d}}^{(n+m)}}.
\>
We will not give the explicit leading order expression for $\mathcal{H}$ here; it can be found in \cite{Beisert:2007sk}. 

Finally, the leading order expression for $\gen{h}$ was obtained in \cite{Zwiebel:2005er}
\[
\gen{h}_0 \state{\phi_a^{(n)}} = \half S(n) \state{\phi_a^{(n)}}, \quad \gen{h}_0 \state{\psi_{\mathfrak{a}}^{(n)}} = \half S(n+1) \state{\psi_{\mathfrak{a}}^{(n)}}, 
\]
where $S(n)$ is the $n$-th (ordinary) Harmonic number. There and in further analysis in \cite{Zwiebel:2007th} it was shown that $\gen{h}_0$ satisfies the two requirements of the ansatz, (\ref{eq:eqforh}). Substitution into (\ref{eq:definex}) yields the leading order expression for $\gen{X}$, which is sufficient to give the NLO $\alg{psu}(1|1)^2$ supercharges and the two-loop dilatation generator, as originally found in \cite{Zwiebel:2005er}.  We will review this result in Section \ref{sec:Dilatation}.

\subsection{Next-to-leading order representation \label{sec:NLO}}
Assuming the algebraic ansatz, the only new ingredient required at next-to-leading order is the $\mathcal{O}(\lambda^1)$ term of the expansion for $\gen{h}$. To solve the conditions (\ref{eq:eqforh}) at this order, $\gen{h}_1$ must act on two adjacent sites. This makes obtaining the explicit form of $\gen{h}_1$ much more challenging. We will not show the lengthy intermediate steps involved in obtaining the solution. Instead we will explain the method used and describe the solution, which is given explicitly in Appendix \ref{sec:h1solution}. 

$\gen{R}$ and $\gen{B}$ symmetry and Hermiticity imply that it is sufficient to obtain a solution for $\gen{h}$ to, for example,
\<
\acomm{\hat{\gen{S}}^<}{\comm{\gen{Q}^{1+<}}{\gen{h}}} \eq 0, \notag \\
 \acomm{\gen{Q}^{1+<}}{\comm{\gen{Q}^{2->}}{\gen{h}}}\eq \half  \gen{R}^{12} - \half \gen{B}^{<>}  - \quarter  \gen{L} + \lambda  \gen{X}^{<>}. 
\>
The $\mathcal{O}(\lambda)$ term of these equations is 
\< \label{eq:eqtosolve}
\acomm{\hat{\gen{S}}^<_0}{\comm{\gen{Q}^{1+<}_0}{\gen{h}_1}}  + \acomm{\hat{\gen{S}}^<_0}{\comm{\gen{Q}^{1+<}_1}{\gen{h}_0}} + \acomm{\hat{\gen{S}}^<_1}{\comm{\gen{Q}^{1+<}_0}{\gen{h}_0}} \eq 0, \notag \\
 \acomm{\gen{Q}^{1+<}_0}{\comm{\gen{Q}^{2->}_0}{\gen{h}_1}} + \acomm{\gen{Q}^{1+<}_0}{\comm{\gen{Q}^{2->}_1}{\gen{h}_0}} + \acomm{\gen{Q}^{1+<}_1}{\comm{\gen{Q}^{2->}_0}{\gen{h}_0}}\eq   \gen{X}^{<>}_0. 
\>
Note that according to the ansatz, $\gen{Q}^{1+<}_1$, $\hat{\gen{S}}^<_1$ and $\gen{Q}^{2->}_1$ are known, and are simply commutators of leading order generators with $\gen{X}_0$. Furthermore, $\gen{h}_1$ appears linearly through only one term in each equation.  Rather than solving for $\gen{h}_1$ directly, it is much simpler to first obtain a solution for $\comm{\gen{Q}^{1+<}_0}{\gen{h}_1}$. We do this by writing $\comm{\gen{Q}^{1+<}_0}{\gen{h}_1}$ in terms of $12$ unknown component functions of three arguments, which parameterize the possible parity-even interactions on two sites\footnote{The possible interactions are limited since they must have the right $\gen{R}$ and $\gen{B}$ charges and be consistent with Hermiticity and with the vanishing commutator between $\gen{h}$ and $\gen{J}^{+-}_0$.}. The arguments are the total number of derivatives initially, the number of derivatives on the first site initially, and the number of derivatives on the first site finally\footnote{There are other possibilities for the arguments, of course.}. Substituting into (\ref{eq:eqtosolve}) and expanding in terms of components  yields a system of 46 equations for these $12$ functions\footnote{There are a few linear dependences between the equations, which are useful for consistency checks.}.  One subtlety is that the first equation of (\ref{eq:eqtosolve}) is only satisfied up to gauge transformations, but this is straightforward to account for with two functions parameterizing these gauge transformations.  Furthermore, we obtain 7 additional equations from the following identities,
\[
\comm{\gen{R}^{11}}{\comm{\gen{Q}^{1+<}_0}{\gen{h}_1} }=0, \quad \comm{\gen{B}^{<<}}{\comm{\gen{Q}^{1+<}_0}{\gen{h}_1} }=0, \acomm{\gen{Q}^{1+<}_0}{\comm{\gen{Q}^{1+<}_0}{\gen{h}_1} }=0.
\]

While the 12 component functions enter these ($46+7$) equations linearly, these are very nontrivial equations to solve. The component functions are coupled nontrivially, and the component functions appear with different arguments and with factors depending on their arguments. It is straightforward to solve for $8$ of the component functions in terms of the other $4$ and (known) commutators that do not involve $\gen{h}_1$. To solve for the final few functions it is most efficient to expand them in terms of a basis of harmonic number functions, and then find the coefficients numerically. The only functions that are needed (allowing for products of functions), are the ordinary harmonic number function
\[ \label{eq:ordinaryS}
S(j) = \sum_{i=1}^j \frac{1}{i}
\]
and the two positive-index degree-two generalized harmonic functions
\[
 S_2(j) = \sum_{i=1}^j \frac{1}{i^2}, \quad S_{1,1}(j) = \sum_{i=1}^j \frac{S(i)}{i}.
 \]
 The set of arguments that enter these harmonic functions must include the total number of derivatives, initially or finally, the number of derivatives on any site initially or finally, or the number of derivatives shifted between the two sites (these arguments may also appear with $\pm 1$). Four of the component functions are given explicitly in Appendix \ref{sec:h1solution}. While we do not prove that we have found a solution for $\comm{\gen{Q}^{1+<}_0}{\gen{h}_1}$, we have checked up to 20 total derivatives that all of the equations are satisfied\footnote{Note that the leading order spin-chain symmetry generators change the total number of derivatives by at most 1, so it is straightforward to truncate at a given maximum number of derivatives.}. Since we fixed the (integer) coefficients for functions entering the component functions only using equations for up to at most 10 derivatives, this is overwhelming evidence.
 
Now a similar process can be applied directly to $\gen{h}_1$. First $\gen{h}_1$ is written in terms of $7$ component functions. Expanding the commutator with $\gen{Q}^{1+<}_0$ yields 12 equations, and it is straightforward to solve for 4 of the component functions in terms of the other $3$. As before, we expand these final few component functions in terms of a basis of functions and solve for the coefficients numerically. Some further generalizations of the harmonic sums are required for this solution, as explained in Appendix \ref{sec:h1solution}, where the solution is presented. This method works, as in for example \cite{Beccaria:2007cn,Kotikov:2007cy}, because the component functions have definite harmonic degree and integer coefficients. The extra challenge here is that these functions have three arguments rather than just one. On the other hand, with this proposal for the solution it is straightforward to verify analytically that all of the equations are satisfied. We have not done that simply because it is much more efficient to gather convincing numerical evidence.  The proposal for $\gen{h}_1$  satisfies the equations for interactions involving up to 20 derivatives, and again we only needed up to 10 derivatives to find the coefficients. Therefore, it is safe to conclude that expressions given for $\gen{h}_1$ in Appendix \ref{sec:h1solution} satisfies the equations (\ref{eq:eqforh}) required by the algebraic ansatz at NLO.

As explained in Section \ref{sec:verification}, further extensive spectral tests lead us to conclude, with almost certainty, that this is the gauge theory solution for $\gen{h}_1$, and therefore that the algebraic ansatz is general enough to describe the $\alg{psu}(1,1|2)$ sector of $\mathcal{N}=4$ SYM at least to three loops.  It is also intriguing that the (probable) gauge theory solution for $\gen{h}_1$ has harmonic degree three and integer coefficients. Perhaps there is further iterative structure that could greatly simplify the perturbative construction of $\gen{h}$. 

\subsection{Homogeneous solutions \label{sec:homogeneous}}
The solution for $\gen{h}_1$  is actually not uniquely specified by the NLO constraints from (\ref{eq:eqforh}). There are two types of homogeneous solutions that can be added. Importantly, based upon the analysis of \cite{Beisert:2005wv}, it must be that all but one of these solutions for $\gen{h}_1$ breaks integrability, so (assuming integrability) they cannot contribute to the dilatation generator. We consider this further at the end of this section.

We first discuss the simpler type of homogeneous solution. Note that $\gen{h}_1$ only appears in (\ref{eq:eqtosolve}) inside commutators with leading order $\alg{psu}(1,1|2)$ generators. Therefore, any function of the $\alg{psu}(1,1|2)$ quadratic Casimir, acting on two adjacent sites for consistency with the perturbative structure, is a homogeneous solution. The quadratic Casimir $\gen{J}^2$ is given explicitly as
\[
\gen{J}^2 =\varepsilon_{cb} \varepsilon_{ad} \gen{R}^{ab} \gen{R}^{cd} - \varepsilon_{\gamma\beta} \varepsilon_{\alpha \delta } \gen{J}^{\alpha \beta} \gen{J}^{\gamma \delta}- \varepsilon_{ad} \varepsilon_{\beta \gamma} \varepsilon_{\mathfrak{ce}} \gen{Q}^{a\beta \mathfrak{c}}  \gen{Q}^{d\gamma \mathfrak{e}},
\]
and has eigenvalues $j(j+1)$ for $j=0, 1, 2, \ldots$; $j$ is the $\alg{psu}(1,1|2)$ ``spin'' It is possible to write explicitly the interactions for a generator with eigenvalues $c_j$ for spin-$j$ two-site states, as explained in Appendix \ref{sec:homsolution}. Including this contribution to $\gen{h}_1$ with arbitrary $c_j$ gives a countably infinite set of solutions (at this order). To eliminate this freedom we apply the maximal transcendentality principle \cite{Kotikov:2002ab}, which in this case implies that it should be possible to write $\gen{h}_1$ completely in terms of generalized harmonic sums of degree three. Furthermore, assuming that no generalized harmonic sums with $-1$ indices appear \cite{Kotikov:2002ab, Kotikov:2007cy}, there are only 7 coefficients to fix, which we do by substituting $\gen{h}_1$ into the expression for the three-loop dilatation generator and matching the resulting spectrum to known spectral data for twist-two states. In fact, the solution that works only includes generalized harmonic functions with positive indices and with integer coefficients, as is the case for expressions for twist-three anomalous dimensions (to four loops) obtained in \cite{Beccaria:2007cn,Kotikov:2007cy}. 

As mentioned above, there is a second type of homogeneous solution. Solutions $\delta  \gen{h}_1$ of this type have nonvanishing commutators with $\alg{psu}(1,1|2)$ supercharges (including $\gen{Q}^{1+<}_0$), but they still satisfy, for example, 
\< \label{eq:secondhom}
\acomm{\hat{\gen{S}}^<_0}{\comm{\gen{Q}^{1+<}_0}{\delta  \gen{h}_1}}  \eq 0, \notag \\
 \acomm{\gen{Q}^{1+<}_0}{\comm{\gen{Q}^{2->}_0}{\delta  \gen{h}_1}} \eq   0. 
\>
First, consider the following ansatz for $\delta  \gen{h}_1$,
\[\label{eq:ansatzfordeltah1}
\delta  \gen{h}_1 = \sum_{i, j} c_{ij} \big( \mathcal{P}_i \, \gen{h}_0 \mathcal{P}_j \big)_{1,2}
\]
Here $\mathcal{P}_j$ is the projector for two-site states with $\alg{psu}(1,1|2)$ spin $j$. Explicitly, it is given by the solution in Appendix \ref{sec:homsolution} with coefficients $c_i = \delta_{ij}$. The final $1,2$ subscript in (\ref{eq:ansatzfordeltah1}) denotes that all three generators in this product act on the same two adjacent sites, with $\gen{h}_0$ acting through the sum of its one-site interactions. As usual, this two-site structure should be summed homogeneously over the length of the spin chain. Note that the projectors satisfy $\mathcal{P}_i \mathcal{P}_j = \delta_{ij} \mathcal{P}_i$. From the algebra relations for $\gen{h}_0$ and the vanishing commutators between the $\mathcal{P}_i$ and the leading order $\alg{psu}(1,1|2)$ generators (including $\gen{B}$), it then follows that
\[
 \acomm{\gen{Q}^{1+<}_0}{\comm{\gen{Q}^{2->}_0}{\delta  \gen{h}_1}} =  \sum_{i, j} c_{ij} \delta_{ij}\big((\text{leading order generators})\mathcal{P}_i  \big)_{1,2}
 \]
 Therefore, the second equation of (\ref{eq:secondhom}) is satisfied provided $c_{ii}=0$ for all $i$. Also, Hermiticity requires that $c_{ij}=c_{ji}$. Solving  the first equation of (\ref{eq:secondhom}) analytically is more difficult because of the length-changing interactions of $\hat{\gen{S}}^<_0$. However, numerically we find that the unique solution (up to normalization) for the $c_{ij}$ is given by $c_{ij}= |S(i)-S(j)|$. We have checked the uniqueness of this solution for the $c_{ij}$ up to 10 excitations, and confirmed that (\ref{eq:secondhom}) is satisfied up to 20 excitations\footnote{Excitations refers to the number of magnons above the half-BPS vacuum of all $\phi_1^{(0)}$. Equivalently, a $\psi^{(j)}$ contributes  $j+1$ excitations, a $\phi_2^{(k)}$,  $k+1$ excitations, and a $\phi_1^{(m)}$, $m$ excitations.}. 
 
This homogeneous solution for $\gen{h}$ apparently is required for the homogeneous structure that first appears in the gauge theory dilatation generator at four loops, preserving integrability. This homogeneous structure corresponds to the phase contribution to the Bethe equations \cite{Arutyunov:2004vx}, with the coefficient $\beta_{2,3}^{(n)}$ at $n+1$ loops, in the notation of \cite{Beisert:2005wv}.  As is well-known and as mentioned in the introduction, the phase is crucial for obtaining the correct interpolation between gauge and string theory asymptotic Bethe equations \cite{Beisert:2006ez}. As explained in  \cite{Beisert:2005wv}, it is possible to construct this homogeneous structure as soon as the Hamiltonian acts on four adjacent sites, i.e. starting at three loops. However, in gauge theory its appearance is delayed by one loop because of the limited possible interactions that can be built from Feynman diagrams at three loops, as will be discussed further below. 

Explicitly, the proposal for $\delta \gen{h}$ is 
\[ \label{eq:phasesolution}
\frac{(4 \pi)^{2(n-1)}}{n} (\delta \gen{h}_{n-1})_{1,2} =  \beta_{2,3}^{(n)} \Big(\sum_{i=1}^{\infty} S_2 (i) \mathcal{P}_i + \sum_{i, j=0}^{\infty} |S(i) - S(j)| \mathcal{P}_i \, \gen{h}_0  \mathcal{P}_j \Big)_{1,2} 
\]
The factors of $4 \pi$ are simply for conversion between $\lambda$ and $g$, while the factor of $n$ compensates for a factor of $1/n$ from integrating $\lambda^{n-1} \gen{X}_{n-1}$ according to (\ref{eq:step1}). Also, observe that the first term $\sim S_2(i)\mathcal{P}_i$ clearly satisfies the algebra constraints since it is a homogeneous solution of the first type. It is intriguing that the first term would be proportional to the one-loop dilatation generator if $S_2(i)$ were replaced with $S(i)$. In Section \ref{sec:testphase} we describe thorough checks that confirm this homogeneous solution for $\gen{h}$ matches the Bethe ansatz predictions for the $\beta_{2,3}$ contributions. Note, that this solution is only the leading order solution for the $\beta_{2,3}$ contributions, and requires corrections at higher orders in $\lambda$. According to our conjecture that the algebraic ansatz does apply to the gauge theory at all orders, these higher-order corrections can be included through corrections to $\gen{h}$.  

In fact, numerical analysis also suggests that there are infinitely more homogeneous solutions of the second type for $\gen{h}_1$. In particular, truncating at $j$ excitations, we find $[(j+1)/2]$ solutions\footnote{Brackets denote greatest integer less than or equal to $(j+1)/2$.}  to  (\ref{eq:secondhom}), including the single solution given explicitly above.  The natural way to generalize the ansatz (\ref{eq:ansatzfordeltah1}) for such solutions is to replace $\gen{h}_0$ there with a new (two-site) generator $X$  satisfying 
\[
\acomm{\gen{Q}^{1+<}_0}{\comm{\gen{Q}^{2->}_0}{X}} = \sum_{i} d_i \mathcal{P}_i.
\]
The simplest such $X$ is $\gen{X}_0$, which corresponds to $d_i \propto S(i)$. However, further analysis of these apparent additional solutions is beyond the scope of this work. 

As noted above, the three-loop dilatation generator cannot include any of the homogeneous solutions included here, again assuming integrability. The simplest possibility is that these solutions are  an artifact of solving for $\gen{h}$ perturbatively, and that closure of the algebra for $\gen{h}$ at higher orders would eliminate them. Initial numerical analysis indicates otherwise; at two excitations, both physically relevant homogeneous solutions\footnote{Naively there should be three homogeneous solutions of the first type at two excitations, but two of these are gauge equivalent to zero.} are consistent with the algebraic ansatz at NNLO. However, there is a conceptually simple way to eliminate these solutions, which works the same way as in the compact $\alg{su}(2|3)$ sector \cite{Beisert:2003ys}. At $l$ loops, interactions can only be included if they involve at most $l$ permutations of adjacent flavors of scalars. Otherwise, there is no $l$-loop Feynman diagram that can be drawn for such an interaction. We have checked at two excitations that this property eliminates both physical homogeneous solutions at three loops (including the phase solution, as mentioned above). We expect that this property also eliminates all of the remaining homogeneous solutions that appear for more excitations. Checking this would involve straightforward but lengthy calculations of coefficients for these types of interactions, which we leave for the future.  

However, even assuming that all homogeneous solutions are structurally eliminated at three loops, they could still appear at four loops. Therefore, it seems likely that the basic Feynman diagram properties used in this work  and extended $\alg{psu}(1,1|2) \times \alg{psu}(1|1)^2$ symmetry are sufficient to guarantee $\alg{psu}(1,1|2)$ sector integrability through three loops, but not beyond.

\subsection{The Dilatation generator to three loops \label{sec:Dilatation}}

With the solution for $\gen{h}_1$, we now have all the necessary ingredients to expand the dilatation generator to three loops. By definition and by the $\alg{psu}(1|1)^2$ algebra, the complete $\alg{psu}(1,1|2)$ sector dilatation generator is
\[
\gen{D} = \gen{D}_0 + \lambda \mathcal{H}, \quad \mathcal{H} = 2 \acomm{\hat{\gen{Q}}^<}{\hat{\gen{S}}^>}.
\]
Then at $\mathcal{O}(\lambda)$,
\[
\gen{D}_1 = \lambda \, \mathcal{H}_0 = 2 \lambda \, \acomm{\hat{\gen{Q}}^<_0}{\hat{\gen{S}}^>_0}.
\]
Next, consider the derivative with respect to $\lambda$ (all generators are functions of $\lambda$, which is suppressed),
\<
\frac{\partial}{\partial \lambda} \mathcal{H} \eq 2 \acomm{\hat{\gen{S}}^>}{\comm{\hat{\gen{Q}}^<}{\gen{X}}} - 2 \acomm{\hat{\gen{Q}}^<}{\comm{\hat{\gen{S}}^>}{\gen{X}}}
\nln
\eq -2 \acomm{\hat{\gen{S}}^>}{\comm{\hat{\gen{Q}}^<}{\acomm{\hat{\gen{S}}^<}{\comm{\hat{\gen{Q}}^>}{\gen{h}}}}} - 2 \acomm{\hat{\gen{Q}}^<}{\comm{\hat{\gen{S}}^>}{\acomm{\hat{\gen{Q}}^>}{\comm{\hat{\gen{S}}^<}{\gen{h}}}}} 
\nln
\eq 2 \acomm{\hat{\gen{S}}^>}{\comm{\hat{\gen{S}}^<}{\acomm{\hat{\gen{Q}}^<}{\comm{\hat{\gen{Q}}^>}{\gen{h}}}}} + 2 \acomm{\hat{\gen{Q}}^<}{\comm{\hat{\gen{Q}}^>}{\acomm{\hat{\gen{S}}^>}{\comm{\hat{\gen{S}}^<}{\gen{h}}}}},
\>
where we used (\ref{eq:moreqhatcommx}) for expanding the two terms on the first line, and vanishing anticommutators of the $\alg{psu}(1|1)^2$ algebra to reach the ordering of the supercharges on the last line. 

According to the algebraic ansatz, this expression for the derivative holds for general $\lambda$. However, to make contact with the perturbative gauge theory representation, expand this expression at $\lambda=0$. For instance,
\[
\acomm{\hat{\gen{S}}^>}{\comm{\hat{\gen{S}}^<}{\acomm{\hat{\gen{Q}}^<}{\comm{\hat{\gen{Q}}^>}{\gen{h}}}}} = \sum_{n=0}^\infty \lambda^n \, \acomm{\hat{\gen{S}}^>}{\comm{\hat{\gen{S}}^<}{\acomm{\hat{\gen{Q}}^<}{\comm{\hat{\gen{Q}}^>}{\gen{h}}}}} _n
\]
Then integrate with respect to $\lambda$ to obtain the perturbative series for $\delta \gen{D}=\lambda \mathcal{H}$
\<
\delta \gen{D}(\lambda)\eq 2 \lambda \, \acomm{\hat{\gen{Q}}^<_0}{\hat{\gen{S}}^>_0} + 2 \sum_{n=0}^\infty \frac{\lambda^{n+2}}{n+1}  \, \acomm{\hat{\gen{S}}^>}{\comm{\hat{\gen{S}}^<}{\acomm{\hat{\gen{Q}}^<}{\comm{\hat{\gen{Q}}^>}{\gen{h}}}}} _n
\nl
+ 2 \sum_{n=0}^\infty \frac{\lambda^{n+2}}{n+1} \,
 \acomm{\hat{\gen{Q}}^<}{\comm{\hat{\gen{Q}}^>}{\acomm{\hat{\gen{S}}^>}{\comm{\hat{\gen{S}}^<}{\gen{h}}}}}_n
\>
It follows that 
\[  \label{eq:diltwoloops}
\delta \gen{D}_2 = 2 \acomm{\hat{\gen{S}}^>}{\comm{\hat{\gen{S}}^<}{\acomm{\hat{\gen{Q}}^<}{\comm{\hat{\gen{Q}}^>}{\gen{h}}}}}_0 + 2 \acomm{\hat{\gen{Q}}^<}{\comm{\hat{\gen{Q}}^>}{\acomm{\hat{\gen{S}}^>}{\comm{\hat{\gen{S}}^<}{\gen{h}}}}}_0.
\]
Through two loops, the dilatation generator only depends on leading order generators (including $\gen{h}_0$). This expression for $\delta \gen{D}_2$ matches the result given in Section 4.3 of \cite{Zwiebel:2005er}, up to changes of notation and convention, which are explained in Appendix  \ref{sec:notation}. 

The three-loop contribution does depend also on next-to-leading contributions to generators,
\[ \label{eq:dilthreeloops}
\delta \gen{D}_3 =  \acomm{\hat{\gen{S}}^>}{\comm{\hat{\gen{S}}^<}{\acomm{\hat{\gen{Q}}^<}{\comm{\hat{\gen{Q}}^>}{\gen{h}}}}}_1 +  \acomm{\hat{\gen{Q}}^<}{\comm{\hat{\gen{Q}}^>}{\acomm{\hat{\gen{S}}^>}{\comm{\hat{\gen{S}}^<}{\gen{h}}}}}_1.
\]
When expanded, this three-loop contribution is a sum of 10 terms. Each of these terms is a nested commutator of leading order generators and one next-to-leading order correction. For instance, the first term of (\ref{eq:dilthreeloops}) expands as
\begin{gather} 
 \acomm{\hat{\gen{S}}^>_1}{\comm{\hat{\gen{S}}^<_0}{\acomm{\hat{\gen{Q}}^<_0}{\comm{\hat{\gen{Q}}^>_0}{\gen{h}_0}}}} + \acomm{\hat{\gen{S}}^>_0}{\comm{\hat{\gen{S}}^<_1}{\acomm{\hat{\gen{Q}}^<_0}{\comm{\hat{\gen{Q}}^>_0}{\gen{h}_0}}}} + \acomm{\hat{\gen{S}}^>_0}{\comm{\hat{\gen{S}}^<_0}{\acomm{\hat{\gen{Q}}^<_1}{\comm{\hat{\gen{Q}}^>_0}{\gen{h}_0}}}} 
\notag \\
 \label{eq:dilexpand} + \acomm{\hat{\gen{S}}^>_0}{\comm{\hat{\gen{S}}^<_0}{\acomm{\hat{\gen{Q}}^<_0}{\comm{\hat{\gen{Q}}^>_1}{\gen{h}_0}}}}
 + \acomm{\hat{\gen{S}}^>_0}{\comm{\hat{\gen{S}}^<_0}{\acomm{\hat{\gen{Q}}^<_0}{\comm{\hat{\gen{Q}}^>_0}{\gen{h}_1}}}}.
\end{gather}
 For these next-to-leading order generators, recall (\ref{eq:qhatcommx}) and (\ref{eq:moreqhatcommx}),
 \begin{align}
 \hat{\gen{Q}}^<_1 & = -\comm{\hat{\gen{Q}}^<_0}{\acomm{\hat{\gen{S}}^<_0}{\comm{\hat{\gen{Q}}^>_0}{\gen{h}_0}}}, &  \hat{\gen{Q}}^>_1 & = \comm{\hat{\gen{Q}}^>_0}{\acomm{\hat{\gen{S}}^>_0}{\comm{\hat{\gen{Q}}^<_0}{\gen{h}_0}}}, \notag \\
 \hat{\gen{S}}^<_1 & = \comm{\hat{\gen{S}}^<_0}{\acomm{\hat{\gen{Q}}^<_0}{\comm{\hat{\gen{S}}^>_0}{\gen{h}_0}}}, &  \hat{\gen{S}}^>_1 & = -\comm{\hat{\gen{S}}^>_0}{\acomm{\hat{\gen{Q}}^>_0}{\comm{\hat{\gen{S}}^<_0}{\gen{h}_0}}}.
 \end{align}
 Upon substitution into (\ref{eq:dilthreeloops}) (expanded as in (\ref{eq:dilexpand})), the proposed three-loop dilatation generator is expressed completely in terms of leading order generators and $\gen{h}_1$.
 
For direct computations of anomalous dimensions, however, it is convenient to apply a similarity transformation.  Using $U(\gen{X}, \lambda)$, which appears in the (path-)integrated form of the algebraic ansatz (\ref{eq:pathintegral}), it is possible to transform away all of the corrections to the raising generators. This is very useful because it implies that dilatation generator (including its perturbative corrections) commutes with leading order raising generators. Then the dilatation generators mixes states annihilated by a given leading order raising generator only with other states annihilated by the same leading order raising generator. 
 
For the perturbative expansion of $\delta \gen{D}$ of the transformed solution, note that the differential version of this transformed solution is
\[ \label{eq:newderivative}
\frac{\partial}{\partial \lambda}  J^{+}(\lambda) =   2 \comm{J^{+}(\lambda)}{\gen{X} (\lambda)}, \quad \frac{\partial}{\partial \lambda}  J^{-}(\lambda) = 0.
 \]
$\gen{X}$ is defined as above in terms of $\gen{h}$, and formally $\gen{h}$ must satisfy the same equations as previously. However, importantly this similarity transformation requires a different solution for $\gen{h}$. For this work simply note that $\gen{h}_0$ is unaffected and 
 \[ \label{eq:newh}
 \gen{h}_1 \mapsto  \gen{h}_1  + \comm{\gen{h}_0}{\gen{X}_0}.
 \]
 Using (\ref{eq:newderivative}) and repeating steps done earlier in this subsection, we find that in this non-Hermitian basis
 \< \label{eq:hamnonHermitian}
 \delta \gen{D}_2 \eq 4 \acomm{\hat{\gen{S}}^>}{\comm{\hat{\gen{S}}^<}{\acomm{\hat{\gen{Q}}^<}{\comm{\hat{\gen{Q}}^>}{\gen{h}}}}}_0, \nln
 \delta \gen{D}_3 \eq  2 \acomm{\hat{\gen{S}}^>}{\comm{\hat{\gen{S}}^<}{\acomm{\hat{\gen{Q}}^<}{\comm{\hat{\gen{Q}}^>}{\gen{h}}}}}_1,
 \>
 where again (\ref{eq:newh}) must be used for $\gen{h}_1$ appearing in the expression for $\delta \gen{D}_3$. Of course, although the matrix elements of these new expressions for the dilatation generator will be different than for the Hermitian expressions presented earlier, the eigenvalues (anomalous dimensions) will be unchanged.

 \subsection{Wrapping interactions and a possible nonplanar lift \label{sec:wrapping}}
Wrapping interactions, initially discussed in \cite{Beisert:2003ys,Serban:2004jf,Beisert:2004hm,Sieg:2005kd}, correspond to nonplanar interactions that become planar for short states by wrapping around the trace. At three loops, the dilatation generator has wrapping interactions when acting on states of length two or three. In the next section,  we will present very strong evidence that the algebraic ansatz combined with the proposal for $\gen{h}_1$ yields the correct three-loop anomalous dimensions, even for short states. We conclude that this solution correctly incorporates wrapping. In fact, there is no wrapping ambiguity for this proposal because the three-loop dilatation generator is built in terms of leading order supercharges, $\gen{h}_0$, and $\gen{h}_1$. There is no wrapping problem for leading order supercharges (and $\gen{h}_0$) since their interactions involve only one initial site, or only one final site.  Finally, it is sufficient that $\gen{h}_1$  is defined through its actions on two sites because the length-decreasing $\hat{\gen{S}}$ annihilate two-site states (this is guaranteed by spin-chain parity)\footnote{Usually we do not consider one-site states which vanish for $SU(N)$ gauge theories, and for any gauge group are protected descendants of the one-site vacuum $\mathrm{Tr} \fldZ$. Even if we choose to include such states, it is not a problem. Both $\hat{\gen{S}}$ and $\hat{\gen{Q}}$ annihilate one-site states, so $\gen{h}$ can be defined to be zero on one-site states (it actually could take arbitrary finite values without having an effect), and the solution properly gives zero anomalous dimension for these one-site states as well.}. 

For this paragraph, let us assume the iterative ansatz is general enough to give the four-loop dilatation generator. Now $\gen{h}_2$ would appear, and it has three-site interactions. Therefore $\gen{h}_2$ has wrapping interactions on two-site states. This is precisely consistent with the fact that starting at four loops wrapping interactions can invalidate the asymptotic Bethe ansatz for two-site states\footnote{Due to the length-increasing action of the $\hat{\gen{Q}}$, any (non-BPS) two-site states is in the same supermultiplet as a four-site state.}  \cite{Beisert:2003ys, Beisert:2005fw}, as in fact happens \cite{Kotikov:2007cy}. Assume that  $\zeta(5)$ appears only within wrapping interactions, as suggested by recent calculations \cite{Fiamberti:2007rj, Keeler:2008ce}\footnote{There is currently a discrepancy between these two results. Also, see \cite{Eden:2007rd}.}. This would correspond to $\zeta(5)$ only appearing in (the homogeneous part of) the wrapping interactions of $\gen{h}_2$.  Of course, it would still remain a challenge to fix coefficients for the infinitely many homogeneous two-site solutions for $\gen{h}$, described in Section \ref{sec:homogeneous}. 

Also, it is possible that the algebraic ansatz is not general enough to describe the gauge theory wrapping interactions. An intermediate possibility is that a more general algebraic ansatz, like the ones presented in Section \ref{sec:addalg}, is required for wrapping interactions.

In \cite{Zwiebel:2005er}, it was argued that the two-loop solution for the dilatation generator has a natural nonplanar generalization. The leading order $\alg{psu}(1|1)^2$ supercharges have a unique lift to the nonplanar theory, as explained there. Since the one-site generator $\gen{h}_0$ certainly also has a unique lift, one can simply substitute the nonplanar generalizations for the supercharges and $\gen{h}_0$ into the algebraic expression for the two-loop dilatation generator to obtain a nonplanar generalization. This nonplanar expression still satisfies all the symmetry constraints because the proof only depended on algebraic properties. Given that wrapping effects are properly taken into account, it is reasonable to conjecture that this is the two-loop nonplanar dilatation generator.
 
 At three loops the new ingredient for the dilatation generator is $\gen{h}_1$. Since this acts on two sites, it is nontrivial to obtain the three-loop nonplanar dilatation generator, even if one assumes that the iterative ansatz still applies.  In fact, if there is a nonplanar solution consistent with the algebraic ansatz, it must be unique. Since this argument is similar to Beisert's argument for the one-loop dilatation generator \cite{Beisert:2003jj}, we first review that. 
 
 The nonplanar one-loop dilatation generator can be written in terms of one normal-ordered nonplanar structure,
 \[ \label{eq:oneloopnonplanar}
 \frac{\lambda}{N} \, C^{\mathcal{AB}}_{\mathcal{CD}} : \mathrm{Tr} \comm{\mathcal{W}_\mathcal{A}}{\hat{\mathcal{W}}^\mathcal C} \comm{\mathcal{W}_\mathcal{B}}{\hat{\mathcal{W}}^\mathcal{D}}:.
 \]
Here the $\mathcal{W}_\mathcal{A}$ represent ``fields,'' which are covariant derivatives of scalars or fermions in the $\alg{psu}(1,1|2)$ sector, taking values in the Lie algebra of the gauge group, and $\hat{\mathcal{W}}_\mathcal{A}$ represents variations with respect to these ``fields''. Beisert showed using gauge invariance and a Jacobi identity that the three possible types of field theory interactions that could arise at one loop could all be combined into this one type of interaction. The coefficients were fixed by considering the planar limit, and from these coefficients one immediately obtains the complete nonplanar expression for $\gen{D}_1 = \lambda \mathcal{H}_0$. 

Similarly, it must be possible to write all interactions of the nonplanar $\gen{h}_1$ (if the algebraic ansatz still applies) as
 \[ \label{eq:nonplanarh}
 (\gen{h}_1)_{\text{nonplanar}}= \frac{\lambda}{N} \, h^{\mathcal{AB}}_{\mathcal{CD}} :\mathrm{Tr} \comm{\mathcal{W}_\mathcal {A}}{\hat{\mathcal{W}}^\mathcal{C}} \comm{\mathcal{W}_\mathcal{B}}{\hat{\mathcal{W}}^\mathcal{D}}:,
 \]
because the interactions come with only two powers of $g_{\text{YM}}$, like the one-loop dilatation generator. As before, the planar limit uniquely fixes the coefficients $ h^{\mathcal{AB}}_{\mathcal{CD}}$.

The same type of argument applies to the second equation for $\gen{h}$ in (\ref{eq:eqforh}), at next-to-leading order.  This equation for $\gen{h}_1$ consists of three terms that do not include $\gen{h}_1$. These terms are commutators of one $\hat{\gen{Q}}_0$, one $\hat{\gen{S}}_0$, and one-site generators. These one-site generators take the form
 \[
C^\mathcal{A}_\mathcal{B} \mathrm{Tr} \mathcal{W}_\mathcal{A}\hat{\mathcal{W}}^\mathcal{B}.
 \]
 It follows that they do not change the gauge group structure; they simply replace a single field with another field with the same gauge group indices. Furthermore, the anticommutators of (Hermitian conjugate pairs of) $\hat{\gen{Q}}$ and $\hat{\gen{S}}$ generate precisely the interactions of the one-loop dilatation generator. Therefore, the terms of the second equation of (\ref{eq:eqforh}) not including $\gen{h}_1$ can be written using new coefficients $\tilde{C}^{\mathcal{AB}}_{\mathcal{CD}}$ multiplying the same nonplanar structure that appears in (\ref{eq:oneloopnonplanar}). Because $\gen{h}_1$ also appears only inside commutators of one-site generators, this equation for $\gen{h}$ is solved by  (\ref{eq:nonplanarh}) with coefficients fixed by the planar limit.

Importantly, it remains to be shown that this ``solution'' for $\gen{h}_1$ (\ref{eq:nonplanarh}) also satisfies the first requirement of (\ref{eq:eqforh}). This requires further investigation because $\gen{h}_1$ appears inside a commutator with a length-changing generator. If the nonplanar generalization for $\gen{h}_1$ satisfies this equation as well, our ansatz will have been lifted to a solution of all symmetry constraints at three loops for the $\alg{psu}(1,1|2)$ sector. Again, since wrapping effects are properly incorporated, this would be a strong candidate for the nonplanar generalization. On the other hand, it is certainly possible that this ansatz cannot apply starting at three (or two) loops to the nonplanar theory. Testing this almost certainly requires rigorous two- and three-loop nonplanar calculations of anomalous dimensions.

\section{Verification of the three-loop proposal \label{sec:verification}}
In this section we test the proposal first by considering Hamiltonian interactions (within the $\alg{su}(2)$ subsector), then by computing the two-magnon S-matrix, and finally through direct diagonalization for multiple-magnon states. In the final part of this section we repeat these checks for the phase solution. For simplifying comparisons with previous results we use the coupling constant $g$, which is related to  the 't Hooft coupling  as\footnote{Previous works have sometimes used different conventions, most often an 8 rather than a 16 in the denominator.}
\[
g^2=\frac{\lambda}{16 \pi^2}.
\]

\subsection{$\alg{su}(2)$ subsector interactions}

Restricting to $\alg{psu}(1,1|2)$ sector states without fermions or derivatives yields the $\alg{su}(2)$ subsector. Now the single-site module is spanned by $\phi_a, \, a=1,2$. The three-loop $\alg{su}(2)$ subsector Hamiltonian was first proposed in \cite{Beisert:2003tq} based on the assumptions of BMN scaling and integrability. That proposal was later proved correct using algebraic and diagrammatic constraints for the larger $\alg{su}(2|3)$ sector \cite{Beisert:2003ys} and a rigorous field theory computation of the three-loop anomalous dimension of the Konishi operator \cite{Eden:2005ve}. Note that BMN scaling is not present beyond three loops. Using the permutation notation introduced in \cite{Beisert:2003tq}, 
\[
\{p_1,p_2,\ldots\}=\sum_{p=1}^L\mathcal{P}_{p+p_1,p+p_1+1} \mathcal{P}_{p+p_2,p+p_2+1}\ldots
\]
where $\mathcal{P}_{i,i+1}$ permutes adjacent sites of the spin chain,
the explicit $\alg{su}(2)$ sector Hamiltonian to three loops is 
\< \label{eq:su2ham}
\delta \gen{D}_2 \eq 2\{\} -2 \{1\}, \nln
\delta \gen{D}_4 \eq -8\{\} + 12\{1\} - 2 (\{1, 2\} + \{2, 1\}), \nln
\delta \gen{D}_6 \eq 60 \{\} - 104 \{1\} + 4 \{1, 3\} + 24 (\{1, 2\} + \{2, 1\}) - 4(\{1, 2, 3\} + \{3, 2, 1\}).
\>
Note that the subscripts now refer to powers of $g=\sqrt{\lambda}/4\pi$. Expanding the expressions (\ref{eq:diltwoloops}) and (\ref{eq:dilthreeloops}) in terms of interactions (using $\texttt{Mathematica}$), restricting to interactions contained within the $\alg{su}(2)$ subsector, and then eliminating interactions that act as chain derivatives on cyclic (or periodic) chains, we find perfect agreement with (\ref{eq:su2ham}). Note that we used the Hermitian form for the proposed $\alg{psu}(1,1|2)$ dilatation generator, since (\ref{eq:su2ham}) is presented in a Hermitian basis. 

While this is a nice and relatively simple check of our proposal, it only checks interactions involving at most two magnons. The reason is that the three-loop dilatation generator only acts on at most 4 sites, and by $\alg{su}(2)$ symmetry any interaction (that affects the spectrum) of $n$ magnons has the same coefficient as the interaction with all $\alg{su}(2)$ spins flipped, which is a $(4-n)$-magnon interaction. In principle, one could repeat such a Hamiltonian comparison within larger subsectors, but the only other subsector with known three-loop dilatation generator is the compact $\alg{su}(1|2)$ subsector (and its  $\alg{su}(1|1)$ subsector), which is the intersection of the $\alg{psu}(1,1|2)$ sector and the $\alg{su}(2|3)$ sector. However, the three-loop dilatation generator for the $\alg{su}(2|3)$ sector \cite{Beisert:2003ys} has not been given explicitly. In any case, such a check will be made almost redundant by the other tests discussed below.

\subsection{S-matrix checks}
In this subsection, we will consider only infinite-length states. First, consider the $\alg{su}(1|1)$ subsector, with module spanned by $\phi_1^{(0)}$ and $\psi_<^{(0)}$. Starting from the ferromagnetic vacuum of a chain of $\phi_1^{(0)}$, the one-magnon states are 
\[
\state{\Psi_p} = \sum_{x = 1}^{\infty} e^{i p x} \state{x},
\]
where $\state{x}$ refers to the state with a $\psi_<^{(0)}$ on the $x$-th site and all other sites still occupied by $\phi_1^{(0)}$. Applying the one-loop, two-loop (\ref{eq:diltwoloops}) and three-loop (\ref{eq:dilthreeloops}) dilatation generator to this state yields 
\[
\delta \gen{D}  \state{\Psi_p} = (8 g^2 \sin^2 (\frac{p}{2}) - 32 g^4 \sin^4 (\frac{p}{2})  + 256 g^6 \sin^6 (\frac{p}{2}) + \mathcal{O}(g^8)) \state{\Psi_p},
\]
which matches the perturbative expansion of the well-known expression for one-magnon dispersion relation 
\[
E(p) = \sqrt{1 + 16 g^2 \sin^2 (\frac{p}{2})}-1.
\]
By $\alg{psu}(1,1|2)$ symmetry, it follows that the proposal yields the correct dispersion relation for all flavors of magnons within the $\alg{psu}(1,1|2)$ sector.  For later use, note that this and closure of the Lie algebra imply that the solution gives the correct action of the Lie algebra generators on asymptotic one-magnon states, at NNLO. 
 
Next, consider the spectral problem for two-magnon states. Again using the proposal for the dilatation generator restricted to the $\alg{su}(1|1)$ sector, one finds eigenstates that are products of one-particle states up to a S-matrix factor and other local terms (to three loops)
\<
\state{\Psi_{p_1,p_2}} \eq \sum_{1 \leq x_1 < x_2 \leq \infty} \bigg( e^{i p_1 x_1 + i p_2 x_2}  + e^{i p_1 x_2 + i p_2 x_1} \Big(S(p_2, p_1) +  c^{(1)}(p_1, p_2)\delta_{x_2(x_1+1)} \nl
+   c^{(2)}(p_1, p_2)\delta_{x_2(x_1+2)}\Big)\bigg) \state{x_1 x_2}.
\>
$\state{x_1 x_2}$ has $\psi_<^{(0)}$ on sites $x_i$, and $\phi_1^{(0)}$ on all other sites. Since the  dilatation generator to three loops is short-ranged, such a state must and does have eigenvalue equal to the leading three terms of $E(p_1)$ + $ E(p_2)$. Solving for the coefficients $S$ and $c^{(i)}$ in the wavefunction, we find agreement with the known 2-particle S-matrix for this sector, first obtained to three-loops in \cite{Staudacher:2004tk}, using the result of \cite{Beisert:2003ys},
\<
S(p_1, p_2) \eq -1 - 2 i g^2 (\sin(p_1) - \sin(p_1-p_2)  - \sin(p_2)) \nl
+ 4 i g^4  \sin(\frac{p_1}{2}) \sin(\frac{p_2}{2}) \Big(\sin(\frac{p_1 - 3 p_2}{2}) - 7\sin(\frac{p_1 -  p_2}{2}) + \sin(3 \frac{p_1 -  p_2}{2}) \nl
+ \sin(\frac{3 p_1 -  p_2}{2})  - 8 i \sin(\frac{p_1}{2}) \sin(\frac{p_2}{2})  \sin^2( \frac{p_1 -  p_2}{2}) \Big). \nl
+ \mathcal{O}(g^6). 
\>
The other coefficients expand as 
\[
c^{(1)}(p_1, p_2) = g^2 c^{(1)}_0(p_1, p_2) + g^4 c^{(1)}_2(p_1, p_2) + \mathcal{O}(g^6), \quad c^{(2)}(p_1, p_2) = g^4 c^{(2)}_0(p_1, p_2)  + \mathcal{O}(g^6).
\]
However, these coefficients are basis dependent and not relevant to the Bethe ansatz, so we do not write them here. 

As explained for instance in \cite{Beisert:2005fw},  the fundamental magnons for the $\alg{psu}(1,1|2)$ sector can be $\phi_2^{(0)}$, $\psi_<^{(0)}$, and $\psi_>^{(0)}$ (this corresponds to a choice of the simple roots of the Lie algebra).  All states are then built by adding these fundamental excitations to the vacuum state of $\phi_1^{(0)}$. Using manifest $\gen{B}$ symmetry and parity, there are only 5 independent components, 
\[\label{eq:smatrixelements}
 S^{22}_{22}, \, S^{2<}_{2<},\, S^{2<}_{<2}, \, S^{[<>]}_{[<>]}, \, S^{<<}_{<<}. 
 \]
 Since we already checked that the proposal gives the correct Hamiltonian in the $\alg{su}(2)$ sector, it must yield the correct S-matrix element $S^{22}_{22}$ as well as the correct element $S^{<<}_{<<}$, which we just computed. We have also directly checked that the correct three-loop $\alg{sl}(2)$  sector S-matrix follows from the proposed solution. Given these multiple checks, and the spectral results below which are consistent with the known S-matrix and  factorized scattering, there should be no doubt that the solution generates the correct two-particle S-matrix.  Therefore we do not explicitly check the remaining S-matrix components\footnote{These components are constrained by the fact that the supercharges act properly on asymptotic one-particle states, and this asymptotic action must commute with the S-matrix. This should fix much if not all of the remaining freedom.}.

\subsection{Spectral tests} 
While the above results constitute a thorough test for two-magnon states, we have not yet tested the proposal on states involving more than two magnons. We do so in this subsection by direct evaluation of eigenvalues of the proposed dilatation generator. First we explain some key aspects of these calculations.  As discussed at the end of Section \ref{sec:Dilatation}, for this purpose it is convenient to use the expressions (\ref{eq:hamnonHermitian}), for which states annihilated by a given leading-order raising generator only mix with other states annihilated by that leading-order raising generator. Due to the algebra relations of Section \ref{sec:comm}, the dilatation generator also only mixes states that have the same $\gen{R}$ and $\gen{B}$ charges, length, and classical dimension. Finally, the dilatation generator is spin-chain parity even, so it only mixes states of the same parity. Consequently, since we only calculate eigenvalues for states with relatively small charges, typically we encounter mixing between two or fewer states (at most four), so that mixing is not a significant problem.

However, computation time for acting with the dilatation generator on even a single state increases rapidly with length and with the number of magnons, especially derivatives. To counter this problem there is another useful shortcut. First let us consider the case with no mixing. An energy eigenstate $\state{E}$ is given as a a linear combination  of `'position (eigen)states,'' which have specific elements of the $\alg{psu}(1,1|2)$ module on each site ($\state{x}$ or $\state{x_1x_2}$ encountered in the previous subsection are simple examples),
\[
\state{E} = c_i \state{\mathbf{x}_i}
\]
The coefficients $c_i$ are found using a leading-order calculation. Since by assumption there is no mixing with $\state{E}$,
\[
\delta \gen{D} \state{E}  = \delta \! D \state{E} =   \delta \!D c_i \state{\mathbf{x}_i}.
\]
It follows, that it is enough to compute the coefficient of a single position state in  $\delta \gen{D} \state{E}$. In practice, that does not save much computation time. Instead, by using the Hermitian conjugate to $\delta \gen{D}$ (recall that we are using a non-Hermitian basis) we can obtain the eigenvalue by acting on just a single position state\footnote{The scalar product is one for normalized states that are identical (up to cyclic permutation), and it is zero otherwise.}
\[
\costate{E} \delta \gen{D}^\dagger  \state{\mathbf{x}_1} = \delta \!D  c_1.
\]
 This is enough to infer the change in eigenvalue $\delta \!D$. This usually results in significant time saving because of fewer terms generated by the action of the (conjugate) dilatation generator.  It is straightforward to generalize this procedure for mixing. If there is mixing between $n$ states, it is necessary to compute the action of $ \delta \gen{D}^\dagger$ on $n$ position states, which again typically yields significant time saving over acting directly on the leading order energy eigenstates. 
 
We have checked the spectrum for many two magnon states, with complete agreement with previous results. This is a redundant check, however, due to the S-matrix results of the previous subsection. Next, as mentioned in Section \ref{sec:homogeneous}, we use twist-two states to fix the first type of homogeneous freedom for $\gen{h}_1$. The twist-two spectrum is given in terms of generalized harmonic sums, which are defined as
\[ \label{eq:generalizedS}
S_a(n) = \sum_{i=1}^n \frac{(\text{sgn}(a))^i}{i^{|a|}}, \quad S_{a_1, a_2, \ldots a_m}(n) = \sum_{i=1}^n \frac{ (\text{sgn}(a_1))^i \,  S_{a_2, \ldots a_{m}}(i)}{i^{|a_1|}}.
\]

The twist-two states' three-loop anomalous dimensions are \cite{Kotikov:2004er} 
\< \label{eq:twisttwo}
D(j)\eq j+2 + 8 \, g^2 S_1 - 16 \,g^4 \big(S_3 + S_{-3} - 2\, S_{-2, 1} + 2\, S_1(S_2+S_{-2}) \big)
\nl
- 64\, g^6 \bigg( 2 \,S_{-3} S_2 - S_5 - 2\, S_{-2}S_{3} - 3 \,S_{-5} + 24\, S_{-2, 1, 1, 1} 
\nl
+6(S_{-4, 1} + S_{-3, 2} + S_{-2, 3}) - 12 (S_{-3, 1, 1} + S_{-2, 1, 2} + S_{-2, 2, 1}) 
\nl
-(S_2+2 S_1^2)(3 \,S_{-3} + S_3 - 2\, S_{-2, 1}) - S_1\Big(8 \,S_{-4} + (S_{-2})^2 
\nl
+ 4\, S_2 S_{-2} + 2 \,(S_2)^2 + 3\, S_4 - 12\, S_{-3, 1} - 10 \,S_{-2, 2} + 16 \, S_{-2, 1, 1} \Big) \bigg)  + \mathcal{O}(g^8), \nl
{}
\>
where all harmonic sums are evaluated at (even) $j$, the  $\alg{psu}(1,1|2)$ spin. It should be noted that these anomalous dimensions follow from a rigorous QCD calculation \cite{Moch:2004pa} combined with the assumption of maximal transcendentality (at three loops). Since a rigorous calculation confirms the $j=2$ value \cite{Eden:2005ve}, and the Bethe ansatz gives the same values  for $j=2, 4, 6, 8, \ldots$ \cite{Staudacher:2004tk}, (\ref{eq:twisttwo}) is almost certainly correct.  We have checked that the solution for $\gen{h}_1$ given in Appendix \ref{sec:h1solution} yields these same three-loop anomalous dimensions, for (even) $j \leq 14$, which is sufficient to fix the first type of homogeneous freedom, again assuming maximal transcendentality. 

For further consistency tests, we first use more states within the $\alg{sl}(2)$ sector. Given that the proposal yields the correct two-particle S-matrix, such tests can be viewed equivalently as tests of consistency with factorized scattering, with integrability, or with the Bethe ansatz. For every state tested, see Table \ref{tab:sl2}, the eigenvalue of the proposed three-loop dilatation generator is in complete agreement with the prediction of the Bethe ansatz. 
 \begin{table}\centering
 $\begin{array}{|l|l|l|}\hline
(D_0; R, L, B) &\bigbrk{D_2,D_4, D_6}^P & \delta D_{\text{phase}}\\\hline
(5; \frac{3}{2},3, 0)& \begin{array}{l} (8,-24,136)^- \end{array} & \begin{array}{l} -32^-   \end{array}\\
\hline
(6; \frac{3}{2}, 3, 0)& \begin{array}{l} (15, -\frac{225}{4}, \frac{3195}{8})^\pm \end{array} & \begin{array}{l} -\frac{135}{4}^{\pm}  \end{array} \\ \hline
 (7; \frac{3}{2}, 3, 0)& \begin{array}{l} (12, -39, \frac{957}{4})^-  \end{array} &  \begin{array}{l} -54^-  \end{array} \\ \hline
 (7; 2, 4, 0)& \begin{array}{l} (12, -42, 288)^\pm \end{array} & \begin{array}{l} -36^\pm  \end{array} \\ \hline
(8;\frac{3}{2}, 3, 0)& \begin{array}{l} (\frac{35}{2}, -\frac{18865}{288}, \frac{1068515}{2304})^\pm \end{array} & \begin{array}{l} -\frac{11515}{288}^{\pm}  \end{array} \\ \hline
(8;2, 4, 0)& \begin{array}{l} (8.76554, -21.001, 100.672)^+ \\ (16.7185, -64.272, 475.048)^+ \\ (23.1826, -92.4124, 668.959)^+ \\ (\frac{46}{3}, -\frac{1331}{27}, \frac{76973}{243})^\pm \end{array} & \begin{array}{l} -23.233^+ \\ -31.9153^+ \\ -91.3516^+ \\ -\frac{257}{27}^\pm \end{array} \\ \hline
(8;\frac{5}{2}, 5, 0)& \begin{array}{l} (9.45862,-28.0586, 169.594)^\pm \\ (15.5414, -57.6916, 423.155)^\pm \end{array} & \begin{array}{l} -22.4301^\pm \\ -15.3199^\pm \end{array} \\ \hline 
(9;\frac{3}{2}, 3, 0)& \begin{array}{l} (\frac{44}{3}, -\frac{443}{9}, \frac{303115}{972})^- \\ (\frac{227}{10}, -\frac{1107503}{12000}, \frac{4837443107}{7200000} )^{\pm} \end{array} & \begin{array}{l} -\frac{5522}{81}^- \\ -\frac{2346601}{24000}^{\pm} \end{array} \\ \hline 
 (10; \frac{3}{2}, 3, 0)& \begin{array}{l} (\frac{581}{30}, -\frac{2606009}{36000}, \frac{99502062989}{194400000})^\pm \end{array} & \begin{array}{l} -\frac{29607907}{648000}^{\pm}  \end{array} \\ \hline\end{array}$
\caption{$\alg{sl}(2)$ sector(s) states. Here and in Table \ref{tab:anotab} the dimension of the states are given to three-loops by $D_0 + g^2 D_2 + g^4 D_4+ g^6 D_6$, $R$ and $B$ correspond to the (absolute value of) $\gen{R}^{<>}$ and $\gen{B}^{<>}$ eigenvalues, and $L$ gives the length of the state. The $P$ exponent of the anomalous dimensions denotes parity. The last column gives the shift in anomalous dimension due to the $\delta \gen{h}_{n-1}$ structure (\ref{eq:phasesolution}), equivalently the leading $\beta_{2,3}^{(n)}$ phase contribution, in units of $\beta_{2,3}^{(n)} g^{2n+2}$.  } 
\label{tab:sl2}
\end{table}
Finally, we perform similar tests for states including fermions, within the $\alg{su}(1|1)$ subsector and the fermionic $\alg{sl}(2)$ sector (with modules spanned by $\psi_>^{(n)}$). The results, listed in Table \ref{tab:anotab}, are again in complete agreement with the Bethe ansatz.
\begin{table} \centering
$\begin{array}{|l|l|l|}\hline
(D_0; R, L, B) &\bigbrk{D_2,D_4, D_6}^P &\delta D_{\text{phase}} \\\hline
(7;\frac{1}{2},5, 2)& \begin{array}{l} (20,-80, 580)^- \end{array} & \begin{array}{l} -80^- \end{array} \\
\hline 
 (\frac{15}{2};\frac{3}{2},6, \frac{3}{2})& \begin{array}{l} (16,-56, 392)^\pm \end{array} & \begin{array}{l} -16^\pm  \end{array}\\
\hline 
(8; 1,6, 2)& \begin{array}{l} (16,-56, 368)^+ \end{array} & \begin{array}{l}-64^+ \end{array} \\
\hline 
(\frac{17}{2};2,7, \frac{3}{2})& \begin{array}{l} (14,-48, 332)^\pm \end{array} & \begin{array}{l} -26^\pm  \end{array} \\
\hline 
(9;\frac{3}{2},7, 2)& \begin{array}{l} (  12.7922, -37.5972,  216.187 )^- \\ (18.2198, -68.4112, 507.403)^- \\ (24.988, -97.9916, 708.41)^- \end{array} & \begin{array}{l} -40.7052^- \\ -24.488^- \\ -86.8068^-  \end{array} \\ \hline
(\frac{19}{2}; \frac{5}{2},8, \frac{3}{2})& \begin{array}{l} (12,-38, 247)^\pm \\ (16, -58, 427)^\pm \end{array} & \begin{array}{l} -24^\pm \\ -8 ^\pm   \end{array} \\
\hline
(\frac{15}{2}; 0,3, \frac{3}{2})& \begin{array}{l} (20,-\frac{245}{3}, \frac{21475}{36})^\pm \end{array} & \begin{array}{l}-\frac{260}{3}^\pm  \end{array} \\
\hline
(\frac{19}{2}; 0,3, \frac{3}{2})& \begin{array}{l} (\frac{133}{6},-\frac{131117}{1440}, \frac{1039405829}{1555200})^\pm \end{array} & \begin{array}{l}-\frac{849121}{8640}^\pm \end{array} \\
\hline  \end{array}$
\caption{States including fermions. The three-loop dimensions for the $\alg{su}(1|1)$ subsector states ($R \neq 0$) actually follow from the known Hamiltonian for this subsector \cite{Beisert:2003ys}, and the fermionic $\alg{sl}(2)$ sector three-loop dimensions were computed using Baxter equation methods in \cite{Belitsky:2007zp}.  }
\label{tab:anotab}
\end{table}
These many tests of the proposal lead us to conclude, with almost complete certainty, that the three-loop solution for the planar dilatation generator presented here is the field theory solution.

\subsection{Tests of the proposal for the leading phase contribution \label{sec:testphase}}

To test the proposal for $\delta \gen{h}$ (\ref{eq:phasesolution}) corresponding to the $\beta_{2,3}$ structure, we simply repeat the tests of the previous sections for the corresponding contribution to the dilatation generator, which at $\mathcal{O}(\lambda^{n+1})$ is 
\[ \label{eq:phaseham}
2\frac{\lambda^{n+1}}{n} \acomm{\hat{\gen{S}}^>_0}{\comm{\hat{\gen{S}}^<_0}{\acomm{\hat{\gen{Q}}^<_0}{\comm{\hat{\gen{Q}}^>_0}{\delta \gen{h}_{n-1}}}}} + 2 \frac{\lambda^{n+1}}{n}  \acomm{\hat{\gen{Q}}^<_0}{\comm{\hat{\gen{Q}}^>_0}{\acomm{\hat{\gen{S}}^>_0}{\comm{\hat{\gen{S}}^<_0}{\delta \gen{h}_{n-1}}}}}.
\]
Again, we emphasize that this only includes the leading $\beta_{2,3 }$ structure, and not subleading corrections.

Restricting to the $\alg{su}(2)$ sector, and switching back to coupling constant $g$,  we find that the homogeneous solution contributes
\< \label{eq:su2phaseham}
\delta \gen{D} \eq g^{2 n +2} \beta^{(n)}_{2,3} \Big(-4 \{\} + 12 \{1\} -6  \{1, 3\} - 4  (\{1, 2\} + \{2, 1\})
\nl
+  4(\{1, 3, 2\} + \{2, 1, 3\}) -  2 \{2, 1, 3, 2\} \Big).
\>
This is in perfect agreement with the $\alg{su}(2)$ sector contribution for $\beta_{2,3}$ \cite{Beisert:2007hz}.  

Next, again we consider S-matrix elements. As above, we have computed both the $\alg{su}(1|1)$ S-matrix and $\alg{sl}(2)$ S-matrix, and the $\alg{su}(2)$ sector S-matrix must be correct since we just checked that sector's dilatation generator. We find that contribution to the S-matrix from $\delta \gen{h}$ (\ref{eq:phasesolution}) exactly matches the leading (nontrivial) contribution from the $2,3$ component of the phase \cite{Arutyunov:2004vx,Beisert:2005wv}, 
\begin{eqnarray}
S(p_1, p_2) & \mapsto & e^{2 i \theta_{2,3}(p_1, p_2)} S(p_1, p_2) 
\notag \\
\theta_{2,3}^{(n)}(p_1, p_2) & = &  g^{(2n)} \beta_{2,3}^{(n)} (q_2(p_1) q_3(p_2) - q_3(p_1) q_2(p_2)).
\end{eqnarray}
Here $q_2$ is the eigenvalue of the dilatation generator (divided by $g^2$), and $q_3$ is the eigenvalue of the next local (parity-odd) charge. For the phase factor, and to leading order, we only need the one-magnon dispersion relation and its analogue for $q_3$, 
\begin{gather}
q_2(p) = 4 \sin^2(\frac{p}{2}) + \mathcal{O}(g^2), \quad q_3(p) = 4 \sin^2(\frac{p}{2}) \sin p + \mathcal{O}(g^2), \notag \\
\theta_{2,3}^{(n)}(p_1, p_2)  =   g^{(2n)}  \beta_{2,3}^{(n)} \Big(16 \sin^2(\frac{p_1}{2})\sin^2(\frac{p_2}{2}) (\sin p_2 - \sin p_1) + \mathcal{O}(g^2)\Big).
\end{gather}
The predictions for the gauge theory coefficients $\beta_{2,3}^{(n)}$ can be found in \cite{Beisert:2006ez}, along with the predictions for all the coefficients for the higher-charge generalizations.

Finally, to confirm that we have found the leading $\beta_{2,3}$ solution for more than two excitations, we compute the contributions of (\ref{eq:phaseham}) to anomalous dimensions, as shown in the last columns of Table \ref{tab:sl2} and Table \ref{tab:anotab}. In all cases the results are in perfect agreement with the predictions from the Bethe ansatz that follow from including the $\beta_{2,3}$ factor. Multiplying the last columns of both tables, for instance, by $4 \zeta(3)$ gives the Bethe ansatz prediction for the transcendental part of four-loop anomalous dimensions \cite{Beisert:2006ez}. 

Especially because of the simple form of the proposal, we conclude from these many successful comparisons that (\ref{eq:phasesolution}) gives the leading $\beta_{2,3}$ solution for $\gen{h}$.

 \section{Additional algebraic considerations \label{sec:addalg}}
 Here we consider the possibility of relaxing some requirements of the algebraic ansatz. A priori, algebraic generalizations may be needed for applications beyond the planar three-loop dilatation generator\footnote{Recall that in the introduction we conjectured that such generalizations are not needed for the planar theory at any order, at least asymptotically.}. Still, we will always assume that there exists some generator(s) of $\pm$ translations in $\lambda$. In the first subsection this generator will still a $\gen{B}$ singlet, while in the second subsection a $\gen{B}$-triplet will be used.
 
 \subsection{A more general ansatz for a $\gen{B}$-singlet $\gen{X}$}
For this subsection, still assume the first step of the algebraic ansatz, the existence of a generator of $\lambda$-translations, $\gen{X}$. However, do not assume $\gen{X}$ is built iteratively from $\alg{psu}(1|1)^2$ generators and an auxiliary generator.  Examining the proof that shifts generated by $\gen{X}$ preserve the $\alg{psu}(1|1)^2$ relation
 \[
 \acomm{\hat{\gen{Q}}^{\{\mathfrak{a}}}{\hat{\gen{S}}^{\mathfrak{b} \}}}=0,
 \]
 which was given after (\ref{eq:btriplet}), we see that it is only necessary for $\gen{X}$ to satisfy
 \[ \label{eq:neweqforx}
\acomm{\hat{\gen{Q}}^{\{\mathfrak{a}}}{\comm{\hat{\gen{S}}^{\mathfrak{b} \}}}{\gen{X}}}=0.
 \]
Next, consider commutators of $\alg{psu}(1|1)^2$ and $\alg{psu}(1,1|2)$ supercharges, which were discussed after (\ref{eq:proofpsu112psu11}). These commutators imply that the first equation of the third part of the algebraic ansatz (\ref{eq:eqforh}) can be relaxed to the equations 
 \[ \label{eq:newfirsteqforh}
 \acomm{\gen{Q}^{a+\mathfrak{b}}}{\comm{\hat{\gen{S}}^{\mathfrak{c}}}{\gen{X}}}=0,
 \]
 and the Hermitian conjugate equations.

Finally, we derive the necessary condition for the commutators between $\alg{psu}(1,1|2)$ generators. The commutators between raising and lowering supercharges are 
\[ \label{eq:originalrelation}
\acomm{\gen{Q}^{a+\mathfrak{c}}(\lambda)}{\gen{Q}^{b-\mathfrak{d}}(\lambda)}= -\lambda \, \varepsilon^{ab} \acomm{\hat{\gen{Q}}^{\mathfrak{c}}(\lambda)}{\hat{\gen{S}}^{\mathfrak{d}}(\lambda)} + \text{$\lambda$-independent}.
\]
Applying $\partial/\partial \lambda$  yields (all generators are functions of $\lambda$, which we suppress),
\<
\acomm{\gen{Q}^{b-\mathfrak{d}}}{\comm{ \gen{Q}^{a+\mathfrak{c}}}{ \gen{X} }} - \acomm{ \gen{Q}^{a+\mathfrak{c}}}{\comm{ \gen{Q}^{b-\mathfrak{d}}}{  \gen{X} }} \eq -\varepsilon^{ab} \, \acomm{\hat{\gen{Q}}^{\mathfrak{c}}}{\hat{\gen{S}}^{\mathfrak{d}}} - \lambda \varepsilon^{ab}\, \acomm{\hat{\gen{S}}^{\mathfrak{d}}}{\comm{ \hat{\gen{Q}}^{\mathfrak{c}}}{ \gen{X} }} \nl
+ \lambda \varepsilon^{ab} \, \acomm{\hat{\gen{Q}}^{\mathfrak{c}}}{\comm{ \hat{\gen{S}}^{\mathfrak{d}}}{  \gen{X} }}.
\>
After using the Jacobi identity and substituting (\ref{eq:originalrelation}), the left hand side simplifies to
\[
-\lambda \, \varepsilon^{ab} \, \comm{ \acomm{\hat{\gen{Q}}^{\mathfrak{c}}}{\hat{\gen{S}}^{\mathfrak{d}}}}{\gen{X}}  - 2  \acomm{ \gen{Q}^{a+\mathfrak{c}}}{\comm{ \gen{Q}^{b-\mathfrak{d}}}{  \gen{X} }}.
\]
Expanding the first commutator and using the $\alg{psu}(1|1)^2$ commutation relations, we can combine these equations into the remarkably simple form
\[
 \acomm{\gen{Q}^{a+\mathfrak{c}}}{\comm{ \gen{Q}^{b-\mathfrak{d}}}{  \gen{X} }} + \lambda  \, \varepsilon^{ab} \,  \acomm{\hat{\gen{Q}}^{\mathfrak{c}}}{\comm{ \hat{\gen{S}}^{\mathfrak{d}}}{  \gen{X} }} = \quarter \,  \varepsilon^{ab} \varepsilon^{\mathfrak{cd}} \, \mathcal{H}. \label{eq:generalxconjugate}
 \]

The three equations (\ref{eq:neweqforx}), (\ref{eq:newfirsteqforh}) (and Hermitian conjugate), and (\ref{eq:generalxconjugate}) are necessary and sufficient conditions for $\gen{X}$. Consider again the case of the planar $\alg{psu}(1,1|2)$ sector.  Assuming integrability, it is now immediately apparent that $\gen{X}$ can be shifted by linear combinations of any of the local higher charges $\mathcal{Q}_{(i)}$  that commute with the dilatation generator $\lambda \, \mathcal{H}$ and all of the extended  $\alg{psu}(1,1|2) \times \alg{psu}(1|1)^2$ generators. However, these transformations of $\gen{X}$ have no physical consequence precisely because $\gen{X}$ only appears through commutators with the ordinary symmetry generators. It is an open problem whether there are  solutions for $\gen{X}$ that both satisfy these more general equations, and are not physically equivalent to an iterative solution in terms of $\alg{psu}(1|1)^2$ generators and some auxiliary $\gen{h}$.  If no more general solutions exist, the conjecture that the iterative ansatz applies at all orders to the gauge theory will automatically be satisfied.

\subsection{A solution using a $\gen{B}$-triplet}
We now present an alternative consistent algebraic ansatz for the solution in terms of a $\gen{B}$-triplet $\tilde{\gen{X}}^{\mathfrak{ab}}$. We then show that the iterative three-loop solution for the $\alg{psu}(1,1|2)$ sector using $\gen{h}$ can be put in this form via a similarity transformation.

The ansatz depends on auxiliary generators $\tilde{\gen{X}}^{\mathfrak{ab}}$ that satisfy
\[
\comm{\tilde{\gen{X}}^{\mathfrak{ab}}}{\gen{B}^{\mathfrak{cd}}} = \varepsilon^{\mathfrak{cb}} \tilde{\gen{X}}^{\mathfrak{ad}}  - \varepsilon^{\mathfrak{ad}} \tilde{\gen{X}}^{\mathfrak{cb}}. 
\]
They also commute with $\gen{R}$ and $\gen{J}^{+-}_0$. The $\tilde{\gen{X}}^{\mathfrak{ab}}$  generate  translations in $\lambda$ of the supercharges as 
\begin{align}
\comm{\gen{Q}^{\mathfrak{a}}}{\tilde{\gen{X}}^{\mathfrak{bc}}} & =  \half \varepsilon^{\mathfrak{ba}} \frac{\partial}{\partial \lambda} \gen{Q}^{\mathfrak{c}} + \half \varepsilon^{\mathfrak{ca}}  \frac{\partial}{\partial \lambda} \gen{Q}^{\mathfrak{b}}, & & \notag \\ \comm{\hat{\gen{Q}}^{\mathfrak{a}}}{\tilde{\gen{X}}^{\mathfrak{bc}}} & = -\half \varepsilon^{\mathfrak{ba}}  \frac{\partial}{\partial \lambda} \hat{\gen{Q}}^{\mathfrak{c}} - \half \varepsilon^{\mathfrak{ca}}  \frac{\partial}{\partial \lambda} \hat{\gen{Q}}^{\mathfrak{b}}, & \comm{\hat{\gen{S}}^{\mathfrak{a}}}{\tilde{\gen{X}}^{\mathfrak{bc}}} & = -\half \varepsilon^{\mathfrak{ba}}  \frac{\partial}{\partial \lambda} \hat{\gen{S}}^{\mathfrak{c}} - \half \varepsilon^{\mathfrak{ca}}  \frac{\partial}{\partial \lambda} \hat{\gen{S}}^{\mathfrak{b}}. \label{eq:btripletansatz}
\end{align}
We have suppressed the first two indices of the $\alg{psu}(1,1|2)$ supercharges since these relation do not depend on them. The $\alg{sl}(2)$ subalgebra generators not included here follow from closure of the $\alg{psu}(1,1|2)$ algebra. Note that these relations imply 
\[
 \frac{\partial}{\partial \lambda} \gen{Q}^{\mathfrak{a}} = -\sfrac{2}{3} \varepsilon_{\mathfrak{bc}} \comm{ \gen{Q}^{\mathfrak{b}}}{\tilde{\gen{X}}^{\mathfrak{ca}}}, \quad  \frac{\partial}{\partial \lambda} \hat{\gen{Q}}^{\mathfrak{a}} = \sfrac{2}{3} \varepsilon_{\mathfrak{bc}} \comm{ \hat{\gen{Q}}^{\mathfrak{b}}}{\tilde{\gen{X}}^{\mathfrak{ca}}}, \quad  \frac{\partial}{\partial \lambda} \hat{\gen{S}}^{\mathfrak{a}} = \sfrac{2}{3} \varepsilon_{\mathfrak{bc}} \comm{ \hat{\gen{S}}^{\mathfrak{b}}}{\tilde{\gen{X}}^{\mathfrak{ca}}}.
 \]
For a realization of this ansatz to satisfy all commutation relations one must only check, for example,
\begin{gather}
\comm{\gen{Q}^{a\beta>}}{\tilde{\gen{X}}^{>>}} = 0, \quad \comm{\hat{\gen{Q}}^{>}}{\tilde{\gen{X}}^{>>}}=0, \quad \comm{\hat{\gen{S}}^{>}}{\tilde{\gen{X}}^{>>}} =0, \label{eq:bconsistency}\\
\acomm{\gen{Q}^{a+<}}{\comm{\gen{Q}^{b->}}{\tilde{\gen{X}}^{<>}}} + \lambda \varepsilon^{ab} \acomm{\hat{\gen{S}}^{>}}{\comm{\hat{\gen{Q}}^{<}}{\tilde{\gen{X}}^{<>}}} = -\sfrac{1}{8} \varepsilon^{ab}  \mathcal{H} \label{eq:btripconjugate}.
\end{gather}
 The equations on the first line ensure that the derivatives of the supercharges transform properly with respect to $\gen{B}$ and that (\ref{eq:btripletansatz}) can be satisfied. Using $\gen{B}$ symmetry, nilpotency of supercharges, and the Jacobi identity, one can then show that the $\alg{psu}(1|1)^2$ algebra is satisfied, that the $\alg{psu}(1,1|2)$ and $\alg{psu}(1|1)^2$ generators commute, and that commutation relations for two $\gen{Q}^{+}$ or for two $\gen{Q}^{-}$ are satisfied. Since this check is straightforward and involves similar steps to those used previously, we leave it to the reader. To obtain the necessary condition for the remaining requirements for $\alg{psu}(1,1|2)$ generators, one can use similar steps to those leading to (\ref{eq:generalxconjugate}) combined with (\ref{eq:btripletansatz}). This results in (\ref{eq:btripconjugate}). Again, the details are left as an exercise for the reader. 
 
 Now, we again consider the (seemingly) less general ansatz presented earlier in terms of $\gen{h}$. A similarity transformation maps that solution to a solution of the form (\ref{eq:btripletansatz}) as follows.  For all  Lie algebra generators $J$, 
 \[
 \frac{\partial}{\partial \lambda} J \mapsto \frac{\partial}{\partial \lambda} J +  \half \comm{J}{\comm{\gen{h}}{\mathcal{H}}}. \label{eq:simtransform}
 \]
 This results in a ``new'' solution of the form (\ref{eq:btripletansatz}) with 
 \[
 \tilde{\gen{X}}^{\mathfrak{ab}} = \acomm{\hat{\gen{Q}}^{\{ \mathfrak{a} }}{\comm{\hat{\gen{S}}^{\mathfrak{b} \}}}{\gen{h}}}.
 \]
 Of course, since in this case the two solutions are related by a similarity transformation, the spectrum of the dilatation generator is unchanged. Also, the consistency condition for $ \tilde{\gen{X}}^{\mathfrak{ab}}$ (\ref{eq:bconsistency}) is satisfied because of the nilpotency of supercharges and the first equation for $\gen{h}$ of (\ref{eq:eqforh}).

We explicitly check this relationship for $\hat{\gen{S}}^>$ and for $\gen{Q}^{a-<}$, as examples. First consider $\hat{\gen{S}}^>$. In the original form of the solution, 
\<
 \frac{\partial}{\partial \lambda} \hat{\gen{S}}^> \eq -\comm{\hat{\gen{S}}^>}{\gen{X}}
 \nln
 \eq -\comm{\hat{\gen{S}}^>}{\acomm{\hat{\gen{Q}}^>}{\comm{\hat{\gen{S}}^<}{\gen{h}}}},
 \>
 where we used (\ref{eq:moreqhatcommx}) to reach the second line. Adding the similarity transformation (\ref{eq:simtransform}) and using the $\alg{psu}(1|1)^2$ relations and the Jacobi identity yields
 \<
 -\comm{\hat{\gen{S}}^>}{\acomm{\hat{\gen{Q}}^>}{\comm{\hat{\gen{S}}^<}{\gen{h}}}} + \half \comm{\hat{\gen{S}}^>}{\comm{\gen{h}}{\mathcal{H}}} \eq -\comm{\hat{\gen{S}}^>}{\acomm{\hat{\gen{Q}}^>}{\comm{\hat{\gen{S}}^<}{\gen{h}}}} - \comm{\hat{\gen{S}}^>}{\comm{\gen{h}}{\acomm{\hat{\gen{Q}}^>}{\hat{\gen{S}}^<}}}
 \nln
 \eq \comm{\hat{\gen{S}}^>}{\acomm{\hat{\gen{S}}^<}{\comm{\hat{\gen{Q}}^>}{\gen{h}}}}
 \nln
 \eq  \comm{\hat{\gen{S}}^<}{\acomm{\hat{\gen{Q}}^>}{\comm{\hat{\gen{S}}^>}{\gen{h}}}}
  \nln
 \eq   \comm{\hat{\gen{S}}^<}{\tilde{\gen{X}}^{>>}},
 \>
 as in (\ref{eq:btripletansatz}). Next we show that the similarity transformation also works for  $\gen{Q}^{a-<}$. In the original solution  
 \<
  \frac{\partial}{\partial \lambda} \gen{Q}^{a-<} \eq - \comm{ \gen{Q}^{a-<}}{\gen{X}}
  \nln
  \eq - \comm{ \gen{Q}^{a-<}}{\acomm{\hat{\gen{Q}}^>}{\comm{\hat{\gen{S}}^<}{\gen{h}}}} - \comm{ \gen{Q}^{a-<}}{\acomm{\hat{\gen{S}}^>}{\comm{\hat{\gen{Q}}^<}{\gen{h}}}}
  \nln
  \eq - \comm{ \gen{Q}^{a-<}}{\acomm{\hat{\gen{Q}}^>}{\comm{\hat{\gen{S}}^<}{\gen{h}}}}
  \>
  We used the expression for $\gen{X}$ (\ref{eq:xsimp>}) and the first equation of (\ref{eq:eqforh}). Adding the similarity transformation, and again using the $\alg{psu}(1|1)^2$ relations, the Jacobi identity, and the first equation of (\ref{eq:eqforh}) results in
  \<
 - \comm{ \gen{Q}^{a-<}}{\acomm{\hat{\gen{Q}}^>}{\comm{\hat{\gen{S}}^<}{\gen{h}}}} + \half \comm{\gen{Q}^{a-<}}{\comm{\gen{h}}{\mathcal{H}}} \eq    \comm{ \gen{Q}^{a-<}}{\acomm{\hat{\gen{S}}^<}{\comm{\hat{\gen{Q}}^>}{\gen{h}}}}
  \nln
  \eq  -\comm{ \gen{Q}^{a->}}{\acomm{\hat{\gen{S}}^<}{\comm{\hat{\gen{Q}}^<}{\gen{h}}}}
  \nln
  \eq \comm{ \gen{Q}^{a->}}{\acomm{\hat{\gen{Q}}^<}{\comm{\hat{\gen{S}}^<}{\gen{h}}}} 
  \nln
  \eq  \comm{ \gen{Q}^{a->}}{\tilde{\gen{X}}^{<<}},
  \>
in agreement with the $\gen{B}$-triplet ansatz (\ref{eq:btripletansatz}). Similar checks can be performed for $\gen{Q}^{a+<}$ and $\hat{\gen{Q}}^>$ (for example), and then $\gen{B}$ symmetry ensures that the similarity transformation indeed maps the original solution into the form of (\ref{eq:btripletansatz}).
 
It is possible that there are solutions of triplet form that cannot be related by a similarity transformation to singlet ansatz solutions. If that is the case, such solutions may be  relevant to the $\alg{psu}(1,1|2)$ sector of the gauge theory at higher loops, potentially falsifying our conjecture.  The results of this work imply that at least to three loops in the planar $\alg{psu}(1,1|2)$ sector, and for the leading $\beta_{2,3}$ phase contribution,  there is no need to consider the $\gen{B}$-triplet ansatz. Also, extending the type of iterative solution found in this work to larger sectors of the gauge theory may require an ansatz more similar to this one then to the singlet ansatz; it may be necessary for the auxiliary generators to carry a charge with respect to one or more Cartan generators. 
 
\section{Conclusions and discussion \label{sec:conc}} 
 We have presented an algebraic solution for $\lambda$-dependent representations of the extended $\alg{psu}(1,1|2) \times \alg{psu}(1|1)^2$ algebra. This solution depends on a generator of $\lambda$-translations, $\gen{X}$. In turn, $\gen{X}$ is built simply from supercharges and from one auxiliary generator $\gen{h}$ that must satisfy certain Serre-relation-like equations. We applied this ansatz to the $\alg{psu}(1,1|2)$ sector of $\mathcal{N}=4$ SYM, extending the results of \cite{Zwiebel:2005er} to three loops. Strong evidence implies that the new solution for $\gen{h}$ at NLO gives the three-loop planar dilatation generator of this sector, with wrapping interactions included naturally.  Also, we identified two types of homogeneous solutions for $\gen{h}_1$. While we used maximal transcendentality to identify the apparent gauge theory solution, we gave evidence and expect that a simple gauge theory structural constraint eliminates this homogeneous freedom at three loops. Also, one of these new homogeneous solutions corresponds to the leading phase contribution of the Beth ansatz, which appears starting at four loops. 
 
As stated in the introduction, these successes of the algebraic ansatz lead us to conjecture that it is satisfied by the gauge theory at all orders. Of course, the conjecture would be automatically true if the algebraic ansatz includes all solutions of the Lie algebra and basic structural constraints. Such uniqueness of the algebraic ansatz is plausible partly because the ansatz automatically ensures the correct multiplet structure for the length-changing supercharges.  Perhaps representation theory analysis can answer this question of uniqueness.  
 
 Clearly there is more to be understood about the algebraic solution, even independently of its gauge theory realization. Adding $\gen{h}$ to the  set of $\alg{psu}(1,1|2) \times \alg{psu}(1|1)^2$ generators does not yield a closed algebra. Instead, it is reasonable to believe that this enlarged set of generators can be embedded (usefully) within a closed algebraic structure, but this hypothesized algebraic structure remains mysterious. Perhaps the simplest possibility is to use the maximally extended version of $\alg{psu}(1,1|2) \times \alg{psu}(1|1)^2$, which includes a $\gen{B}$-triplet of central charges that vanish for the gauge theory representation. Another possibility is the exceptional superalgebra $\mathfrak{d}(2, 1; \epsilon)$, considered in the context of AdS/CFT first in \cite{Beisert:2005tm}, and more recently in \cite{Heckenberger:2007ry, Matsumoto:2008ww}.  

 Even if the algebraic ansatz gives the most general solution for the $\alg{psu}(1,1|2)$ spin chain, beyond three loops it appears that there still will be vast freedom for homogeneous solutions. The most direct way to proceed is to consider larger sectors then $\alg{psu}(1,1|2)$~\footnote{The $\alg{psu}(1,2|3)$ sector or the full $\alg{psu}(2,2|4)$ spin chain are probably the only useful choices.}, which may eliminate the integrability-breaking homogeneous solutions. This is a challenging problem because larger sectors have greater complexity due to symmetry generators that expand in all integer powers of $g$ with more general length-changing interactions. Still, the remarkably simple algebraic form of the $\alg{psu}(1,1|2)$ sector solution  provides hope that significant progress is possible. More concretely, this solution gives an important constraint since any solution for a larger sector must be compatible with it. Also, understanding how strong Lie algebra constraints are for the full theory is an important problem. It is possible that understanding this would be sufficient to prove integrability. If instead the full $\alg{psu}(2,2|4)$ spin chain has homogeneous (structurally allowed) solutions that break integrability, it would lead to the interesting question of what further gauge theory properties are required for integrability.
 
 Alternatively one could fix the homogeneous freedom by instead assuming integrability. The resulting solutions would hopefully then lead to a better understand of the gauge theory origin of integrability. For integrable solutions, the Lie algebra symmetry is enhanced by generators that act bilocally on the spin chain, which in turn generate an infinite-dimensional Yangian symmetry \cite{Drinfeld:1985rx}.  Beyond the leading order result for the full $\mathcal{N}=4$ planar theory \cite{Dolan:2003uh,Dolan:2004ps}, previous work\footnote{Also, the Hopf algebra structure and Yangian symmetry of the AdS/CFT S-matrix has been investigated in \cite{Gomez:2006va,Plefka:2006ze,Beisert:2006qh,Gomez:2007zr,
Torrielli:2007mc,Beisert:2007ds,Young:2007wd,Moriyama:2007jt,Matsumoto:2007rh,Beisert:2007ty,Matsumoto:2008ww,Spill:2008tp,deLeeuw:2008dp,Torrielli:2008wi}. Also see \cite{Ihry:2008gm}.} has focused on sectors of rank one \cite{Serban:2004jf,Agarwal:2004sz,Agarwal:2005ed,Beisert:2007jv} or rank two \cite{Zwiebel:2006cb}, which may be too small to reveal iterative structure.   For the $\alg{psu}(1,1|2)$ sector, an exciting prospect is that some auxiliary generator(s) of $\lambda$-translations can also be used to obtain corrections to the nonlocal Yangian generators, including Yangian generators corresponding to the $\alg{su}(2)$ automorphism \cite{Beisert:2007sk}. Another possibility is that the iterative structure of the local spin-chain generators described here can be extended to the higher local charges associated with integrability. In any case, constructing Yangian generators or higher local charges would reveal additional constraints on $\gen{h}$, of course eliminating any integrability-breaking homogeneous solutions. Related to integrability, it would also be very interesting to find the three-loop $\alg{psu}(1,1|2)$ sector Baxter operator, which may be a step toward a R-matrix formulation of the long-range asymptotic spin chain.

There are other exciting possible directions for further research. For instance, the algebraic solution's iterative structure may allow for a generalization of the relation of the $\alg{su}(2)$ subsector to the Hubbard model. Up to wrapping interactions, the  rational part of the $\alg{su}(2)$ sector dilatation generator corresponds to a strong coupling expansion of the Hubbard model \cite{Rej:2005qt}. Finding a generalization of the Hubbard model related in a parallel way to the $\alg{psu}(1,1|2)$ sector would likely give a more efficient way to compute (the rational part of) $\gen{h}$, and of course would be of great interest for multiple other reasons. The regular transcendentality pattern and integer coefficients appearing in $\gen{h}$ at the first two orders are perhaps hints of such a possible relation.  Alternatively, consider the relationship between anomalous dimensions and the BFKL equation \cite{Lipatov:1976zz,Kuraev:1977fs,Balitsky:1978ic}, which constrains the singularities of twist-two anomalous dimensions analytically continued for negative integer spin, as analyzed in \cite{Kotikov:2007cy}.  Perhaps the connection to the BFKL equation can give useful constraints directly for $\gen{h}$. Finding relations to a generalization of the Hubbard model or to BFKL physics also may provide information about the wrapping interactions for $\gen{h}$, which are needed starting at four loops if the iterative algebraic structure is to apply beyond the asymptotic regime.

One may wonder about a strong coupling expansion of the algebraic ansatz. It seems that such an expansion cannot be related simply and directly to the string dual because there is no closed $\alg{psu}(1,1|2)$ sector at strong coupling. To make contact in this way, it is necessary to generalize such an algebraic ansatz to all of $\alg{psu}(2,2|4)$. On the other hand, the successful interpolations from weak to strong coupling mentioned in the introduction depended crucially on the phase. Therefore, it may be instructive to find the higher homogeneous solutions for $\gen{h}$ corresponding to the leading $\beta_{r,s}$ phase contributions (for $s>3$). Since these homogeneous solutions only need to satisfy equations involving leading order Lie algebra generators, this should be a tractable problem, especially given the simple form of the $\beta_{2,3}$ solution (\ref{eq:phasesolution}).
 
 Finally, there are interesting possible generalizations beyond planar $\mathcal{N}=4$ SYM. As discussed in Section \ref{sec:wrapping}, it is straightforward to generalize the algebraic ansatz to the nonplanar theory precisely because the ansatz is algebraic. Furthermore, the limited number of independent nonplanar interaction structures at low orders suggest that a naive lifting of the solution  may give the nonplanar three-loop dilatation generator. It would be wonderful if this could be verified. Also, related to the above discussion of the connection to the BFKL equation, it would be very interesting to find other representations of the extended algebra that realize the algebraic ansatz. Particularly interesting are those with continuous values of $\alg{su}(1,1)$ spin, which is a feature of certain nonlocal gauge theory operators \cite{Balitsky:1987bk,Braun:2003rp} that have been investigated recently \cite{Hofman:2008ar}.
 
 \subsection*{Acknowledgements}
I thank Niklas Beisert, Andrei Belitsky, Nick Dorey, and Alessandro Torrielli for interesting discussions and insightful suggestions.
 
\appendix
\section{Relations to previous notations \label{sec:notation}}
First we review the restriction of the full $\mathcal{N}=4$ SYM $\alg{psu}(2,2|4)$ spin chain to the $\alg{psu}(1,1|2)$ sector, including the relation of the $\alg{psu}(1,1|2)$ symmetry generators and ``fields'' to those of the full spin chain. Then we present the mapping between the notation and conventions of this work and those of \cite{Zwiebel:2005er}, where the two-loop $\alg{psu}(1,1|2)$ sector dilatation generator was first obtained.

States of the $\mathcal{N}=4$ SYM spin chain (gauge-invariant local operators) can be classified according to eigenvalues with respect to $\alg{su}(2)_\text{L} \times \alg{su}(2)_\text{R}$ Lorentz and $\alg{su}(4)$ $\gen{R}$-symmetry generators, which we denote by $L^\alpha{}_\beta$, $\dot{L}^{\dot{\alpha}}{}_{\dot{\beta}}$, and $R^a{}_b$. Then  the $\alg{psu}(1,1|2)$ sector contains the states of the full $\alg{psu}(2,2|4)$ spin chain with classical dimension $D_0$ satisfying \cite{Beisert:2004ry, Zwiebel:2007th}
\<
D_0 = L^1{}_1 - L^2{}_2 - 2 R^4{}_4, \nln
D_0 = \dot{L}^1{}_1 - \dot{L}^2{}_2 + 2 R^3{}_3.
\>

The $\alg{psu}(1,1|2)$ symmetry generators are related simply to generators of the $\alg{psu}(2,2|4)$ chain. The relation to the notation of \cite{Beisert:2004ry} is as in \cite{Beisert:2007sk}:
\begin{align}
\gen{Q}^{a+>} & = \gen{Q}^a{}_1, & \gen{Q}^{a+<} & = \varepsilon^{ab}\dot{\gen{Q}}_{1b}, & \gen{R}^{ab} &= \half \varepsilon^{ac} \gen{R}^b{}_c +  \half \varepsilon^{bc} \gen{R}^a{}_c,
\notag \\
\gen{Q}^{a-<} & = \varepsilon^{ab}\gen{S}^1{}_b, & \gen{Q}^{a->} & = \dot{\gen{S}}^{a1}, &   &
\notag \\
\gen{J}^{++} & = \gen{P}_{11}, & \gen{J}^{--} & = \gen{K}^{11}, & \gen{J}^{+-} & = \half \gen{D} + \half \gen{L}^1{}_1 + \half \dot{\gen{L}}^1{}_1.
\end{align}
However, the $\alg{psu}(1|1)^2$ generators have been rescaled by $\sqrt{\lambda}$,
\begin{align}
\sqrt{\lambda} \, \hat{\gen{Q}}^< & = \dot{\gen{Q}}_{23}, & \sqrt{\lambda} \, \hat{\gen{Q}}^> & = -\gen{Q}^4{}_2, 
\notag \\
\sqrt{\lambda} \, \hat{\gen{S}}^> & = \dot{\gen{S}}^{32}, & \sqrt{\lambda} \, \hat{\gen{S}}^< & = \gen{S}^2{}_4. 
\end{align}
This rescaling is possible because these generators, within the $\alg{psu}(1,1|2)$ sector and before the rescaling, expand at $\lambda=0$ in only odd powers of $\sqrt{\lambda}$ .
Again using the notation of \cite{Beisert:2004ry}, the elements of the $\alg{psu}(1,1|2)$ module are related to the ``fields'' of $\mathcal{N}=4$ SYM as \cite{Beisert:2007sk},
\[
\state{\phi_a^{(n)}} \simeq \frac{1}{n!}(\mathcal{D}_{11})^n \Phi_{a3}, \quad \state{\psi_<^{(n)}} \simeq \frac{1}{n!\sqrt{n+1}}(\mathcal{D}_{11})^n \dot{\Psi}^4{}_1, \quad \state{\psi_>^{(n)}} \simeq \frac{1}{n!\sqrt{n+1}}(\mathcal{D}_{11})^n \Psi_{13}.
\]

Next we give the mapping from the symmetry generators and states of this work to those of \cite{Zwiebel:2005er}. In addition to some sign changes, the $\gen{R}$ index values $1,2$ have to be replaced by $\updownarrow$, and the $\gen{B}$ index values $<, >$ are mapped to $\leftrightarrow$. More precisely,
\begin{align}
\gen{Q}^{1+>} & = -\overrightarrow{\gen{Q}}^{+\downarrow}, & \gen{Q}^{2+>} & = -\overrightarrow{\gen{Q}}^{+\uparrow}, & \gen{Q}^{1+<} & = \overleftarrow{\gen{Q}}^{+\downarrow}, & \gen{Q}^{2+<} & = -\overleftarrow{\gen{Q}}^{+\uparrow},
\notag \\
\gen{Q}^{1->} & = -\overrightarrow{\gen{Q}}^{-\downarrow}, & \gen{Q}^{2->} & = -\overrightarrow{\gen{Q}}^{-\uparrow}, & \gen{Q}^{1-<} & = \overleftarrow{\gen{Q}}^{-\downarrow}, & \gen{Q}^{2-<} & = -\overleftarrow{\gen{Q}}^{-\uparrow},
\notag \\
\gen{R}^{11} & = \gen{R}^{\downarrow \downarrow}, & \gen{R}^{22} & = -\gen{R}^{\uparrow \uparrow}, & \gen{R}^{12} & = -\half \gen{R}^{0}, & &
\notag \\
\gen{J}^{++} & = \gen{J}^{++}, & \gen{J}^{--} & = \gen{J}^{--}, & \gen{J}^{+-} & = \half \gen{J}^{0}, &  &
\notag \\
\sqrt{\lambda} \, \hat{\gen{Q}}^> & = -\overrightarrow{\gen{T}}^{+}, & \sqrt{\lambda} \, \hat{\gen{Q}}^< & = -\overleftarrow{\gen{T}}^{+}, & \sqrt{\lambda} \, \hat{\gen{S}}^> & = -\overrightarrow{\gen{T}}^{-}, & \sqrt{\lambda} \, \hat{\gen{S}}^< & =  \overleftarrow{\gen{T}}^{-}.
\end{align}
For the module elements the relations are
\[
\state{\phi_1^{(n)}} = \state{\phi_n^{\uparrow}}, \quad \state{\phi_2^{(n)}} = \state{\phi_n^{\downarrow}}, \quad \state{\psi_<^{(n)}} = -\state{\overrightarrow{\psi}_n}, \quad \state{\psi_>^{(n)}} = \state{\overleftarrow{\psi}_n}.
\]
Finally, note that \cite{Zwiebel:2005er} used $g$ as the coupling constant, with the normalization $g^2 = \lambda / (8 \pi^2)$.

\section{A Chevalley-Serre basis for $\alg{psu}(1,1|2) \ltimes \Real^3$ \label{sec:Chevalley}}

Presenting the centrally extended $\alg{psu}(1,1|2)$ algebra in a (fermionic) Chevalley-Serre basis provides a simple way to check that the proof of Section \ref{sec:proof} includes all generators of this (sub-)algebra, as will be explained below. We need the central extension to represent the algebra in this basis, but these central charges vanish in the gauge theory realization and in the algebra considered throughout the rest of this work. Since this algebra only differs by a change of signature from the $\alg{psu}(2|2) \ltimes \Real^3$ algebra recently considered in \cite{Spill:2008tp} in relation to the Yangian of the AdS/CFT S-matrix, we will use the same (standard) presentation and notation for the basis, commutation relations, and Serre relations.

The basis consists of three Cartan elements $\mathfrak{H}_i$ and three pairs of fermionic elements $\mathfrak{E}^\pm_i$ satisfying
\<
\comm{\mathfrak{H}_i}{\mathfrak{H}_j} \eq 0, \label{eq:HiHj} \\
\comm{\mathfrak{H}_i}{\mathfrak{E}_j^\pm} \eq \pm a_{ij} \mathfrak{E}_j^\pm, \label{eq:HiEj} \\
\acomm{\mathfrak{E}_i^+}{\mathfrak{E}_j^-} \eq \delta_{ij} \mathfrak{H}_i. \label{eq:EiEj}
\>
The Cartan matrix $a_{ij}$ is 
\[
a_{ij} = \begin{pmatrix} 0 & -1 & 1 \\ -1 & 0 & 0 \\ 1 & 0 & 0 \end{pmatrix}.
\]
Furthermore, there are Serre relations for the $\mathfrak{E}^\pm_i$ 
\[
\comm{\mathfrak{E}_1^{\pm}}{\acomm{\mathfrak{E}_1^{\pm}}{\mathfrak{E}_2^{\pm}}}=\comm{\mathfrak{E}_2^{\pm}}{\acomm{\mathfrak{E}_2^{\pm}}{\mathfrak{E}_1^{\pm}}}=0, \quad \acomm{\mathfrak{E}_2^{\pm}}{\mathfrak{E}_3^{\pm}} = \text{central}. \label{eq:Serre}
\]

The above presentation matches that of \cite{Spill:2008tp}. However, the difference now appears through the realization of the  $\mathfrak{H}_i$ and $\mathfrak{E}^\pm_i$ in terms of generators. To match the commutation relations given in Section \ref{sec:comm}, we can choose for instance 
\begin{align}
\mathfrak{E}^+_1 & = \gen{Q}^{2+>}, & \mathfrak{E}^-_1 & = \gen{Q}^{1-<}, & \mathfrak{H}_1 & = -\gen{R}^{12}-\gen{J}^{+-} -\gen{C}^{<>}, \notag \\
\mathfrak{E}^+_2 & = \gen{Q}^{1+<}, & \mathfrak{E}^-_2 & = \gen{Q}^{2->}, & \mathfrak{H}_2 & = \gen{R}^{12}-\gen{J}^{+-} +\gen{C}^{<>}, \notag \\
\mathfrak{E}^+_3 & = \gen{Q}^{2-<}, & \mathfrak{E}^-_3 & = \gen{Q}^{1+>}, & \mathfrak{H}_3 & = -\gen{R}^{12}+\gen{J}^{+-} +\gen{C}^{<>}.
\end{align}
Here $\gen{C}^{<>}$ is the central element of the triplet of central charges that extend $\alg{psu}(1,1|2)$.

We now briefly explain how the proof in Section \ref{sec:proof} covers all of the commutation and Serre relations given above. Recall that the first itemized step of that proof showed that commutation relations between any two lowering (raising) generators are satisfied for the algebraic solution, and the last itemized step showed that all of the commutation relations between a lowering and a raising $\alg{psu}(1,1|2)$ supercharge are satisfied. Now (\ref{eq:HiHj}) follows from the vanishing of the central charges and the preservation of manifest $\gen{R}$ symmetry. (\ref{eq:HiEj}) follows for the above reasons combined with the vanishing commutator between $\gen{X}$ and $\gen{J}^{+-}_0$ and the centrality of $\delta \gen{D}$. Instances of (\ref{eq:EiEj}) are either included within the set of commutators between raising and lowering $\alg{psu}(1,1|2)$ supercharges, or within the set of commutators between two raising (lowering) generators.  Finally, the first two Serre relations follow from the proof for commutators between two raising (lowering) generators, and the last Serre relation (with ``central''$=0$) is included within the set of commutators between raising and lowering $\alg{psu}(1,1|2)$ generators. 

\section{Additional details of proof of Section \ref{sec:solution} \label{sec:extradetails}}
Here we prove that (\ref{eq:q+q-}) remains satisfied after shifts in $\lambda$ generated by $\gen{X}$.
Recall that (\ref{eq:q+q-}) is
\[
\acomm{\gen{Q}^{1+>}}{\gen{Q}^{2-<}}=\gen{R}^{12} + \gen{J}^{+-}. \label{eq:q+q-again}
\]
Taking the derivative, using the Jacobi identity, and substituting (\ref{eq:q+q-again}) back into the resulting equation leads to

\<
\frac{\partial}{\partial \lambda} \acomm{\gen{Q}^{1+>}(\lambda)}{\gen{Q}^{2_<}(\lambda)}_{|\lambda=\lambda_0} \eq  \acomm{\comm{\gen{Q}^{1+>}(\lambda_0)}{\gen{X}(\lambda_0)}}{\gen{Q}^{2-<}(\lambda_0)} 
\nl
- \acomm{\gen{Q}^{1+>}(\lambda_0)}{\comm{\gen{Q}^{2-<}(\lambda_0)}{\gen{X}(\lambda_0)}} 
\nln
\eq \comm{\gen{J}^{+-}(\lambda_0)}{\gen{X}(\lambda_0)} - 2  \acomm{\gen{Q}^{1+>}(\lambda_0)}{\comm{\gen{Q}^{2-<}(\lambda_0)}{\gen{X}(\lambda_0)}}. 
\notag \\
\label{eq:twoterms} \>
Now we will simplify the two terms of the last line separately. For the first term, applying (\ref{eq:sharedcentralcharge}) and (\ref{eq:psu11squaredalg}) yields
\<
 \comm{\gen{J}^{+-}(\lambda_0)}{\gen{X}(\lambda_0)} \eq \lambda_0 \comm{\acomm{\hat{\gen{Q}}^<(\lambda_0)}{\hat{\gen{S}}^>(\lambda_0)}}{\gen{X}(\lambda_0)} \nln
 \eq \lambda_0 \acomm{\comm{\hat{\gen{Q}}^<(\lambda_0)}{\gen{X}(\lambda_0)}}{\hat{\gen{S}}^>(\lambda_0)} + \lambda_0 \acomm{\hat{\gen{Q}}^<(\lambda_0)}{\comm{\hat{\gen{S}}^>(\lambda_0)}{\gen{X}(\lambda_0)}} \nln
 \eq - \lambda_0 \acomm{\comm{\hat{\gen{Q}}^<(\lambda_0)}{\acomm{\hat{\gen{S}}^<(\lambda_0)}{\comm{\hat{\gen{Q}}^>(\lambda_0)}{\gen{h}(\lambda_0)}}}}{\hat{\gen{S}}^>(\lambda_0)} \notag \\
 & + & \,  \lambda_0 \acomm{\hat{\gen{Q}}^<(\lambda_0)}{\comm{\hat{\gen{S}}^>(\lambda_0)}{\acomm{\hat{\gen{Q}}^>(\lambda_0)}{\comm{\hat{\gen{S}}^<(\lambda_0)}{\gen{h}(\lambda_0)}}}}. \label{eq:firsttermsimp}
\>
 The last equality follows from (\ref{eq:moreqhatcommx}). 
 
 For the second term of the last line of (\ref{eq:twoterms}), we first replace $\gen{X}$ with a single nested commutator, using the expression for $\gen{X}$ (\ref{eq:xsimp>}), the vanishing anticommutators between $\alg{psu}(1,1|2)$ and $\alg{psu}(1|1)^2$ supercharges, and the first equation for $\gen{h}$ of (\ref{eq:eqforh}). Then  vanishing anticommutators between $\alg{psu}(1,1|2)$ and $\alg{psu}(1|1)^2$ supercharges allow us to apply the second equation of (\ref{eq:eqforh}). 
\<
  \acomm{\gen{Q}^{1+>}(\lambda_0)}{\comm{\gen{Q}^{2-<}(\lambda_0)}{\gen{X}(\lambda_0)}} \eq     \acomm{\gen{Q}^{1+>}(\lambda_0)}{\comm{\gen{Q}^{2-<}(\lambda_0)}{\acomm{\hat{\gen{Q}}^> (\lambda_0)}{\comm{\hat{\gen{S}}^<(\lambda_0)}{\gen{h}(\lambda_0)}}}}  \nln
\eq   \acomm{\hat{\gen{Q}}^> (\lambda_0)}{\comm{\hat{\gen{S}}^<(\lambda_0)}{-\half \gen{B}^{<>} + \quarter \gen{L} + \lambda_0 \gen{X}^{><}(\lambda_0)}} \nln
\eq  \half \acomm{\hat{\gen{Q}}^>(\lambda_0)}{\hat{\gen{S}}^<(\lambda_0)} +
\lambda_0 \acomm{\hat{\gen{Q}}^> (\lambda_0)}{\comm{\hat{\gen{S}}^<(\lambda_0)}{ \gen{X}^{><}(\lambda_0)}} \nln
\eq   -\quarter \mathcal{H}(\lambda_0) 
\nl
 + \lambda_0 \acomm{\hat{\gen{Q}}^> (\lambda_0)}{\comm{\hat{\gen{S}}^<(\lambda_0)}{\acomm{ \hat{\gen{Q}}^<(\lambda_0)}{\comm{\hat{\gen{S}}^>(\lambda_0)}{\gen{h}(\lambda_0)}}}}. \notag \\
\> 
 The simplification of the last two steps depends on commutation relations of the ordinary extended Lie algebra generators given in Section \ref{sec:comm}, as well as substitution for $\gen{X}^{<>}$ using (\ref{eq:definex}).

The next step is to substitute this result and (\ref{eq:firsttermsimp}) back into (\ref{eq:twoterms}), and to combine and cancel terms. Then (\ref{eq:moreqhatcommx}) allows nested commutators to be replaced with  $\gen{X}$. The resulting expression can be identified with a derivative with respect to $\lambda$, using the $\alg{psu}(1|1)^2$ commutation relations (\ref{eq:psu11squaredalg}).
\<
\frac{\partial}{\partial \lambda} \acomm{\gen{Q}^{1+>}(\lambda)}{\gen{Q}^{2_<}(\lambda)}_{|\lambda=\lambda_0} \eq  \half \mathcal{H}(\lambda_0)  
\nl
-  \lambda_0 \acomm{\hat{\gen{Q}}^< (\lambda_0)}{\comm{\hat{\gen{S}}^>(\lambda_0)}{\acomm{ \hat{\gen{Q}}^>(\lambda_0)}{\comm{\hat{\gen{S}}^<(\lambda_0)}{\gen{h}(\lambda_0)}}}} \notag \\
& - & \,    \lambda_0 \acomm{\hat{\gen{S}}^> (\lambda_0)}{\comm{\hat{\gen{Q}}^<(\lambda_0)}{\acomm{ \hat{\gen{S}}^<(\lambda_0)}{\comm{\hat{\gen{Q}}^>(\lambda_0)}{\gen{h}(\lambda_0)}}}} \nln
\eq \half \mathcal{H}(\lambda_0)  -  \lambda_0 \acomm{\hat{\gen{Q}}^< (\lambda_0)}{\comm{\hat{\gen{S}}^>(\lambda_0)}{\gen{X}(\lambda_0)}} 
\nl
+ \lambda_0 \acomm{\comm{\hat{\gen{Q}}^< (\lambda_0)}{\gen{X}(\lambda_0)}}{\hat{\gen{S}}^>(\lambda_0)} \nln
\eq \frac{\partial}{\partial \lambda} \lambda \acomm{\hat{\gen{Q}}^< (\lambda)}{\hat{\gen{S}}^> (\lambda)} _{|\lambda=\lambda_0} \nln
\eq \frac{\partial}{\partial \lambda} (\gen{J}^{+-}(\lambda) - \gen{J}^{+-}_0 )_{|\lambda=\lambda_0}  \nln
\eq \frac{\partial}{\partial \lambda} \gen{J}^{+-}(\lambda)_{|\lambda=\lambda_0}  \nln
\eq \frac{\partial}{\partial \lambda} (\gen{R}^{12} + \gen{J}^{+-}(\lambda))_{|\lambda=\lambda_0}. 
\>
The third to last step follows from (\ref{eq:psu11squaredalg}) and the identification between $\alg{psu}(1,1|2)$ and $\alg{psu}(1|1)^2$  central charge(s) (\ref{eq:sharedcentralcharge}), and the remaining steps use the $\lambda$-(in)dependence of generators. Since the last expression is the derivative of the right side of the initial equation (\ref{eq:q+q-}), the proof is now complete.

\section{The solution for $\gen{h}_1$ \label{sec:h1solution}}
We now give the explicit form of $\gen{h}_1$ acting on two adjacent sites. Like the one-loop dilatation generator \cite{Beisert:2007sk}, it can be written in terms of seven coefficient functions:
 \<
 (4 \pi)^2 \, \gen{h}_1\state{\phi_a^{(j)} \phi_b^{(n-j)}} \eq
\sum_{k=0}^{n}  f_1 (n, j,k) \state{\phi_a^{(k)} \phi_b^{(n-k)}}
 +  \sum_{k=0}^{n} f_2 (n, j,k)\state{\phi_b^{(k)} \phi_a^{(n-k)}} 
\nl
+ \sum_{k=0}^{n-1} \sqrt{\frac{k+1}{n-k}} f_3 (n, j,k)\varepsilon_{ab} \varepsilon^{\mathfrak{cd}}  \state{\psi_{\mathfrak{c}}^{(k)} \psi_{\mathfrak{d}}^{(n-1-k)}},
  \nln
(4 \pi)^2 \, \gen{h}_1 \state{\phi_a^{(j)} \psi_{\mathfrak{b}}^{(n-j)}} \eq
\sum_{k=0}^{n} \sqrt{\frac{n-k+1}{n-j+1}}f_4 (n, j,k) \state{\phi_a^{(k)} \psi_{\mathfrak{b}}^{(n-k)}} 
\nl
+\sum_{k=0}^{n}  \sqrt{\frac{k+1}{n-j+1}} f_5 (n, j,k)\state{\psi_{\mathfrak{b}}^{(k)} \phi_a^{(n-k)}},
  \nln
 (4 \pi)^2 \,  \gen{h}_1 \state{\psi_{\mathfrak{a}}^{(j)} \phi_b^{(n-j)}} \eq
\sum_{k=0}^{n} \sqrt{\frac{k+1}{j+1}}  f_4 (n,n-j,n-k) \state{\psi_{\mathfrak{a}}^{(k)} \phi_b^{(n-k)}}
\nl
 +\sum_{k=0}^{n}  \sqrt{\frac{n-k+1}{j+1}} f_5 (n,n-j,n-k)\state{\phi_b^{(k)} \psi_{\mathfrak{a}}^{(n-k)}},
  \nln
(4 \pi)^2 \, \gen{h}_1 \state{\psi_{\mathfrak{a}}^{(j)} \psi_{\mathfrak{b}}^{(n-j)}} \eq
\sum_{k=0}^{n} \sqrt{\frac{(k+1)(n-k+1)}{(j+1)(n-j+1)}}f_6 (n, j,k) \state{\psi_{\mathfrak{a}}^{(k)} \psi_{\mathfrak{b}}^{(n-k)}}
 \nl
 +
\sum_{k=0}^{n}  \sqrt{\frac{(k+1)(n-k+1)}{(j+1)(n-j+1)}} f_7 (n, j,k)\state{\psi_{\mathfrak{b}}^{(k)} \psi_{\mathfrak{a}}^{(n-k)}}
\nl
+ \sum_{k=0}^{n+1} \sqrt{\frac{j+1}{n-j+1}} f_3 (n+1,k,j)\varepsilon_{\mathfrak{ab}} \varepsilon^{cd} \state{\phi_c^{(k)} \phi_d^{(n+1-k)}}.
\nl
{}
  \>
The $f_n$ are built out of a few ingredients. First, $\theta(n)$ is the step function (one for $n \geq 0$ and 0 otherwise).  Also, the ordinary and generalized harmonic numbers, (\ref{eq:ordinaryS}) and (\ref{eq:generalizedS}), appear repeatedly. The last ingredients are further generalizations of the harmonic sums to two arguments:
\[ \label{eq:definesg}
\sg(m, n) = \sum_{i=1 }^n \frac{S(i+m)}{i}, \quad  \tilde{S}_{2,1}(m, n) = \sum_{i=1 }^n \frac{S(i+m)}{i^2}.
\]
The first new function $\sg$ appears multiple times. It is similar to $S_{1,1}$, but note that the argument in the numerator is shifted by the first argument of $\sg$. The second new function, $\tilde{S}_{2,1}$, only appears once explicitly in the expressions below (in the $\delta_{jk}$ term of $\tilde{f}_1$). 

Now we are ready to give the explicit form of the $f_n$. Using symmetries under interchanges of arguments (partly due to Hermiticity and parity), we can write relatively compact expressions. The bosonic coefficient $f_2$ takes the simplest form,
\<
 f_2(n, j, k) \eq \frac{1}{4} \Big(\tilde{f}_2(n, j, k) +  \tilde{f}_2(n, k, j) + \tilde{f}_2(n, n-j, n-k) +  \tilde{f}_2(n, n-k, n-j)  \Big)\,,
 \nln
  \tilde{f}_2(n, j, k) \eq \frac{1}{(n+1)} \Big( S_2(j) - S_{1,1}(n) - S(j)S(k) - S(j)S(n-k) + 2 S(j) S(n+1) \Big) \,.
  \nl
 \>
We  write the other purely bosonic coefficient, $f_1$, in terms of $f_2$ and two new functions, $f_{1,0}$ and $\tilde{f}_1$. $f_{1,0}$  governs interactions with the same initial and final states, and $\tilde{f}_1$ applies otherwise. 
  \<
  f_1(n, j, k) \eq -f_2(n, j, k) + \frac{1}{2} \delta_{jk}(f_{1,0}(n, j) + f_{1,0}(n, n-j)) 
  \nl
   + \frac{\delta_{j\neq k}}{4} \Big(\tilde{f}_1(n, j, k) + \tilde{f}_1(n, k, j)  + \tilde{f}_1(n, n-j, n-k) + \tilde{f}_1(n, n-k, n-j)\Big)\, ,
   \nln
  f_{1,0}(n, j) \eq 2 S_3(j) - 5 S_{2,1}(j) - 2 S_{1,2}(j)   +  2 S_{1,1,1}(j)  -  S_2(j)S(n-j)  +  \tilde{S}_{2,1}(j, n-j) \, ,
  \nln
  \tilde{f}_1(n, j, k) \eq \frac{1}{j-k} \Big( -S_2(j) + 2 S_{1,1}(j) + S(j)S(n-k) + 2 \sg(j, n-j) \Big)
  \nl
  +
   \frac{1}{|j-k|} \Big(-S_{1,1}(|j-k|-1) - 2 S(j)S(|j-k|) + 4 \sg(j, k-j) \Big) \, .
   \>
   Note that $\sg$ is zero if its second argument is negative or zero, by the definition (\ref{eq:definesg}). Next we give $f_5$, one of the mixed boson-fermion functions.  
 \<
  f_5(n, j, k) \eq \frac{1}{4(k+1)} \Big(\tilde{f}_5(n, j, k) + \tilde{f}_5(n, n-k, n-j) \Big) \,,
  \nln
 \tilde{f}_5(n, j, k) \eq  S_2(j) + 2 S_{1,1}(k+1) - S_{1,1}(n+1) - S(j)S(k+1) 
 \nl
 + S(j)S(n-j+1) -S(j)S(n-k) + S(j)S(n+1)
  \nl
  + S(k+1)S(n+1) - 2 \sg(j, n-j+1)   
  \nl
  + \theta(j-k-1) \Big( S_{1,1}(j-k-1) + S(j)S(j-k-1) 
  \nl
  + S(k+1)S(j-k-1) - 2 \sg(k+1,j-k-1) \Big)
  \nl
  + \theta(k-j)\Big( 2 S(k+1)/(k+1) -  S_{1,1}(k-j) -S(j)S(k-j) 
  \nl
  - S(k+1)S(k-j) + 2 \sg(j,k-j) \Big)  \,. 
 \>
 \newpage
 Finally, the remaining four  functions are given most efficiently as sums of the above three functions and additional functions, $g_i$, 
 \< \label{eq:last4f}
 f_3(n, j, k) \eq f_2(n, j, k) - f_5(n, j, k) - \frac{g_1(n, j, k)}{4(k+1)}  \, ,
 \nln
 f_4(n, j, k) \eq - \frac{k }{n-k+1}f_3(n, j, k-1) + f_1(n, j, k) - \frac{g_2(n, j, k)}{4(n-k+1)}   \, ,
 \nln
 f_6(n, j, k) \eq \phantom{-} \frac{(n-j+1)}{n-k+1}f_3(n+1, j, k) + f_4(n, j, k) - \frac{1}{4} g_{3,1}(n, j, k)  - \frac{g_{3,2}(n, j, k)}{4(n-k+1)}  \, ,
 \nln
  f_7(n, j, k) \eq -\frac{(n-j+1)}{n-k+1}f_3(n+1, j, k) - f_5(n, j, k) - \frac{g_{4, 1}(n, j, k)}{4(n+2)}  - \frac{g_{4, 2}(n, j, k)}{4(n-k+1)} \, .
  \nl
  {}
  \>
 The $g_i$ appear in the expansion of  $\comm{\gen{Q}^{1+<}_0}{\gen{h}_1}$ as 
 \<
 (4 \pi)^2\comm{\gen{Q}^{1+<}_0}{\gen{h}_1} \state{\phi_2^{(j)}\phi_1^{(n-j)}} \eq  \frac{g_1 (n, j, k)}{4 \sqrt{k+1}} \state{\psi_>^{(k)} \phi_2^{(n-k)}} + \frac{g_2 (n, j, k)}{4 \sqrt{n-k+1}}  \state{\phi_2^{(k)}\psi_>^{(n-k)}},
 \nln
  (4 \pi)^2\comm{\gen{Q}^{1+<}_0}{\gen{h}_1} \state{\phi_1^{(j)}\psi_<^{(n-j)}} \eq \Big(\frac{\sqrt{k+1} \sqrt{n-k+1}}{4\sqrt{n-j+1}}  g_{3,1}(n, j, k) 
 \nl
  + \frac{\sqrt{k+1} }{4 \sqrt{n-j+1}\sqrt{n-k+1}} g_{3,2}(n, j, k) \Big) \state{\psi_>^{(k)}\psi_<^{(n-k)}}
 \nl
 + \Big(\frac{\sqrt{k+1} \sqrt{n-k+1}}{4 \sqrt{n-j+1}(n+2)}  g_{4,1}(n, j, k)  
 \nl
 + \frac{\sqrt{k+1} }{4 \sqrt{n-j+1}\sqrt{n-k+1}} g_{4,2}(n, j, k) \Big) \state{\psi_<^{(k)}\psi_>^{(n-k)}} 
 \nl
 + \ldots
 \> 
The explicit expressions for the $g_i$ are  
 \<
 g_1(n, j, k) \eq  \frac{-S(j)+S(n+1)}{n-j+1} + \frac{2 S(j) + 2 S(n-j) -1/(k+1)}{n+1} 
 \nl
 + \theta(j-k-1) \Big(\frac{-2 S(n-k)}{n-k} + \frac{-S(n-k)+S(j-k-1)}{n-j+1} \Big) 
 \nl
 + \theta(k-j) \frac{-4 S(n-j+1) + S(n-k) + S(k-j)}{n-j+1}, \nln
 g_2(n, j, k) \eq  \frac{S(j)-S(n+1)}{n-j+1} + \frac{-2 S(j) - 2 S(n-j) + 1/(n-k+1)}{n+1} 
 \nl
 + \delta_{jk} \Big(-S_2(n-j+1) + 3 S_2(n-j) + 2 S_{1,1}(n-j+1)- 4 S_{1,1}(n-j) \Big) 
 \nl
 + \theta(j-k-1) \Big( \frac{S(n-k) -S(j-k)}{n-j+1} + \frac{2 S(j-k)}{j-k} \Big) 
 \nl
 + \theta(k-j) \Big( \frac{4 S(n-j+1)  - S(n-k+1) -S(k-j)}{n-j+1} \Big)
 \nl
 + \theta(k-j-1) \Big( \frac{S(j) - S(n-j+1) - S(k) + S(n-k)}{k-j} + \frac{2 S(k)}{k} \Big),   \notag
  \>
  \newpage
  \<
  g_{3,1}(n, j, k) \eq \frac{S(n-j+1)-S(n+1) + \theta(j-k)\Big(4 S(j+1)-S(k+1)\Big)}{(j+1)(k+1)}
 \nl
 + \frac{ \theta(k-j) S(k)}{(j+1)(k+1)} + \delta_{jk} \frac{2 S_2(j) + S_{1,1}(j+1) - 3 S_{1,1}(j)}{j+1} 
 \nl
 + \frac{\theta(j-k-1)}{(j-k)} \Big( \frac{S(j-k)}{(j+1)} 
 \nl
 + \frac{-S(j+1) + S(n-j+1) + S(k) - S(n-k+1) - S(j-k)}{(k+1)} \Big) 
 \nl
 + \frac{\theta(k-j-1)}{(k-j)} \Big(\frac{-S(k-j)}{(j+1)} + \frac{3 S(k-j)}{(k+1)} \Big),
 \nln
  g_{3,2}(n, j, k) \eq    \frac{2 S(j+1) - S(j) + S(n-j+1) - S(k+1) - S(n-k) + 2 S(n+2) }{n+2}
 \nl
  + \frac{-2 S(j+1) - 2 S(n-j+1) - S(k)+S(n-k)}{k+1} 
 \nl
 + \theta(j-k-1) \frac{S(k+1) + S(n-k+1)}{k+1}
 \nl
  + \theta(k-j) \frac{S(j)-S(n-j+1) + 2 S(k-j)}{k+1} ,
  \nln
 g_{4, 1}(n, j, k) \eq \frac{-S(j+1)-S(n-j+1)-S(k+1)-S(n-k+1)+2 S(n+2)}{k+1} 
 \nl
 + \frac{-2 S(j+1)-2 S(n-j+1)+1 /(n-k+1)}{n-k+1},
 \nln
  g_{4, 2}(n, j, k) \eq  \frac{S(n-j+1)-S(n+2)}{j+1} + \frac{2 S(k+1)}{k+1} 
 \nl
 + \theta(j-k-1) \Big( \frac{4 S(j+1) - S(k+1) - S(j-k)}{j+1} + \frac{-S(j) - S(k+1) }{k+1}
 \nl
 + \frac{ S(n-j+1) - S(n-k+1) + S(j-k) + S(j-k-1)}{k+1} \Big)
 \nl
 + \theta(k-j) \frac{3 S(k+1) - 2 S(k) - S(k-j)}{j+1} .
 \>
Besides $g_{3,1}$, which is degree 3, the remaining $g_i$ are degree 2. It follows from (\ref{eq:last4f}) that all of the $f_i$ are degree three. The degree of the $g_i$ could be made manifest by replacing all the factors of the form $1/(i+1)$ with $S(i+1)-S(i)$,  but this would lead to much lengthier expressions. 

\section{Homogeneous solutions for $\gen{h}_1$ \label{sec:homsolution}}
Recall that the $\alg{psu}(1,1|2)$ quadratic Casimir $\gen{J}^2$ has eigenvalues $j(j+1)$ for all nonnegative integer $\alg{psu}(1,1|2)$ spin $j$. The first type of homogeneous solution for $\gen{h}_1$ is specified by its eigenvalues $c_j$ on spin-$j$ states. However, for acting on general spin-chain states, we need to change to a ``position'' basis. This can be done by expanding the spin-$j$ states in terms of two-site position states. The change of basis yields a representation of $\gen{h}_1$ in terms of shifts $\delta \! f_l$ of the seven coefficient functions $f_l$, which were given in the previous section. Again, $\alg{psu}(1,1|2)$ symmetry relates the coefficients, so all components can be written in terms of $\delta \! f_1$, $\delta \! f_2$, and  $\delta \! f_4$. After significant simplification, we obtain the following. First the bosonic components $\delta \! f_l$ for $l=1,2$ are given by
\begin{gather}
\delta \! f_l(n, j, k) = \half (n-k)!(n-j)! \sum_{i=0}^n \bigg( \frac{(n-i)!}{(n+i+1)!} C_l(i, c_i) \notag \\
\times {}_3F^{\text{reg}}_2(-j, -i, -i; 1, n-j-i+1;1) \, {}_3F^{\text{reg}}_2(-k, -i, -i; 1, n-k-i+1;1) \bigg) , \notag \\
C_1(i, c_i) = \phantom{-}i  \,c_{i-1}+ (2 \, i + 1)\, c_i + (i+1) \,c_{i+1}, \notag \\
 C_2(i, c_i) = -i\,c_{i-1}+ (2 \, i + 1)\, c_i - (i+1) \,c_{i+1}.
\end{gather}
The $j$ appearing here is the number of derivatives initially on the first site, and should not be confused with a $\alg{psu}(1,1|2)$ spin. Next, the mixed boson-fermion interactions coefficient $\delta \! f_4$ is similar (but note that some arguments are shifted by 1),
\begin{gather}
\delta \! f_4(n, j, k) =   (n-k)!(n-j+1)! \sum_{i=0}^n \bigg( \frac{(n-i)!}{(n+i+2)!} (i + 1) (c_i + c_{i+1}) \notag \\
\times {}_3F^{\text{reg}}_2(-j, -i-1, -i; 1, n-j-i+1;1) \, {}_3F^{\text{reg}}_2(-k, -i-1, -i; 1, n-k-i+1;1) \bigg) . \notag \\
{}
\end{gather}
Here we use the  $\texttt{Mathematica}$ definition of the regularized hypergeometric function, which is a ratio of the ordinary hypergeometric function and gamma functions. In particular,
\[
 {}_3F^{\text{reg}}_2(a_1,a_2, a_3; b_1, b_2; z)=\frac{{}_3F_2(a_1,a_2, a_3; b_1, b_2; z)}{\Gamma(b_1) \Gamma(b_2)}.
 \]
 The remaining components of this homogeneous solution are then given by
 \<
 \delta \! f_3(n, j, k) \eq \frac{n-k}{(k+1)}\Big(\delta \! f_1(n, j, k)- \delta  \! f_4(n, j, k+1)\Big),
 \nln
 \delta \! f_5(n, j, k) \eq \delta \! f_2(n, j, k)- \theta(n-k-1) \delta  \! f_3(n, j, k), 
 \nln
 \delta \! f_6(n, j, k) \eq \phantom{-} \frac{(n+1-j)}{n-k+1} \delta \! f_3(n+1, j, k)+  \delta  \! f_4(n, j, k), 
 \nln
 \delta \! f_7(n, j, k) \eq -\frac{(n+1-j)}{n-k+1} \delta \! f_3(n+1, j, k)-  \delta \! f_5(n, j, k).
 \>

One can check this expression, at least numerically, by evaluating its commutator with leading order $\alg{psu}(1,1|2)$ generators, which vanishes. Also, substituting the known eigenvalues for the one-loop dilatation generator, $c_j = 4 S(j)$, yields perfect agreement with the $f_l$ given in Section 3.4 of \cite{Beisert:2007sk}. Note that the expressions for the $f_l$ given in \cite{Beisert:2007sk} include square root factors that we have factored out here.


\begin{thebibliography}{100}
\ifx\href\asklfhas\newcommand{\href}[2]{#2}\fi
\raggedright
\small
\parskip 0pt

\bibitem{Minahan:2002ve}
J.~A.~Minahan and K.~Zarembo,
\textit{``The Bethe-ansatz for {$\mathcal{N}=\mathord{}$4} super Yang-Mills''},
\textsf{JHEP~0303,~013~(2003)},
\href{http://arXiv.org/abs/arXiv:hep-th/0212208}{\texttt{arXiv:hep-th/0212208}%
}.
%
\bibitem{Bena:2003wd}
I.~Bena, J.~Polchinski and R.~Roiban,
\textit{``Hidden symmetries of the {$AdS_5\times S^5$} superstring''},
\textsf{Phys.~Rev.~D69,~046002~(2004)},
\href{http://arXiv.org/abs/arXiv:hep-th/0305116}{\texttt{arXiv:hep-th/0305116}%
}.
%
\bibitem{Beisert:2003tq}
N.~Beisert, C.~Kristjansen and M.~Staudacher,
\textit{``The Dilatation Operator of {$\mathcal{N}=\mathord{}$4} Conformal
  Super Yang-Mills Theory''},
\textsf{Nucl.~Phys.~B664,~131~(2003)},
\href{http://arXiv.org/abs/arXiv:hep-th/0303060}{\texttt{arXiv:hep-th/0303060}%
}.
%
\bibitem{Beisert:2003yb}
N.~Beisert and M.~Staudacher,
\textit{``The {$\mathcal{N}=\mathord{}$4} SYM Integrable Super Spin Chain''},
\textsf{Nucl.~Phys.~B670,~439~(2003)},
\href{http://arXiv.org/abs/arXiv:hep-th/0307042}{\texttt{arXiv:hep-th/0307042}%
}.
%
\bibitem{Beisert:2004hm}
N.~Beisert, V.~Dippel and M.~Staudacher,
\textit{``A Novel Long Range Spin Chain and Planar {$\mathcal{N}=\mathord{}$4}
  Super Yang-Mills''},
\textsf{JHEP~0407,~075~(2004)},
\href{http://arXiv.org/abs/arXiv:hep-th/0405001}{\texttt{arXiv:hep-th/0405001}%
}.
%
\bibitem{Arutyunov:2004vx}
G.~Arutyunov, S.~Frolov and M.~Staudacher,
\textit{``Bethe ansatz for quantum strings''},
\textsf{JHEP~0410,~016~(2004)},
\href{http://arXiv.org/abs/arXiv:hep-th/0406256}{\texttt{arXiv:hep-th/0406256}%
}.
%
\bibitem{Staudacher:2004tk}
M.~Staudacher,
\textit{``The factorized S-matrix of CFT/AdS''},
\textsf{JHEP~0505,~054~(2005)},
\href{http://arXiv.org/abs/arXiv:hep-th/0412188}{\texttt{arXiv:hep-th/0412188}%
}.
%
\bibitem{Beisert:2005fw}
N.~Beisert and M.~Staudacher,
\textit{``Long-Range PSU(2,2$|$4) Bethe Ans{\"a}tze for Gauge Theory and
  Strings''},
\textsf{Nucl.~Phys.~B727,~1~(2005)},
\href{http://arXiv.org/abs/arXiv:hep-th/0504190}{\texttt{arXiv:hep-th/0504190}%
}.
%
\bibitem{Beisert:2005tm}
N.~Beisert,
\textit{``The SU(2$|$2) Dynamic S-Matrix''},
\href{http://arXiv.org/abs/arXiv:hep-th/0511082}{\texttt{arXiv:hep-th/0511082}%
}.
%
\bibitem{Arutyunov:2006ak}
G.~Arutyunov, S.~Frolov, J.~Plefka and M.~Zamaklar,
\textit{``The Off-shell Symmetry Algebra of the Light-cone $AdS_5\times S^5$
  Superstring''},
\textsf{J.~Phys.~A40,~3583~(2007)},
\href{http://arXiv.org/abs/arXiv:hep-th/0609157}{\texttt{arXiv:hep-th/0609157}%
}.
%
\bibitem{Janik:2006dc}
R.~A.~Janik,
\textit{``The $AdS_5\times S^5$ superstring worldsheet S-matrix and crossing
  symmetry''},
\textsf{Phys.~Rev.~D73,~086006~(2006)},
\href{http://arXiv.org/abs/arXiv:hep-th/0603038}{\texttt{arXiv:hep-th/0603038}%
}.
%
\bibitem{Beisert:2006ib}
N.~Beisert, R.~Hern\'andez and E.~L\'opez,
\textit{``A Crossing-Symmetric Phase for $AdS_5 \times S^5$ Strings''},
\textsf{JHEP~0611,~070~(2006)},
\href{http://arXiv.org/abs/arXiv:hep-th/0609044}{\texttt{arXiv:hep-th/0609044}%
}.
%
\bibitem{Beisert:2006ez}
N.~Beisert, B.~Eden and M.~Staudacher,
\textit{``Transcendentality and Crossing''},
\textsf{J.~Stat.~Mech.~07,~P01021~(2007)},
\href{http://arXiv.org/abs/arXiv:hep-th/0610251}{\texttt{arXiv:hep-th/0610251}%
}.
%
\bibitem{Bern:2006ew}
Z.~Bern, M.~Czakon, L.~J.~Dixon, D.~A.~Kosower and V.~A.~Smirnov,
\textit{``The Four-Loop Planar Amplitude and Cusp Anomalous Dimension in
  Maximally Supersymmetric Yang-Mills Theory''},
\textsf{Phys.~Rev.~D75,~085010~(2007)},
\href{http://arXiv.org/abs/arXiv:hep-th/0610248}{\texttt{arXiv:hep-th/0610248}%
}.
%
\bibitem{Eden:2006rx}
B.~Eden and M.~Staudacher,
\textit{``Integrability and transcendentality''},
\textsf{J.~Stat.~Mech.~06,~P11014~(2006)},
\href{http://arXiv.org/abs/arXiv:hep-th/0603157}{\texttt{arXiv:hep-th/0603157}%
}.
%
\bibitem{Benna:2006nd}
M.~K.~Benna, S.~Benvenuti, I.~R.~Klebanov and A.~Scardicchio,
\textit{``A Test of the AdS/CFT Correspondence Using High-Spin Operators''},
\textsf{Phys.~Rev.~Lett.~98,~131603~(2007)},
\href{http://arXiv.org/abs/arXiv:hep-th/0611135}{\texttt{arXiv:hep-th/0611135}%
}.
%
\bibitem{Kotikov:2006ts}
A.~V.~Kotikov and L.~N.~Lipatov,
\textit{``On the highest transcendentality in {$\mathcal{N}=\mathord{}$4}
  SUSY''},
\textsf{Nucl.~Phys.~B769,~217~(2007)},
\href{http://arXiv.org/abs/arXiv:hep-th/0611204}{\texttt{arXiv:hep-th/0611204}%
}.
%
\bibitem{Cachazo:2006az}
F.~Cachazo, M.~Spradlin and A.~Volovich,
\textit{``Four-loop cusp anomalous dimension from obstructions''},
\textsf{Phys.~Rev.~D75,~105011~(2007)},
\href{http://arXiv.org/abs/arXiv:hep-th/0612309}{\texttt{arXiv:hep-th/0612309}%
}.
%
\bibitem{Alday:2007qf}
L.~F.~Alday, G.~Arutyunov, M.~K.~Benna, B.~Eden and I.~R.~Klebanov,
\textit{``On the strong coupling scaling dimension of high spin operators''},
\textsf{JHEP~0704,~082~(2007)},
\href{http://arXiv.org/abs/arXiv:hep-th/0702028}{\texttt{arXiv:hep-th/0702028}%
}.
%
\bibitem{Kostov:2007kx}
I.~Kostov, D.~Serban and D.~Volin,
\textit{``{Strong coupling limit of Bethe ansatz equations}''},
\textsf{Nucl.~Phys.~B789,~413~(2008)},
\href{http://arXiv.org/abs/arXiv:hep-th/0703031}{\texttt{arXiv:hep-th/0703031}%
}.
%
\bibitem{Beccaria:2007tk}
M.~Beccaria, G.~F.~De~Angelis and V.~Forini,
\textit{``The scaling function at strong coupling from the quantum string Bethe
  equations''},
\textsf{JHEP~0704,~066~(2007)},
\href{http://arXiv.org/abs/arXiv:hep-th/0703131}{\texttt{arXiv:hep-th/0703131}%
}.
%
\bibitem{Roiban:2007jf}
R.~Roiban, A.~Tirziu and A.~A.~Tseytlin,
\textit{``Two-loop world-sheet corrections in $AdS_5\times S^5$ superstring''},
\textsf{JHEP~0707,~056~(2007)},
\href{http://arXiv.org/abs/arXiv:0704.3638}{\texttt{arXiv:0704.3638}} \texttt{[hep-th]}.
%
\bibitem{Casteill:2007ct}
P.~Y.~Casteill and C.~Kristjansen,
\textit{``{The Strong Coupling Limit of the Scaling Function from the Quantum
  String Bethe Ansatz}''},
\textsf{Nucl.~Phys.~B785,~1~(2007)},
\href{http://arXiv.org/abs/arXiv:0705.0890}{\texttt{arXiv:0705.0890}} \texttt{[hep-th]}.
%
\bibitem{Basso:2007wd}
B.~Basso, G.~P.~Korchemsky and J.~Kotanski,
\textit{``{Cusp anomalous dimension in maximally supersymmetric Yang- Mills
  theory at strong coupling}''},
\textsf{Phys.~Rev.~Lett.~100,~091601~(2008)},
\href{http://arXiv.org/abs/arXiv:0708.3933}{\texttt{arXiv:0708.3933}} \texttt{[hep-th]}.
%
\bibitem{Roiban:2007dq}
R.~Roiban and A.~A.~Tseytlin,
\textit{``{Strong-coupling expansion of cusp anomaly from quantum
  superstring}''},
\textsf{JHEP~0711,~016~(2007)},
\href{http://arXiv.org/abs/arXiv:0709.0681}{\texttt{arXiv:0709.0681}} \texttt{[hep-th]}.
%
\bibitem{Belitsky:2007kf}
A.~V.~Belitsky,
\textit{``{Strong coupling expansion of Baxter equation in $\mathcal{N}=4$
  SYM}''},
\textsf{Phys.~Lett.~B659,~732~(2008)},
\href{http://arXiv.org/abs/arXiv:0710.2294}{\texttt{arXiv:0710.2294}} \texttt{[hep-th]}.
%
\bibitem{Kostov:2008ax}
I.~Kostov, D.~Serban and D.~Volin,
\textit{``{Functional BES equation}''},
\href{http://arXiv.org/abs/arXiv:0801.2542}{\texttt{arXiv:0801.2542}} \texttt{[hep-th]}.
%
\bibitem{Maldacena:2006rv}
J.~Maldacena and I.~Swanson,
\textit{``Connecting giant magnons to the pp-wave: An interpolating limit of
  $AdS_5\times S^5$''},
\textsf{Phys.~Rev.~D76,~026002~(2007)},
\href{http://arXiv.org/abs/arXiv:hep-th/0612079}{\texttt{arXiv:hep-th/0612079}%
}.
%
\bibitem{Klose:2007rz}
T.~Klose, T.~McLoughlin, J.~A.~Minahan and K.~Zarembo,
\textit{``{World-sheet scattering in $AdS_5\times S^5$ at two loops}''},
\textsf{JHEP~0708,~051~(2007)},
\href{http://arXiv.org/abs/arXiv:0704.3891}{\texttt{arXiv:0704.3891}} \texttt{[hep-th]}.
%
\bibitem{Puletti:2007hq}
V.~Giangreco Marotta~Puletti, T.~Klose and O.~Ohlsson~Sax,
\textit{``{Factorized world-sheet scattering in near-flat $AdS_5\times
  S^5$}''},
\textsf{Nucl.~Phys.~B792,~228~(2008)},
\href{http://arXiv.org/abs/arXiv:0707.2082}{\texttt{arXiv:0707.2082}} \texttt{[hep-th]}.
%
\bibitem{Dorey:2006dq}
N.~Dorey,
\textit{``Magnon bound states and the AdS/CFT correspondence''},
\textsf{J.~Phys.~A39,~13119~(2006)},
\href{http://arXiv.org/abs/arXiv:hep-th/0604175}{\texttt{arXiv:hep-th/0604175}%
}.
%
\bibitem{Dorey:2007xn}
N.~Dorey, D.~M.~Hofman and J.~Maldacena,
\textit{``On the singularities of the magnon S-matrix''},
\textsf{Phys.~Rev.~D76,~025011~(2007)},
\href{http://arXiv.org/abs/arXiv:hep-th/0703104}{\texttt{arXiv:hep-th/0703104}%
}.
%
\bibitem{Dorey:2007an}
N.~Dorey and K.~Okamura,
\textit{``{Singularities of the Magnon Boundstate S-Matrix}''},
\textsf{JHEP~0803,~037~(2008)},
\href{http://arXiv.org/abs/arXiv:0712.4068}{\texttt{arXiv:0712.4068}} \texttt{[hep-th]}.
%
\bibitem{Belitsky:2006en}
A.~V.~Belitsky, A.~S.~Gorsky and G.~P.~Korchemsky,
\textit{``{Logarithmic scaling in gauge / string correspondence}''},
\textsf{Nucl.~Phys.~B748,~24~(2006)},
\href{http://arXiv.org/abs/arXiv:hep-th/0601112}{\texttt{arXiv:hep-th/0601112}%
}.
%
\bibitem{Frolov:2006qe}
S.~Frolov, A.~Tirziu and A.~A.~Tseytlin,
\textit{``{Logarithmic corrections to higher twist scaling at strong coupling
  from AdS/CFT}''},
\textsf{Nucl.~Phys.~B766,~232~(2007)},
\href{http://arXiv.org/abs/arXiv:hep-th/0611269}{\texttt{arXiv:hep-th/0611269}%
}.
%
\bibitem{Alday:2007mf}
L.~F.~Alday and J.~M.~Maldacena,
\textit{``{Comments on operators with large spin}''},
\textsf{JHEP~0711,~019~(2007)},
\href{http://arXiv.org/abs/arXiv:0708.0672}{\texttt{arXiv:0708.0672}} \texttt{[hep-th]}.
%
\bibitem{Roiban:2007ju}
R.~Roiban and A.~A.~Tseytlin,
\textit{``{Spinning superstrings at two loops: strong-coupling corrections to
  dimensions of large-twist SYM operators}''},
\textsf{Phys.~Rev.~D77,~066006~(2008)},
\href{http://arXiv.org/abs/arXiv:0712.2479}{\texttt{arXiv:0712.2479}} \texttt{[hep-th]}.
%
\bibitem{Freyhult:2007pz}
L.~Freyhult, A.~Rej and M.~Staudacher,
\textit{``{A Generalized Scaling Function for AdS/CFT}''},
\href{http://arXiv.org/abs/arXiv:0712.2743}{\texttt{arXiv:0712.2743}} \texttt{[hep-th]}.
%
\bibitem{Basso:2008tx}
B.~Basso and G.~P.~Korchemsky,
\textit{``{Embedding nonlinear O(6) sigma model into $\mathcal{N}=4$
  super-Yang- Mills theory}''},
\href{http://arXiv.org/abs/arXiv:0805.4194}{\texttt{arXiv:0805.4194}} \texttt{[hep-th]}.
%
\bibitem{Bombardelli:2008ah}
D.~Bombardelli, D.~Fioravanti and M.~Rossi,
\textit{``{Large spin corrections in ${\cal N}=4$ SYM sl$(2)$: still a linear
  integral equation}''},
\href{http://arXiv.org/abs/arXiv:0802.0027}{\texttt{arXiv:0802.0027}} \texttt{[hep-th]}.
%
\bibitem{Fioravanti:2008rv}
D.~Fioravanti, P.~Grinza and M.~Rossi,
\textit{``{Strong coupling for planar ${\cal N}=4$ SYM theory: an all-order
  result}''},
\href{http://arXiv.org/abs/arXiv:0804.2893}{\texttt{arXiv:0804.2893}} \texttt{[hep-th]}.
%
\bibitem{Fioravanti:2008ak}
D.~Fioravanti, P.~Grinza and M.~Rossi,
\textit{``{The generalised scaling function: a note}''},
\href{http://arXiv.org/abs/arXiv:0805.4407}{\texttt{arXiv:0805.4407}} \texttt{[hep-th]}.
%
\bibitem{Gromov:2008en}
N.~Gromov,
\textit{``{Generalized Scaling Function at Strong Coupling}''},
\href{http://arXiv.org/abs/arXiv:0805.4615}{\texttt{arXiv:0805.4615}} \texttt{[hep-th]}.
%
\bibitem{Arutyunov:2003za}
G.~Arutyunov, J.~Russo and A.~A.~Tseytlin,
\textit{``Spinning strings in {$AdS_5\times S^5$}: New integrable system
  relations''},
\textsf{Phys.~Rev.~D69,~086009~(2004)},
\href{http://arXiv.org/abs/arXiv:hep-th/0311004}{\texttt{arXiv:hep-th/0311004}%
}.
%
\bibitem{Park:2005ji}
I.~Y.~Park, A.~Tirziu and A.~A.~Tseytlin,
\textit{``Spinning strings in $AdS_5\times S^5$: One-loop correction to energy
  in SL(2) sector''},
\textsf{JHEP~0503,~013~(2005)},
\href{http://arXiv.org/abs/arXiv:hep-th/0501203}{\texttt{arXiv:hep-th/0501203}%
}.
%
\bibitem{Schafer-Nameki:2005is}
S.~Sch{\"a}fer-Nameki and M.~Zamaklar,
\textit{``Stringy sums and corrections to the quantum string Bethe ansatz''},
\textsf{JHEP~0510,~044~(2005)},
\href{http://arXiv.org/abs/arXiv:hep-th/0509096}{\texttt{arXiv:hep-th/0509096}%
}.
%
\bibitem{SchaferNameki:2006ey}
S.~Schafer-Nameki, M.~Zamaklar and K.~Zarembo,
\textit{``{How accurate is the quantum string Bethe ansatz?}''},
\textsf{JHEP~0612,~020~(2006)},
\href{http://arXiv.org/abs/arXiv:hep-th/0610250}{\texttt{arXiv:hep-th/0610250}%
}.
%
\bibitem{Kotikov:2007cy}
A.~V.~Kotikov, L.~N.~Lipatov, A.~Rej, M.~Staudacher and V.~N.~Velizhanin,
\textit{``{Dressing and Wrapping}''},
\textsf{J.~Stat.~Mech.~0710,~P10003~(2007)},
\href{http://arXiv.org/abs/arXiv:0704.3586}{\texttt{arXiv:0704.3586}} \texttt{[hep-th]}.
%
\bibitem{Ambjorn:2005wa}
J.~Ambj{\o}rn, R.~A.~Janik and C.~Kristjansen,
\textit{``Wrapping interactions and a new source of corrections to the
  spin-chain / string duality''},
\textsf{Nucl.~Phys.~B736,~288~(2006)},
\href{http://arXiv.org/abs/arXiv:hep-th/0510171}{\texttt{arXiv:hep-th/0510171}%
}.
%
\bibitem{Arutyunov:2007tc}
G.~Arutyunov and S.~Frolov,
\textit{``{On String S-matrix, Bound States and TBA}''},
\textsf{JHEP~0712,~024~(2007)},
\href{http://arXiv.org/abs/arXiv:0710.1568}{\texttt{arXiv:0710.1568}} \texttt{[hep-th]}.
%
\bibitem{Hofman:2006xt}
D.~M.~Hofman and J.~M.~Maldacena,
\textit{``Giant magnons''},
\textsf{J.~Phys.~A39,~13095~(2006)},
\href{http://arXiv.org/abs/arXiv:hep-th/0604135}{\texttt{arXiv:hep-th/0604135}%
}.
%
\bibitem{Arutyunov:2006gs}
G.~Arutyunov, S.~Frolov and M.~Zamaklar,
\textit{``Finite-size Effects from Giant Magnons''},
\textsf{Nucl.~Phys.~B778,~1~(2007)},
\href{http://arXiv.org/abs/arXiv:hep-th/0606126}{\texttt{arXiv:hep-th/0606126}%
}.
%
\bibitem{Janik:2007wt}
R.~A.~Janik and T.~Lukowski,
\textit{``{Wrapping interactions at strong coupling -- the giant magnon}''},
\textsf{Phys.~Rev.~D76,~126008~(2007)},
\href{http://arXiv.org/abs/arXiv:0708.2208}{\texttt{arXiv:0708.2208}} \texttt{[hep-th]}.
%
\bibitem{Hatsuda:2008gd}
Y.~Hatsuda and R.~Suzuki,
\textit{``{Finite-Size Effects for Dyonic Giant Magnons}''},
\href{http://arXiv.org/abs/arXiv:0801.0747}{\texttt{arXiv:0801.0747}} \texttt{[hep-th]}.
%
\bibitem{Minahan:2008re}
J.~A.~Minahan and O.~Ohlsson~Sax,
\textit{``{Finite size effects for giant magnons on physical strings}''},
\textsf{Nucl.~Phys.~B801,~97~(2008)},
\href{http://arXiv.org/abs/arXiv:0801.2064}{\texttt{arXiv:0801.2064}} \texttt{[hep-th]}.
%
\bibitem{Heller:2008at}
M.~P.~Heller, R.~A.~Janik and T.~Lukowski,
\textit{``{A new derivation of Luscher F-term and fluctuations around the giant
  magnon}''},
\href{http://arXiv.org/abs/arXiv:0801.4463}{\texttt{arXiv:0801.4463}} \texttt{[hep-th]}.
%
\bibitem{Gromov:2008ie}
N.~Gromov, S.~Schafer-Nameki and P.~Vieira,
\textit{``{Quantum Wrapped Giant Magnon}''},
\href{http://arXiv.org/abs/arXiv:0801.3671}{\texttt{arXiv:0801.3671}} \texttt{[hep-th]}.
%
\bibitem{Beisert:2003jj}
N.~Beisert,
\textit{``The Complete One-Loop Dilatation Operator of
  {$\mathcal{N}=\mathord{}$4} Super Yang-Mills Theory''},
\textsf{Nucl.~Phys.~B676,~3~(2004)},
\href{http://arXiv.org/abs/arXiv:hep-th/0307015}{\texttt{arXiv:hep-th/0307015}%
}.
%
\bibitem{Beisert:2003ys}
N.~Beisert,
\textit{``The SU(2$|$3) Dynamic Spin Chain''},
\textsf{Nucl.~Phys.~B682,~487~(2004)},
\href{http://arXiv.org/abs/arXiv:hep-th/0310252}{\texttt{arXiv:hep-th/0310252}%
}.
%
\bibitem{Beisert:2007hz}
N.~Beisert, T.~McLoughlin and R.~Roiban,
\textit{``The Four-Loop Dressing Phase of {$\mathcal{N}=\mathord{}$4} SYM''},
\textsf{Phys.~Rev.~D76,~046002~(2007)},
\href{http://arXiv.org/abs/arXiv:0705.0321}{\texttt{arXiv:0705.0321}} \texttt{[hep-th]}.
%
\bibitem{Belitsky:2005bu}
A.~V.~Belitsky, G.~P.~Korchemsky and D.~M{\"u}ller,
\textit{``Integrability of two-loop dilatation operator in gauge theories''},
\textsf{Nucl.~Phys.~B735,~17~(2006)},
\href{http://arXiv.org/abs/arXiv:hep-th/0509121}{\texttt{arXiv:hep-th/0509121}%
}.
%
\bibitem{Zwiebel:2005er}
B.~I.~Zwiebel,
\textit{``{$\mathcal{N}=\mathord{}$4} SYM to two loops: Compact expressions for
  the non-compact symmetry algebra of the $\alg{su}(1,1|2)$ sector''},
\textsf{JHEP~0602,~055~(2006)},
\href{http://arXiv.org/abs/arXiv:hep-th/0511109}{\texttt{arXiv:hep-th/0511109}%
}.
%
\bibitem{Beisert:2007sk}
N.~Beisert and B.~I.~Zwiebel,
\textit{``{On Symmetry Enhancement in the $\alg{psu}(1,1|2)$ Sector of
  $\mathcal{N}=4$ SYM}''},
\textsf{JHEP~0710,~031~(2007)},
\href{http://arXiv.org/abs/arXiv:0707.1031}{\texttt{arXiv:0707.1031}} \texttt{[hep-th]}.
%
\bibitem{Zwiebel:2007th}
B.~I.~Zwiebel,
\textit{``The $\alg{psu}(1,1|2)$ Spin Chain of $\mathcal{N} = 4$ Supersymmetric
  Yang-Mills Theory''},
UMI-32-67426,
 \href{http://www.physics.princeton.edu/www/jh//grad/grad_theses/thesis_zwiebel_benjamin.pdf}{www.physics.princeton.edu/www/jh//grad/grad\_theses/thesis\_zwiebel\_benjami%
n.pdf}$\,$.
%
\bibitem{Beisert:2004ry}
N.~Beisert,
\textit{``The Dilatation Operator of {$\mathcal{N}=\mathord{}$4} Super
  Yang-Mills Theory and Integrability''},
\textsf{Phys.~Rept.~405,~1~(2004)},
\href{http://arXiv.org/abs/arXiv:hep-th/0407277}{\texttt{arXiv:hep-th/0407277}%
}.
%
\bibitem{Beccaria:2007cn}
M.~Beccaria,
\textit{``Anomalous dimensions at twist-3 in the $\alg{sl}(2)$ sector of
  {$\mathcal{N}=\mathord{}$4} SYM''},
\textsf{JHEP~0706,~044~(2007)},
\href{http://arXiv.org/abs/arXiv:0704.3570}{\texttt{arXiv:0704.3570}} \texttt{[hep-th]}.
%
\bibitem{Beisert:2005wv}
N.~Beisert and T.~Klose,
\textit{``Long-Range $\alg{gl}(n)$ Integrable Spin Chains and Plane-Wave Matrix
  Theory''},
\textsf{J.~Stat.~Mech.~06,~P07006~(2006)},
\href{http://arXiv.org/abs/arXiv:hep-th/0510124}{\texttt{arXiv:hep-th/0510124}%
}.
%
\bibitem{Kotikov:2002ab}
A.~V.~Kotikov and L.~N.~Lipatov,
\textit{``DGLAP and BFKL equations in the {$\mathcal{N}=\mathord{}$4}
  supersymmetric gauge theory''},
\textsf{Nucl.~Phys.~B661,~19~(2003)},
\href{http://arXiv.org/abs/arXiv:hep-ph/0208220}{\texttt{arXiv:hep-ph/0208220}%
}.
%
\bibitem{Serban:2004jf}
D.~Serban and M.~Staudacher,
\textit{``Planar {$\mathcal{N}=\mathord{}$4} gauge theory and the Inozemtsev
  long range spin chain''},
\textsf{JHEP~0406,~001~(2004)},
\href{http://arXiv.org/abs/arXiv:hep-th/0401057}{\texttt{arXiv:hep-th/0401057}%
}.
%
\bibitem{Sieg:2005kd}
C.~Sieg and A.~Torrielli,
\textit{``{Wrapping interactions and the genus expansion of the 2-point
  function of composite operators}''},
\textsf{Nucl.~Phys.~B723,~3~(2005)},
\href{http://arXiv.org/abs/arXiv:hep-th/0505071}{\texttt{arXiv:hep-th/0505071}%
}.
%
\bibitem{Fiamberti:2007rj}
F.~Fiamberti, A.~Santambrogio, C.~Sieg and D.~Zanon,
\textit{``{Wrapping at four loops in $\mathcal{N}=4$ SYM}''},
\href{http://arXiv.org/abs/arXiv:0712.3522}{\texttt{arXiv:0712.3522}} \texttt{[hep-th]}.
%
\bibitem{Keeler:2008ce}
C.~A.~Keeler and N.~Mann,
\textit{``{Wrapping Interactions and the Konishi Operator}''},
\href{http://arXiv.org/abs/arXiv:0801.1661}{\texttt{arXiv:0801.1661}} \texttt{[hep-th]}.
%
\bibitem{Eden:2007rd}
B.~Eden,
\textit{``{Boxing with Konishi}''},
\href{http://arXiv.org/abs/arXiv:0712.3513}{\texttt{arXiv:0712.3513}} \texttt{[hep-th]}.
%
\bibitem{Eden:2005ve}
B.~Eden, C.~Jarczak, E.~Sokatchev and Y.~S.~Stanev,
\textit{``Operator mixing in {$\mathcal{N}=\mathord{}$4} SYM: The Konishi
  anomaly revisited''},
\textsf{Nucl.~Phys.~B722,~119~(2005)},
\href{http://arXiv.org/abs/arXiv:hep-th/0501077}{\texttt{arXiv:hep-th/0501077}%
}.
%
\bibitem{Kotikov:2004er}
A.~V.~Kotikov, L.~N.~Lipatov, A.~I.~Onishchenko and V.~N.~Velizhanin,
\textit{``Three-loop universal anomalous dimension of the Wilson operators in
  {$\mathcal{N}=\mathord{}$4} SUSY Yang-Mills model''},
\textsf{Phys.~Lett.~B595,~521~(2004)},
\href{http://arXiv.org/abs/arXiv:hep-th/0404092}{\texttt{arXiv:hep-th/0404092}%
}.
%
\bibitem{Moch:2004pa}
S.~Moch, J.~A.~M.~Vermaseren and A.~Vogt,
\textit{``The three-loop splitting functions in QCD: The non-singlet case''},
\textsf{Nucl.~Phys.~B688,~101~(2004)},
\href{http://arXiv.org/abs/arXiv:hep-ph/0403192}{\texttt{arXiv:hep-ph/0403192}%
}.
%
\bibitem{Belitsky:2007zp}
A.~V.~Belitsky,
\textit{``{Baxter equation for long-range SL(2$|$1) magnet}''},
\textsf{Phys.~Lett.~B650,~72~(2007)},
\href{http://arXiv.org/abs/arXiv:hep-th/0703058}{\texttt{arXiv:hep-th/0703058}%
}.
%
\bibitem{Heckenberger:2007ry}
I.~Heckenberger, F.~Spill, A.~Torrielli and H.~Yamane,
\textit{``{Drinfeld second realization of the quantum affine superalgebras of
  $D^{(1)}(2,1:x)$ via the Weyl groupoid}''},
\href{http://arXiv.org/abs/arXiv:0705.1071}{\texttt{arXiv:0705.1071}} \texttt{[math.QA]}.
%
\bibitem{Matsumoto:2008ww}
T.~Matsumoto and S.~Moriyama,
\textit{``{An Exceptional Algebraic Origin of the AdS/CFT Yangian Symmetry}''},
\textsf{JHEP~0804,~022~(2008)},
\href{http://arXiv.org/abs/arXiv:0803.1212}{\texttt{arXiv:0803.1212}} \texttt{[hep-th]}.
%
\bibitem{Drinfeld:1985rx}
V.~G.~Drinfeld,
\textit{``Hopf algebras and the quantum Yang-Baxter equation''},
\textsf{Sov.~Math.~Dokl.~32,~254~(1985)}.
%
\bibitem{Dolan:2003uh}
L.~Dolan, C.~R.~Nappi and E.~Witten,
\textit{``A Relation Between Approaches to Integrability in Superconformal
  Yang-Mills Theory''},
\textsf{JHEP~0310,~017~(2003)},
\href{http://arXiv.org/abs/arXiv:hep-th/0308089}{\texttt{arXiv:hep-th/0308089}%
}.
%
\bibitem{Dolan:2004ps}
L.~Dolan, C.~R.~Nappi and E.~Witten,
\textit{``Yangian symmetry in $D=$4 superconformal Yang-Mills theory''},
\href{http://arXiv.org/abs/arXiv:hep-th/0401243}{\texttt{arXiv:hep-th/0401243}%
},
in: \textit{``Quantum theory and symmetries''},
ed.: P.~C.~Argyres et~al.,
World Scientific (2004),
Singapore.
%
\bibitem{Gomez:2006va}
C.~Gomez and R.~Hern\'andez,
\textit{``The magnon kinematics of the AdS/CFT correspondence''},
\textsf{JHEP~0611,~021~(2006)},
\href{http://arXiv.org/abs/arXiv:hep-th/0608029}{\texttt{arXiv:hep-th/0608029}%
}.
%
\bibitem{Plefka:2006ze}
J.~Plefka, F.~Spill and A.~Torrielli,
\textit{``{On the Hopf algebra structure of the AdS/CFT S-matrix}''},
\textsf{Phys.~Rev.~D74,~066008~(2006)},
\href{http://arXiv.org/abs/arXiv:hep-th/0608038}{\texttt{arXiv:hep-th/0608038}%
}.
%
\bibitem{Beisert:2006qh}
N.~Beisert,
\textit{``The Analytic Bethe Ansatz for a Chain with Centrally Extended
  $\alg{su}(2|2)$ Symmetry''},
\textsf{J.~Stat.~Mech.~07,~P01017~(2007)},
\href{http://arXiv.org/abs/arXiv:nlin.SI/0610017}{\texttt{arXiv:nlin.SI/061001%
7}}.
%
\bibitem{Gomez:2007zr}
C.~Gomez and R.~Hern\'andez,
\textit{``Quantum deformed magnon kinematics''},
\textsf{JHEP~0703,~108~(2007)},
\href{http://arXiv.org/abs/arXiv:hep-th/0701200}{\texttt{arXiv:hep-th/0701200}%
}.
%
\bibitem{Torrielli:2007mc}
A.~Torrielli,
\textit{``Classical r-matrix of the $\alg{su}(2|2)$ SYM spin-chain''},
\textsf{Phys.~Rev.~D75,~105020~(2007)},
\href{http://arXiv.org/abs/arXiv:hep-th/0701281}{\texttt{arXiv:hep-th/0701281}%
}.
%
\bibitem{Beisert:2007ds}
N.~Beisert,
\textit{``{The S-Matrix of AdS/CFT and Yangian Symmetry}''},
\textsf{PoS~SOLVAY,~002~(2006)},
\href{http://arXiv.org/abs/arXiv:0704.0400}{\texttt{arXiv:0704.0400}} \texttt{[nlin.SI]}.
%
\bibitem{Young:2007wd}
C.~A.~S.~Young,
\textit{``q-Deformed Supersymmetry and Dynamic Magnon Representations''},
\textsf{J.~Phys.~A40,~9165~(2007)},
\href{http://arXiv.org/abs/arXiv:0704.2069}{\texttt{arXiv:0704.2069}} \texttt{[hep-th]}.
%
\bibitem{Moriyama:2007jt}
A.~Moriyama,~Sanefumi~andTorrielli,
\textit{``A Yangian Double for the AdS/CFT Classical r-matrix''},
\textsf{JHEP~0706,~083~(2007)},
\href{http://arXiv.org/abs/arXiv:0706.0884}{\texttt{arXiv:0706.0884}} \texttt{[hep-th]}.
%
\bibitem{Matsumoto:2007rh}
T.~Matsumoto, S.~Moriyama and A.~Torrielli,
\textit{``{A Secret Symmetry of the AdS/CFT S-matrix}''},
\textsf{JHEP~0709,~099~(2007)},
\href{http://arXiv.org/abs/arXiv:0708.1285}{\texttt{arXiv:0708.1285}} \texttt{[hep-th]}.
%
\bibitem{Beisert:2007ty}
N.~Beisert and F.~Spill,
\textit{``{The Classical r-matrix of AdS/CFT and its Lie Bialgebra
  Structure}''},
\href{http://arXiv.org/abs/arXiv:0708.1762}{\texttt{arXiv:0708.1762}} \texttt{[hep-th]}.
%
\bibitem{Spill:2008tp}
F.~Spill and A.~Torrielli,
\textit{``{On Drinfeld's second realization of the AdS/CFT $\alg{su}(2|2)$
  Yangian}''},
\href{http://arXiv.org/abs/arXiv:0803.3194}{\texttt{arXiv:0803.3194}} \texttt{[hep-th]}.
%
\bibitem{deLeeuw:2008dp}
M.~de~Leeuw,
\textit{``{Bound States, Yangian Symmetry and Classical r-matrix for the
  $AdS_5\times S^5$ Superstring}''},
\href{http://arXiv.org/abs/arXiv:0804.1047}{\texttt{arXiv:0804.1047}} \texttt{[hep-th]}.
%
\bibitem{Torrielli:2008wi}
A.~Torrielli,
\textit{``{Structure of the string R-matrix}''},
\href{http://arXiv.org/abs/arXiv:0806.1299}{\texttt{arXiv:0806.1299}} \texttt{[hep-th]}.
%
\bibitem{Ihry:2008gm}
J.~N.~Ihry,
\textit{``{Yangians in Deformed Super Yang-Mills Theories}''},
\textsf{JHEP~0804,~051~(2008)},
\href{http://arXiv.org/abs/arXiv:0802.3644}{\texttt{arXiv:0802.3644}} \texttt{[hep-th]}.
%
\bibitem{Agarwal:2004sz}
A.~Agarwal and S.~G.~Rajeev,
\textit{``Yangian symmetries of matrix models and spin chains: The dilatation
  operator of {$\mathcal{N}=\mathord{}$4} SYM''},
\textsf{Int.~J.~Mod.~Phys.~A20,~5453~(2005)},
\href{http://arXiv.org/abs/arXiv:hep-th/0409180}{\texttt{arXiv:hep-th/0409180}%
}.
%
\bibitem{Agarwal:2005ed}
A.~Agarwal,
\textit{``Comments on higher loop integrability in the SU(1$|$1) sector of
  {$\mathcal{N}=\mathord{}$4}: Lessons from the SU(2) sector''},
\href{http://arXiv.org/abs/arXiv:hep-th/0506095}{\texttt{arXiv:hep-th/0506095}%
}.
%
\bibitem{Beisert:2007jv}
N.~Beisert and D.~Erkal,
\textit{``{Yangian Symmetry of Long-Range $\alg{gl}(N)$ Integrable Spin
  Chains}''},
\textsf{J.~Stat.~Mech.~0803,~P03001~(2008)},
\href{http://arXiv.org/abs/arXiv:0711.4813}{\texttt{arXiv:0711.4813}} \texttt{[hep-th]}.
%
\bibitem{Zwiebel:2006cb}
B.~I.~Zwiebel,
\textit{``Yangian symmetry at two-loops for the $\alg{su}(2|1)$ sector of
  {$\mathcal{N}=\mathord{}$4} SYM''},
\textsf{J.~Phys.~A40,~1141~(2007)},
\href{http://arXiv.org/abs/arXiv:hep-th/0610283}{\texttt{arXiv:hep-th/0610283}%
}.
%
\bibitem{Rej:2005qt}
A.~Rej, D.~Serban and M.~Staudacher,
\textit{``Planar {$\mathcal{N}=\mathord{}$4} Gauge Theory and the Hubbard
  Model''},
\textsf{JHEP~0603,~018~(2006)},
\href{http://arXiv.org/abs/arXiv:hep-th/0512077}{\texttt{arXiv:hep-th/0512077}%
}.
%
\bibitem{Lipatov:1976zz}
L.~N.~Lipatov,
\textit{``{Reggeization of the Vector Meson and the Vacuum Singularity in
  Nonabelian Gauge Theories}''},
\textsf{Sov.~J.~Nucl.~Phys.~23,~338~(1976)}.
%
\bibitem{Kuraev:1977fs}
E.~A.~Kuraev, L.~N.~Lipatov and V.~S.~Fadin,
\textit{``{The Pomeranchuk Singularity in Nonabelian Gauge Theories}''},
\textsf{Sov.~Phys.~JETP~45,~199~(1977)}.
%
\bibitem{Balitsky:1978ic}
I.~I.~Balitsky and L.~N.~Lipatov,
\textit{``{The Pomeranchuk Singularity in Quantum Chromodynamics}''},
\textsf{Sov.~J.~Nucl.~Phys.~28,~822~(1978)}.
%
\bibitem{Balitsky:1987bk}
I.~I.~Balitsky and V.~M.~Braun,
\textit{``{Evolution Equations for QCD String Operators}''},
\textsf{Nucl.~Phys.~B311,~541~(1989)}.
%
\bibitem{Braun:2003rp}
V.~M.~Braun, G.~P.~Korchemsky and D.~Muller,
\textit{``The uses of conformal symmetry in QCD''},
\textsf{Prog.~Part.~Nucl.~Phys.~51,~311~(2003)},
\href{http://arXiv.org/abs/arXiv:hep-ph/0306057}{\texttt{arXiv:hep-ph/0306057}%
}.
%
\bibitem{Hofman:2008ar}
D.~M.~Hofman and J.~Maldacena,
\textit{``{Conformal collider physics: Energy and charge correlations}''},
\href{http://arXiv.org/abs/arXiv:0803.1467}{\texttt{arXiv:0803.1467}} \texttt{[hep-th]}.
%
\end{thebibliography}

\addtolength\textheight{12pt} \addtolength\topmargin{-6pt}

\end{document}